\documentclass[11pt]{article}
\pdfoutput=1

\usepackage{jheppub}
\usepackage{amsmath,amssymb,amsfonts}
\usepackage{amsthm}
\usepackage{breqn}

\usepackage{esint}
\usepackage{cancel}
\usepackage{graphicx,tikz}
\usetikzlibrary{decorations.pathreplacing}
\usetikzlibrary{decorations.pathmorphing}
\usetikzlibrary{arrows}
 \usepackage{epstopdf}
\usepackage{framed}
\usepackage{xcolor}
\colorlet{shadecolor}{gray!20}
\usepackage{mathtools}
\usepackage{physics}
\usepackage{wrapfig}
\usepackage{enumitem}

\usepackage{mathrsfs}
\usepackage{setspace}
\definecolor{hyperref}{RGB}{026,028,087}

\usepackage{placeins}
\numberwithin{equation}{section}

\usepackage{bbm}
\usepackage[framemethod=TikZ]{mdframed}
\mdfdefinestyle{boxed}{%
linecolor=red,
roundcorner=5pt,
innerleftmargin=0,%
innerrightmargin=0,
backgroundcolor=gray!10,%
}

\def \tdelta {\tilde \delta}
\newcommand{\bfk}{{\bf  k}}
\newcommand{\bfp}{{\bf  p}}

\newcommand{\bfq}{{\bf  q}}
\newcommand{\bfx}{{\bf  x}}

\def\disc{\text{Disc}}

\newcommand{\changg}[1]{{#1}}

\begin{document}

\title{Cosmological Cutting Rules}

\author[a]{Scott Melville}
\affiliation[a]{Department of Applied Mathematics and Theoretical Physics, University of Cambridge, Wilberforce Road, Cambridge CB3 0WA, U.K.}

\author[a]{and Enrico Pajer}

\emailAdd{scott.melville@damtp.cam.ac.uk}
\emailAdd{enrico.pajer@gmail.com}

\abstract{
Primordial perturbations in our universe are believed to have a quantum origin, and can be described by the wavefunction of the universe (or equivalently, cosmological correlators). 
It follows that these observables must carry the imprint of the founding principle of quantum mechanics: unitary time evolution. 
Indeed, it was recently discovered that unitarity implies an infinite set of relations among tree-level wavefunction coefficients, dubbed the Cosmological Optical Theorem. 
Here, we show that unitarity leads to a systematic set of ``Cosmological Cutting Rules'' which constrain wavefunction coefficients for any number of fields and to \textit{any loop order}. 
These rules fix the discontinuity of an $n$-loop diagram in terms of lower-loop diagrams and the discontinuity of tree-level diagrams in terms of tree-level diagrams with fewer external fields. 
Our results apply with remarkable generality, namely for arbitrary interactions of fields of any mass and any spin with a Bunch-Davies vacuum around a very general class of FLRW spacetimes. 
As an application, we show how one-loop corrections in the Effective Field Theory of inflation are fixed by tree-level calculations and discuss related perturbative unitarity bounds.
These findings greatly extend the potential of using unitarity to bootstrap cosmological observables and to restrict the space of consistent effective field theories on curved spacetimes.
}

\maketitle

%%%%%%%%%%%%%%%%%%%%%%%%%%%%%%%%
\section{Introduction}
%%%%%%%%%%%%%%%%%%%%%%%%%%%%%%%%

Unitarity is a central pillar of quantum mechanics. On the one hand, the positive norm of states in the Hilbert space is essential for ensuring that probabilities are positive. On the other hand, a unitary time evolution ensures that the total probability is conserved and hence the theory can make consistent statistical predictions for observables. In quantum field theory on flat spacetime, several general properties and relations are known to follow from unitarity (see e.g. \cite{Schwartz:2013pla}). For example, $n$-point correlators must factorize into products of lower order ones in particular kinematic limits. Through the LSZ reduction formula, this in turn leads to the factorization of amplitudes and the positivity of factorization coefficients. An even more general consequence of unitarity is the Optical Theorem, which constrains amplitudes for generic values of the kinematic variables. The non-linear nature of the Optical Theorem is particularly useful in perturbation theory because it allows one to fix higher-order amplitudes in terms of lower-order ones. In its most basic implementation, this allows one to fix the imaginary part of one-loop diagrams in terms of tree-level ones. 
While the Optical Theorem is a fully non-perturbative result, it is oftentimes useful to know how it is satisfied order by order in perturbation theory---this is given by Cutkosky's Cutting Rules \cite{Cutkosky:1960sp} (see also \cite{tHooft:1973wag,Veltman:1994wz} for a pedagogical derivation). In a nutshell, these rules tell us how to compute the discontinuity of a given loop amplitude across one of its branch cuts using some modified Feynman rules, in which the propagators of particles responsible for the discontinuity are substituted with delta functions that put their four-momenta on-shell. It is important to notice that in all of the above cases, one manages to express the rather formal condition of unitary time evolution in terms of a constraint on physical observables, namely amplitudes in this case.\\

% unitarity in dS and the COT

Somewhat surprisingly, until a few months ago an analogous understanding of the implications of unitarity was missing in the case of cosmological spacetimes and primordial correlators. In this paper\footnote{A complementary discussion of cutting rules in cosmology will appear in \cite{single} with emphasis on extensions to massive and spinning fields beyond de Sitter at tree level, and a number of non-trivial checks.} we fill this gap and derive \emph{Cosmological Cutting Rules}, which, in analogy with their flat space counterpart, consists of a set of unitarity conditions to be satisfied order by order in perturbation theory. The natural observable for which these conditions are formulated is the wavefunction of the universe. If desired these can be translated into constraints on correlators. However, in the most general case (e.g. without restricting to massless scalar field), the wavefunction expressions are much more compact. Our results build upon a recently derived Cosmological Optical Theorem \cite{Goodhew:2020hob}, and the associated conserved quantities of \cite{Cespedes:2020xqq}. The main insight of this work has been to recognize that the Hermitian conjugate time evolution, $U^\dagger$, appearing in the iconic unitarity condition $UU^\dagger=1$, can be related to a specific analytic continuation of the wavefunction of the universe, with the same boundary conditions (the Bunch-Davies vacuum in most practical applications). This is highly non-trivial. Naively one might have expected that the imprint of quantum mechanics limits itself to the non-commutation of fields with their conjugate momenta. If this were the case, unitarity would be a weak constraint because the natural cosmological observables associated with the conjugate momenta decay exponentially with (cosmological) time during inflation and are therefore practically unobservable. Instead, the Cosmological Optical Theorem tells us that \textit{the quantum mechanical origin of perturbations manifests itself in a very specific analytic structure of the wavefunction}. Recall that the boundary wavefunction encodes the correlation of fields at the same time and at separated spatial points. From this point of view there isn't a priori a natural expectation of what unitarity would mean for such an object because time has completely disappeared. This is in stark contrast with what happens in AdS, where the CFT on the boundary still has a standard notion of time and of the associated unitarty evolution (see \cite{Meltzer:2020qbr} for progress on the cutting rules in AdS). It is therefore quite remarkable to finally discover how time evolution is hidden in the spatial correlation at the boundary of de Sitter.\\

%Comparison with Cutkosky's rules
One might hope that cutting rules in cosmology can be derived in complete analogy with flat spacetime, but this is unfortunately not the case beyond tree-level diagrams. In flat space, the cutting rules can be derived from a master ``largest time equation" \cite{Veltman:1963th,tHooft:1973wag,Veltman:1994wz}. An analogous formula can be derived for the bulk-to-bulk propagator appearing in the calculation of the wavefunction of the universe, a close relative of the Feynman propagator. However, such a procedure does not map directly to the standard representation of wavefunction coefficient in terms of bulk time integrals. The main obstacle is that, when computing a wavefunction either in Minkowski or in FLRW spacetimes, we need to adjust the propagator to account for the presence of a boundary corresponding to the time at which the wavefunction is computed. It is the presence of the associated boundary term in the (bulk-to-bulk) propagator that makes the Cosmological Cutting Rule look quantitatively different from their flat spacetime analogue. Away from the boundary, i.e. in the so-called vanishing total energy limit, our cutting rules should reduce to the well-known ones for amplitudes. However, the Cosmological Cutting Rules encode more information. Indeed, as it will be discussed elsewhere, while one kinematical limit of the Cosmological Optical Theorem produces the standard Optical Theorem, a different kinematical limit leads to the factorization theorems at the heart of on-shell methods for amplitudes (see e.g. \cite{Benincasa:2013faa,Elvang:2013cua,Cheung:2017pzi}).\\

%cutting rules, single cutting and derivative representation
It is interesting to ask which \textit{types of functions} can appear in the final result for the wavefunction coefficients in perturbation theory. For comparison, we know that amplitudes at tree level only involve rational functions of the momenta (and the spinor helicity variables for spinning fields). Logarithm, polyogarithm and their associated branch cuts appear at loop level. Things are unfortunately more complicated in cosmology and the reason can be traced back to the absence of time translation invariance (even the maximally symmetric de Sitter does not have a globally defined time-like Killing vector). Indeed, even at tree level, for a diagram with $V$ vertices we have to perform $V$ \textit{nested} integrals in time and even starting with simple mode functions such as those for massless and conformally couples scalar fields (see \eqref{modefct}), we can end up with polylogarithms (see e.g. \cite{Arkani-Hamed:2015bza,Hillman:2019wgh}). From this perspective, the Cosmological Cutting Rules can be thought of as identifying which parts of wavefunction coefficients can be formulated in terms of "simpler" functions. In particular, cutting rules tell us that a specific discontinuity can be computed in terms of diagrams with one or more fewer time integrals, which feature functions with a lower transcendental weight i.e. closer to the starting mode functions (e.g. in the sense of "the symbol" \cite{Arkani-Hamed:2018bjr,Hillman:2019wgh}). Related to this, it would be interesting to see if our relations have a natural avatar in the cosmological polytope representation of the wavefunction \cite{Arkani-Hamed:2017fdk,Arkani-Hamed:2018bjr,Benincasa:2018ssx,Benincasa:2019vqr}.\\

%cutting rules as a practical tool
The Cosmological Cutting Rules we derive are a very useful practical tool to derive certain effects of quantum loops while performing only tree level calculations. This is particularly useful in cosmology where, due to the absence of time translation invariance, calculations become computationally demanding very quickly. For example, starting with \cite{Weinberg:2005vy} much attention has been devoted to loop corrections during inflation. The simplest possible case is a correction to the power spectrum, which at least naively has a chance to be sizable in general class of models that are capture by the Effective Field Theory of inflation \cite{Cheung:2007st}. The cutting rules allow one to preform these calculation with much less effort than with the direct bulk integration, as we will see in Section \ref{sec:applications}.\\

%unitarity and the bootstrap
Dulcis in fundo, we discuss the "\textit{bootstrap}" approach, namely the prospect of using the powerful constraints of unitarity, combined with other basic principles such as locality, the choice of vacuum and symmetries as a computational tool to derive observables, and potentially bypass the traditional bulk in-in calculation. This approach has a demonstrated track record for the calculation of amplitudes \cite{Benincasa:2013faa,Elvang:2013cua,Cheung:2017pzi}, and has gained much traction in the cosmological context. Indeed, in the presence of a high degree of symmetry, such as Poincar\'e invariance in Minkowski, very general results can be derived, such as for example the classification of all consistent cubic amplitude for particles of any spin (see e.g. \cite{Benincasa:2007xk,McGady:2013sga,Arkani-Hamed:2017jhn}). Already in this context, relaxing the amount of symmetry opens the door for many new and relative unexplored possibilities. For example, in \cite{Pajer:2020wnj}, all consistent cubic amplitudes were derived allowing for spontaneously (non-linearly realized) or explicitly broken Lorentz boosts, as relevant for many systems of interest including all conceivable cosmological backgrounds. Similarly, when restricting to the most symmetric spacetime relevant for cosmology, namely de Sitter, very general results can be obtained, as for example various combinations of scalar and graviton correlators \cite{Maldacena:2011nz,Creminelli:2011mw,Kehagias:2012pd,Mata:2012bx,Ghosh:2014kba,Kundu:2014gxa,Kundu:2015xta,Pajer:2016ieg,Arkani-Hamed:2018kmz,Baumann:2019oyu,Baumann:2020dch}. When combined with much progress on the front of perturbative calculations \cite{Bzowski:2013sza,Arkani-Hamed:2015bza,Sleight:2019hfp,Sleight:2020obc,Sleight:2019mgd,Arkani-Hamed:2017fdk,Baumgart:2019clc,Bzowski:2019kwd,Baumgart:2020oby}, these powerful symmetry-based results have given us a much better understanding of general structures that appear in the wavefunction coefficients, such as \textit{singularities} and the \textit{analytic structure}. At the same, if we want to make connection with observations we absolutely need a "\textit{boostless}" bootstrap approach where we relax the requirement of invariance under de Sitter boosts, since such symmetries are incompatible with large non-Gaussianity in single field inflation \cite{Green:2020ebl}. Very promising results in this direction have already been derived using constraints from factorization \cite{Baumann:2020dch}, the formulation of very general boostless \textit{Bootstrap Rules} \cite{Pajer:2020wnj}, and the recently derived Manifestly Local Test and partial-energy recursion relations \cite{MLT}. From this perspective our Cosmological Cutting Rules add a powerful tool to bootstrap in full generality higher order correlators form lower order ones, and in particular exchange and loop diagrams.  \\

%%%%%%%%%%%%%%%%
\subsection{Summary of Results}
%%%%%%%%%%%%%%%%
 For the convenience of the reader, we provide below a summary of our main results.
 \begin{itemize}
     \item We derive Cosmological Cutting Rules for the wavefunction coefficients $\psi_n$ for any number of external legs and to any loop order. The rules are as follows (see Section \ref{sec:theorem} for a more formal discussion):
     \begin{itemize}
         \item For any particular diagram $D$ that contributes $\psi_n^{(D)}$ to a wavefunction coefficient, sum over all possible $2^{I}$ ways to cut its $I$ internal lines. This may divide $D$ into a set of disconnected subdiagrams, each with associated $\psi^{(\rm subdiagram)}$.
         \item Take the discontinuity (defined in \eqref{defdisc}) of all possible subdiagrams by analytically continuing all external legs except those arising from the cutting of an internal line.
         \item For every cut line add a factor of the power spectrum $P$, and then integrate over all cut momenta (which now flow to the boundary).
     \end{itemize}
     Schematically, this procedure results in the following constraints, which we call \textit{Cosmological Cutting Rules} (see \eqref{statement})
     \begin{align} 
   i \, \underset{\substack{ \text{internal} \\ \text{lines} } }{\disc }  \left[ i \, \psi^{(D)} \right]     &=  \sum_{\rm cuts}   \left[ \prod_{\substack{\rm cut \\ \rm momenta}} \int P \right]
   \prod_{\rm subdiagrams} (-i) \underset{\substack{ \text{internal } \& \\  \text{ cut lines} } }{\disc } \left[ i \, \psi^{(\rm subdiagram)} \right]\,,
    \end{align}
    \item Since sometimes a picture is worth a thousand words, we provide an example in the figure below, where we state the Cosmological Cutting Rules graphically. The double vertical red lines denote a cut. The discontinuity is taken of every disconnected diagram with argument given by all highlighted lines (in orange). According to our definition of $\disc$ in \eqref{defdisc}, the arguments of $\disc$ are just spectators, i.e. they are \textit{not} analytically continued. Throughout this work internal lines are never analytically continued.
     \item We provide several explicit examples of the Cosmological Cutting Rules at tree level, for one (see Section \ref{sec:cutting_one} reproducing the Cosmological Optical Theorem of \cite{Goodhew:2020hob}) and two (see Section \ref{sec:cutting_two}) internal lines, as well as two one-loop examples (see Section \ref{sec:cutting_loop}).
     \item Our Cosmological Cutting Rules are valid very generally. In particular, they apply to fields of any mass and spin with arbitrary interactions (provided they are local in time and compatible with Hermitial analytiticy, which is the case for all common interactions). To account for these cases, the above rules can be simply modified by assuming that each internal or external momentum $\bfk$ or $\bfp$ carries additional quantum numbers, such as the type of field, its spin and possible charges. By keeping all the polarization tensors in the vertices, the derivation of the Cosmological Cutting Rules reduces straightforwardly to the case of scalar fields. Indeed, our results are valid in a large class of FLRW spacetimes, including de Sitter, slow-roll inflation and all power law cosmologies, as long as all fields satisfy the Bunch Davies vacuum. These generalizations are reviewed in Section \ref{sec:multiple}, and we refer the reader to \cite{single} for more details. 
     \item As applications of our newly derived relations, in Section \ref{sec:applications} we show how to compute certain loop corrections from tree-level results in a series of physically relevant cases. First, in Minkowski we compute the one-loop correction to the power spectrum from $\lambda \phi^4$ and $\lambda \phi^3$ interactions, respectively. Second, around quasi de Sitter spacetime, we consider the leading cubic interactions in the effective field theory of inflation, and compute the induced one-loop correction to the power spectrum. In the case of $\dot \pi^3$, we confirm the result of \cite{Senatore:2009cf}, while our results for $\dot \pi \partial \pi^2$ are new. 
     \item In the bulk of the paper we use the path integral representation of the wavefunction of the universe, which allows us to find a general result to all loop orders. In Appendix~\ref{app:schrodinger}, we present the connection with the Schr\"{o}dinger picture of \cite{Cespedes:2020xqq}, and show some examples of one-loop cutting rules. 
 \end{itemize}

    \FloatBarrier
\begin{figure}
\centering
\includegraphics[width=1.0\textwidth]{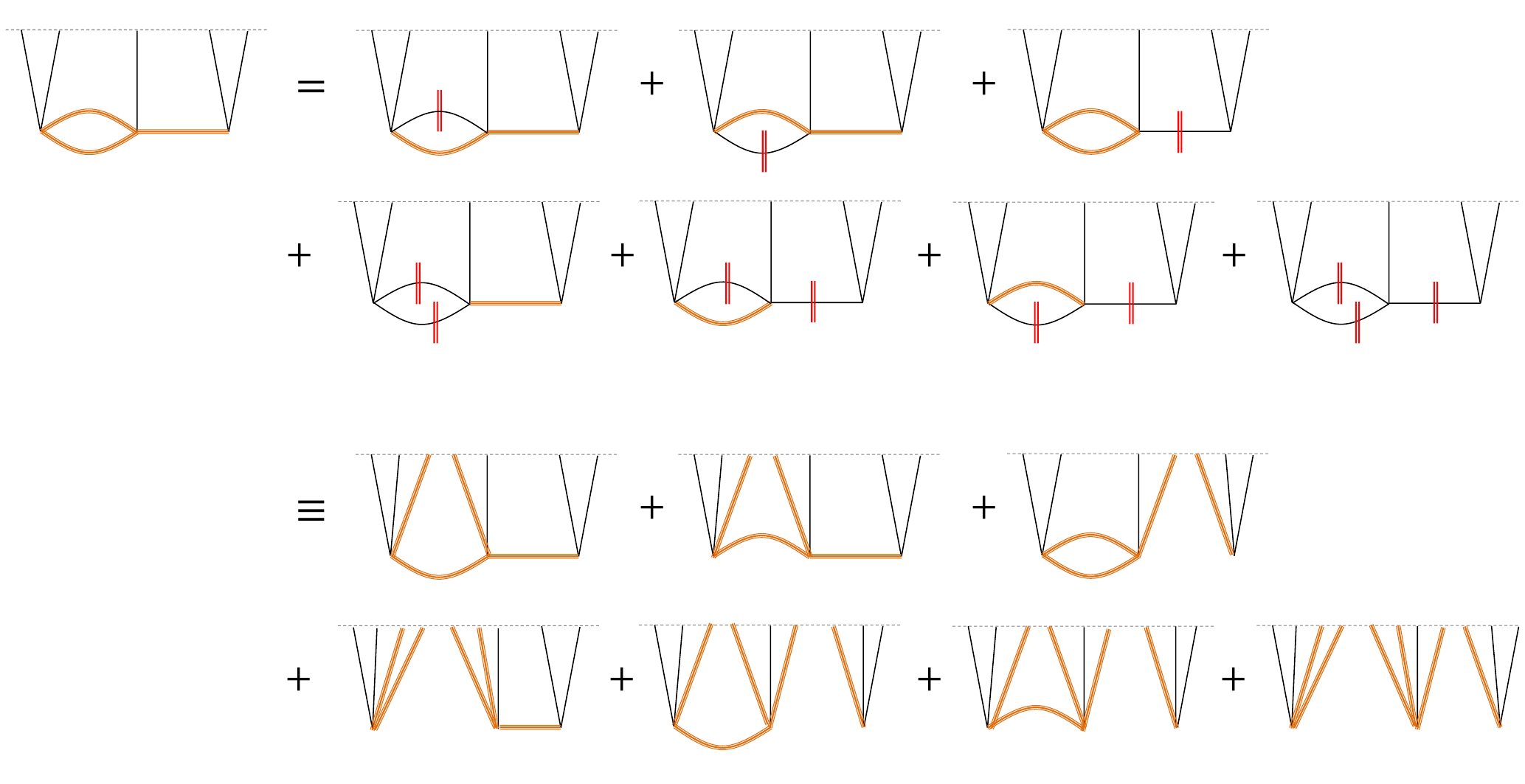}
\caption{An example of the Cosmological Cutting Rules, applied to a particular diagram that contributes to the wavefunction of the universe.}
\end{figure}
\FloatBarrier
%

%%%%%%%%%%%%%%%%
\paragraph{Notation and conventions:} The spatial Fourier transformation is defined by, 
\begin{align}
f(\bfx)&=\int \dfrac{d^3\bfk}{(2\pi)^3}{f}(\bfk)\exp(i\bfk\cdot\bfx)\equiv\int_{\bfk}{f}(\bfk)\exp(i\bfk\cdot\bfx) \,,
\end{align}
and commutes with time derivatives.
Bold letters to refer to vectors, e.g. $\bfk$, and we write their magnitude as $k\equiv |\bfk|$. 
A prime on a wave function coefficient or correlator denotes that we have extracted the overall momentum-conserving delta function, 
\begin{align}
\psi_n (\bfk_1,\dots ,\bfk_n)& \equiv \psi_n' (\bfk_1,\dots ,\bfk_n) \; (2\pi)^3 \delta^3 \left(\sum \bfk_a \right)\\
&\equiv \psi_n' (\bfk_1,\dots ,\bfk_n) \; \tdelta^3 \left(\sum \bfk_a \right) \,.
\end{align}
Note that, unfortunately, our conventions for $\psi_n$ differ from those in \cite{Goodhew:2020hob,single} by a minus sign, $\psi_n^{\text{here}}=-\psi_n^{\text{there}}$. We apologize for the inconvenience that this might cause. 

When discussing functions of four momenta, such as $\psi_4 ( \bfk_1 , \bfk_2 , \bfk_3 , \bfk_4 )$, it will be convenient to use the variables,
\begin{align}
 p_s = | \bfk_1 + \bfk_2 |\,, \;\; p_t = | \bfk_1 + \bfk_3| \,, \;\; p_u = | \bfk_1 + \bfk_4 | \,,
\end{align}
which are related by $p_s^2 + p_t^2 + p_u^2 = \sum_{a=1}^4 k_{a}^2$ (so only six of  the seven variables $k_a, p_s, p_t, p_u$ are independent). 
For general $n$-point wavefunction coefficients, we adopt a convention in which $\bfk_a$ label momenta of external legs, $\bfp_a$ label the momenta of internal legs, and $\bfq_a$ is reserved for dummy integration variables (which arise after performing every cut). This conventions is encoded in the way we write the arguments of the wavefunction coefficients, namely
\begin{align}
    \psi_n&=\psi_n(\text{ external energies ; internal energies ; contractions })\\
    &=\psi_n(k_1,\dots,k_n;p_1,\dots,p_I; \bfk_a\cdot \bfk_b, \bfk_a \cdot (\bfk_b \times \bfk_c), \bfk_a \cdot \epsilon(\bfk_b), \dots )\\
    &\equiv \psi_n(\{k\};\{p\};\{\bfk\})\,,
\end{align}
where the last argument $\{\bfk\}$ denotes rotation-invariant contractions of the external momenta with $\delta_{ij}$, $\epsilon_{ijk}$ or with polarization tensors. We define the power spectrum as 
\begin{align}
    P_{\bfq \bfq'} := \langle \phi_{\bfq} \phi_{\bfq'} \rangle = P_{q} \; \tdelta^3 \left( \bfq + \bfq'  \right) \; .\label{Pk}
\end{align}

%%%%%%%%%%%%%%%%%%%%%%%%%%%%%%%%
\section{Feynman Rules for Wavefunction Coefficients}
\label{sec:rules}
%%%%%%%%%%%%%%%%%%%%%%%%%%%%%%%%

Consider a $(d+1)$-dimensional conformally flat spacetime, $ds^2 = a^2 (\eta) ( - d \eta^2 + d \bfx^2)$. 
We describe the state of the Universe and its fields, denoted collectively by $\phi$, at conformal time $\eta$ using the wavefunction $\Psi_\eta [ \phi ] = \langle \phi | \Psi_{\eta} \rangle$, where $| \phi \rangle$ is a field eigenstate. 
Starting from an initial state $| \Omega \rangle$ at early times, the state at a later time $\eta_0$ is given by $\Psi_{\eta_0} [ \phi ] = \langle \phi  | U (  \eta_0 , - \infty ) | \Omega \rangle$, where $U(\eta_2, \eta_1)$ is the unitary operator that implements time translations from $\eta_1$ to $\eta_2$. This can be computed using the path integral, 
\begin{align}
 \Psi_{\eta_0} [ \phi  ] = \int_{ \Omega  \; \text{at} \; \eta \to - \infty }^{\Phi  (\eta_0) = \phi  } \mathcal{D} \Phi  \; e^{i S [\Phi]}\,,
 \label{eqn:Psi_path_integral}
\end{align}
where $\Phi_{\bfk} (\eta)$ represents paths which coincide with $|\Omega \rangle$ at early times and end on the configuration $\phi$ at $\eta_0$, and $S [ \Phi ]$ is the corresponding classical action.

%%%%
\paragraph{Wavefunction Coefficients:}
%%%%
The wavefunction \eqref{eqn:Psi_path_integral} (a functional of the fields $\phi$) is conveniently represented in terms of \emph{wavefunction coefficients}\footnote{
Explicitly (the choice of sign is the same as in \cite{Maldacena:2002vr}),
\begin{align}
\Psi [ \phi  ] &\propto \exp \left[+ 
 \sum_{n=2}^{\infty} \int_{\bfk_1,..,\bfk_n}\frac{1}{n!} \psi_{ \bfk_1 ... \bfk_n }  \phi_{\bfk_1} ... \phi_{\bfk_n}  \right]\;  .
\end{align}
},
\begin{align}
\psi_{ \bfk_1  \dots \bfk_n } (\eta_0) = \frac{1}{\Psi_{\eta_0} [0] } \frac{\delta^n}{ \delta \phi_{\bfk_1} \dots \delta \phi_{\bfk_n} } \log \, \Psi_{\eta_0} [ \phi ] \; \bigg|_{\phi  = 0 }  \; .
\label{eqn:psi_def}
\end{align}
which are functions of time (and momenta) only. 
For brevity we will not explicitly write the dependence on the time $\eta_0$, at which the state is defined.

%%%%
\paragraph{Propagators:}
%%%%
The wavefunction coefficients \eqref{eqn:psi_def} can be computed perturbatively in a diagrammatic expansion analogous to the usual Feynman diagrams used to compute the partition function (sometimes called Witten diagrams in analogy with the AdS/CFT calculation \cite{Witten:1998qj}).
To do this, one first identifies the classical field configurations (saddle points of $S[\Phi ]$) which dominate the path integral. These solve the equations of motion $\delta S [\Phi ]/\delta \Phi_{\bfk} =0$ subject to the boundary conditions, 
\begin{align}
\lim_{\eta\to -\infty(1-i\epsilon)} \Phi_{\bfk} (\eta) =0 \qquad \text{and} \qquad \Phi_{\bfk} (\eta_0 )= \phi_{\bfk} \, ,
\end{align}
which corresponds to projecting onto the free vacuum in the asymptotic past\footnote{
The free vacuum is annihilated by $\hat{a}_{\bfk}  \propto  \omega_k  \hat{\phi}_{-\bfk} + i \hat{\Pi}_{-\bfk}$, and so this condition can also be written as,
\begin{align}
\lim_{\eta \to - \infty} \left( \Pi_{-\bfk} - i \omega_k \Phi_{-\bfk} \right) = 0 \; ,
\label{eqn:vacuum_condition}
\end{align}
where $ \Pi_{\bfk}$ is the conjugate momentum associated with the path $\Phi_{\bfk}$. For a canonical scalar field of mass $m$,  $\Pi_{\bfk} = a^{d-1} \partial_\eta \Phi_{\bfk}$ and $\omega_k^2  = a^{d+1} ( k^2/a^2 + m^2 )$ in \eqref{eqn:vacuum_condition}, which selects the behaviour $\Phi_{\bfk} (\eta) \sim \text{exp} ( + i k \eta )$ at early times.
}. 
%\SM{Need to explain the $i\epsilon$ prescription. Perhaps best way is to start with \eqref{eqn:vacuum_condition}, and then argue that an equivalent condition (since $\eta \to -\infty$ and $\Phi \sim e^{i k \eta}$) is to require that $\Phi \to 0$ for the deformed $k \eta \to k \eta (1 - i \epsilon)$. In this work we will take $k$ to be slightly complex, in particular $k$ always approaches the real axis from below.}
Writing the variation $\delta S[\Phi] / \delta \Phi_{\bfk}$ as,
\begin{align}
\mathcal{O}_k (\eta) \Phi_{\bfk} (\eta) =  - \frac{ \delta S_{\rm int} [ \Phi ]  }{\delta \Phi_{\bfk} (\eta) } \; , 
\end{align}
where $\mathcal{O}_k (\eta) \Phi_k$ denotes the linearized (exactly solvable) equations of motion and depends only on the magnitude of the momentum, solutions can be constructed perturbatively in the interactions $S_{\rm int}$. This requires two propagators: the \emph{bulk-to-boundary propagator} $K_k$ and the \emph{bulk-to-bulk propagator} $G_k$, which satisfy,
\begin{align}
\mathcal{O}_k ( \eta ) K_k ( \eta , \eta_0 )  =0 \; ,\qquad 
\mathcal{O}_k (\eta) G_k ( \eta ,\eta' , \eta_0) = -\delta (\eta-\eta')\,,
\label{eqn:propagator_def}
\end{align}
subject to the boundary conditions, 
\begin{align}
 &\, \; \lim_{\eta\rightarrow\eta_0} K_k  ( \eta, \eta_0 )=1,&  \, \; \lim_{\eta\to -\infty(1-i\epsilon)}K_k   ( \eta , \eta_0 )=0\,,\\ \nonumber
& \lim_{\eta \rightarrow\eta_0} G_k  (\eta,\eta', \eta_0)=0,&  \lim_{\eta \to -\infty(1-i\epsilon)} G_k   ( \eta,\eta', \eta_0)=0\, ,
\end{align}
and the symmetry condition $G_k  (\eta , \eta' , \eta_0) = G_k  ( \eta', \eta , \eta_0)$. 
The classical field configurations are then defined implicitly by the relation,
\begin{align}
  \Phi_{\bfk} (\eta )  = K_{k}  (\eta , \eta_0 ) \phi_{\bfk} + 
  \int d \eta'  G_{k} ( \eta , \eta' , \eta_0 ) \frac{\delta S_{\rm int} [ \Phi  ] }{\delta \Phi_{\bfk} (\eta' ) } \; . 
\end{align}
which can be solved perturbatively to any desired order in the interactions. As an aside, notice that $K_{k}(\eta)$ is completely analogous to the transfer functions and growth functions used in the study of perturbations of the large scale structures or of the cosmic microwave background. It would be interesting to see if the techniques developed here could be also useful in those lines of research.

%%%%
\paragraph{Connection with Feynman diagrams:}
%%%%
The wavefunction coefficients \eqref{eqn:psi_def} can then be represented as a sum over diagrams, in which vertices correspond to the interactions in $S_{\rm int}$, and lines correspond to factors of either $K_{\bfk} ( \eta, \eta_0)$ (if connected to the final time $\eta_0$) or $G_{\bfk} (\eta , \eta' , \eta_0 )$ (if connected between two earlier times, $\eta$ and $\eta'$ both $< \eta_0$). 
These diagrams are a close analogue of the Feynman diagrams which are used to represent time-ordered correlation functions, 
\begin{align}
\frac{ \langle \Omega | T \; \phi (x_1) ... \phi (x_n)    | \Omega \rangle
}{ \langle \Omega | \Omega \rangle }
= \frac{ (-i)^n }{Z[0]} \frac{\delta^n Z[J]}{\delta J (x_1) ... \delta J(x_n)} \big|_{\phi=0} \qquad \text{where} \;\; Z[J] = \int \mathcal{D} \phi \, e^{i S [\phi] + i \int_x J (x) \phi (x) } \; .
 \label{eqn:greens_fns}
\end{align}
While these matrix elements are obtained by summing over \emph{all} Feynman diagrams, replacing $Z[J]$ with $W[J] = \log Z [J]$ generates instead the \emph{connected} correlation functions, which correspond to summing over only connected Feynman diagrams. 
In these diagrams, vertices are the interactions contained in $S_{\rm int}$, and the edges are either external lines (connected to one of the $x_n$), or internal lines, which correspond to the matrix elements,
\begin{align}
&\text{External:} \;\; &\langle \Omega | \hat{a}_{\bfk} \hat{\phi}_{\bfk} (\eta) | \Omega \rangle' &= f_{k}^{*} ( \eta) \,, \nonumber \\ 
&\text{Internal:} \;\; &\langle \Omega | T \hat{\phi}_{\bfk} (\eta) \hat{\phi}_{-\bfk} (\eta') | \Omega \rangle' &= \Delta_{k} ( \eta, \eta' )\,,
\label{eqn:Feyn_rules_1}
\end{align} 
where $f_{k}^{*} (\eta)$ and $\Delta_{k} (\eta, \eta')$ are the usual mode function and Feynman propagator respectively. These Feynman rules reproduce the more laborious calculation of canonical quantisation using $\hat{\phi}_{\bfk} (\eta) = f_{k} (\eta) \hat{a}_{-\bfk} + f_{k}^{*} (\eta) \hat{a}^{\dagger}_{\bfk}$ in the Heisenberg picture. 
  
Comparing  \eqref{eqn:greens_fns} with \eqref{eqn:psi_def}, we see that $\psi_{\bfk_1 ... \bfk_n}$ corresponds to the connected part of the matrix element, 
\begin{align}
\frac{  \langle  0_{\eta_0} | \hat{\Pi}_{\bfk_1}  ... \hat{\Pi}_{\bfk_n} | \Psi_{\eta_0} \rangle  }{ \langle 0_{\eta_0} | \Psi_{\eta_0} \rangle } = \frac{ (-i)^n }{\Psi_{\eta_0} [ 0 ]} \frac{\delta^n \Psi_{\eta_0} [\phi ]}{ \delta \phi_{\bfk_1}  ... \delta \phi_{\bfk_n}  } \bigg|_{\phi  = 0} \; , 
\end{align}
where $| 0_{\eta_0} \rangle$ is the field eigenstate in which all fields are set to zero at time $\eta_0$. 
Just as the time-ordered correlators \eqref{eqn:greens_fns} can be represented via Feynman diagrams, so too can the wavefunction coefficients. The only difference is that the rules for replacing internal/external lines \eqref{eqn:Feyn_rules_1} must be updated to,
\begin{align}
&\text{Bulk-to-boundary:} \;\; &\langle  0_{\eta_0} | \hat{\Pi}_{\bfk} (\eta_0) \hat{\phi}_{\bfk} (\eta) | \Omega \rangle &= K_k ( \eta  , \eta_0)  = \frac{ f_k^{*} ( \eta ) }{f_k^{*} ( \eta_0 ) }  \,,  \nonumber \\
&\text{Bulk-to-bulk:}  \;\; &\langle 0_{\eta_0} | T  \hat{\phi}_{\bfk} ( \eta ) \hat{\phi}_{-\bfk} ( \eta' ) | \Omega \rangle' &= G_k ( \eta , \eta' , \eta_0)  \; . 
\label{eqn:Feyn_rules_2}
\end{align} 
Note that $G_k (\eta, \eta', \eta_0)$ is similar to the Feynman propagator $\Delta_{\bfk} (\eta, \eta')$, only with the feature that it vanishes if either $\eta$ or $\eta'$ are taken close to $\eta_0$ (due to the zero-field eigenstate bra).  
 
From \eqref{eqn:Feyn_rules_1} and \eqref{eqn:Feyn_rules_2}, we see that the internal (bulk-to-bulk) propagators can be written in terms of the external (bulk-to-boundary) propagators\footnote{This expression for $G_k$ is valid only for real momenta, $k\in \mathbb{R}$. This is sufficient for this paper where we never analytically continue internal energies.},
\begin{align}
\Delta_{k} (\eta_1, \eta_2) &=   \theta( \eta_1 - \eta_2 ) f_k (\eta_1 ) f^{*}_k (\eta_2 ) + (\eta_1 \leftrightarrow \eta_2 ) \, ,  \nonumber \\ 
G_k ( \eta_1, \eta_2, \eta_0) &= 
 2 P_{k}  \left[  \theta( \eta_1 - \eta_2) K_{k} ( \eta_2  )\text{Im} \, K_k ( \eta_1  )  + \left( \eta_1 \leftrightarrow \eta_2 \right)    \right]\; 
 \label{eqn:B2B}\\
 &=iP_k  \left[  \theta(\eta_1-\eta_2) K_k^{\ast}(\eta_1 ) K_k(\eta_2 )+\theta(\eta_2-\eta_1) K^{\ast}(\eta_2)K(\eta_1)-K(\eta_1 )K(\eta_2 )\right] \,,\nonumber
\end{align}
where $P_{k} = \langle \hat{\phi}_{\bfk} \hat{\phi}_{-\bfk} \rangle' =  | f_k ( \eta_0) |^2$ is the power spectrum at the time $\eta_0$ at which the state is defined. Just as with $\psi_{\bfk_1 ... \bfk_n}$, above and in the following we do not explicitly write the dependence on $\eta_0$ in $K_k$, $G_k$ or $P_k$. 

In particular, the bulk-to-bulk propagator differs from the Feynman propagator by a boundary term,
\begin{align}
    G_k ( \eta_1, \eta_2 ) = i \Delta_k (\eta_1, \eta_2 ) -iP_k K_k(\eta_1 )K_k(\eta_2 ) \; . 
\end{align}
The presence of this additional term has a profound meaning and important practical consequences. Its meaning is that we are in the presence of a (conformal) boundary. We would find such a term in Minkowksi as well if we wanted to compute wavefunction coefficients (or correlators) on a constant time hypersurface. It reminds us of the asymmetry between the past and the future, which qualitatively distinguishes cosmology from particle physics. This boundary term is the main obstacle to extend flat space cutting rules to cosmology. While Veltman's largest time equation still holds, it does not map explicitly to a set of relation among observables. In this paper we overcome this difficulty by deriving a bespoke set of Cosmological Cutting Rules.

%%%%
\paragraph{Rules for Computing a Diagram:}
%%%%
In the following we will express various contributions to the wavefunction coefficients diagrammatically, in a way that is analogous to the usual Feynman diagram expansion for amplitudes. The analogue of Feynman rules are the following:
\begin{itemize}
    \item Draw a graph and assign momenta $\bfk_a $ to each of its $n$ external legs and momenta $\bfp_m$ to each of the $|I|$ internal legs in way that respects momentum conservation at every vertex (but not energy conservation). For a diagram with $L$ loops, this fixes all but $L$ internal "loop" momenta. Assign a conformal time $\eta_{A}$ to each of the $V$ vertices. 
    \item Multiply a bulk-to-boundary propagator $K_{k_a}(\eta_A)$ for every external leg (which reaches the (conformal) boundary $\eta_0 \to 0$), and a bulk-to-bulk propagator $G_{p_m}(\eta_A,\eta_B)$ for every internal line (that connects two vertices at times $\eta_A$ and $\eta_B$).
    \item For every vertex at $\eta_A$, add the appropriate factors of momenta corresponding to spatial derivatives and for time derivatives act with $\partial_{\eta_A}$ on the appropriate internal or external line connected to the vertex. Sum over all allowed permutations. There is no factor of $i$ associated with the vertex. For example, the vertex corresponding to $\lambda \phi^n / n!$ is simply $\lambda$.
    \item Multiply by an overall factor of $i^{1-L}$. This could equivalently be viewed as $i^{V-I}$, a factor of $i$ for each vertex and a $(-i)$ for each propagator, which accounts for the fact that our normalisation of $G_k$ in \eqref{eqn:propagator_def} differs from the usual Feynman normalisation.
    \item Integrate over all times $\eta_A$ from $-\infty (1-i \epsilon)$ to $\eta_0 \to 0$ and over all loop momenta $\bfp_l \in \mathbb{R}^3$.
\end{itemize}

%\FloatBarrier
\begin{figure}
    \centering
    \includegraphics[width=1.0\textwidth]{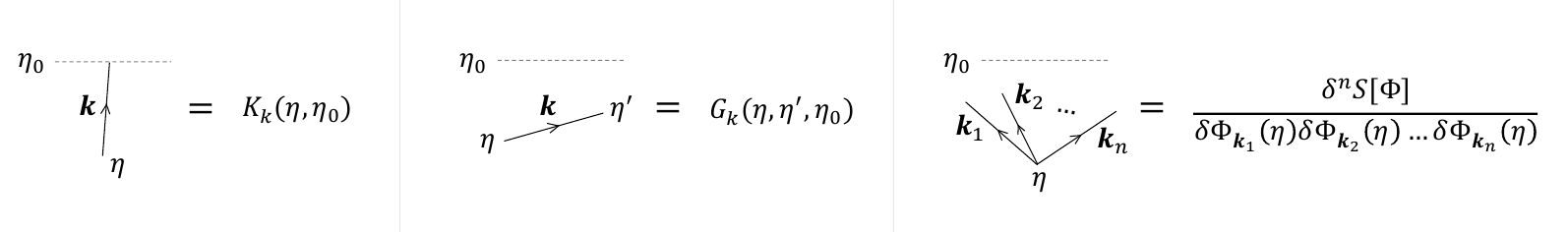}
    \caption{A graphical representation of the Feynman rules to compute the wavefunction of the universe in perturbation theory.}
\end{figure}
%\FloatBarrier
Our strategy will be to first prove the cutting rules for individual diagrams, since they can then be applied to any $\psi_{\bfk_1 ... \bfk_n}$ to any desired order in perturbation theory (see Appendix~\ref{app:schrodinger} for an alternative derivation of the cutting rules directly at the level of the $\psi_{\bfk_1 ... \bfk_n}$).

%%%%%%%%%%%%%%%%
\section{Some Examples of Cutting Wavefunction Diagrams}
\label{sec:examples}
%%%%%%%%%%%%%%%%

Our goal in this section is to present the algebraic structure of simple diagrams and how one can compute certain discontinuities using the Cosmological Cutting Rules. The idea is to see the practical application of these rules in concrete cases before moving on to the move formal proof to all orders in Section~\ref{sec:theorem}. We will start with the simplest case of a single propagator and re-derive the Cosmological Optical Theorem of \cite{Goodhew:2020hob}. Then we will consider in turn a hot to cut two propagators at tree level and at one loop level. A parallel derivation using the Schr\"odinger equation along the lines of \cite{Cespedes:2020xqq} is presented in Appendix \ref{app:schrodinger}.

%%%%%%%%%%%%%%%%
\subsection{Cutting One Propagator}
\label{sec:cutting_one}
%%%%%%%%%%%%%%%%

For our first example, consider a simple cubic interaction interaction, $S_{\rm int} [\phi]  = \int dt d^3 \bfx \, a^3 (t) \lambda \phi^3$. The corresponding wavefunction coefficients (with overall momentum-conserving delta functions removed) are given by,
\FloatBarrier
\begin{figure}[htbp!]
\centering
\includegraphics[width=0.5\textwidth]{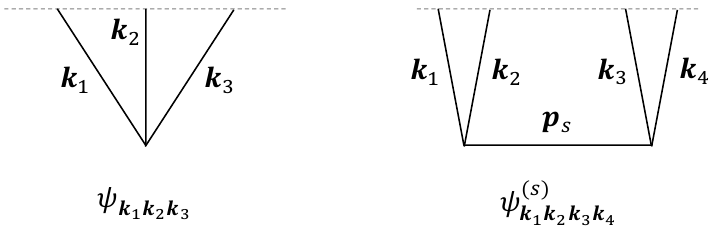}
\end{figure}
\FloatBarrier
\begin{align}
 \psi_{\bfk_1 \bfk_2 \bfk_3}' &=  i \lambda \int^{t_0}_{- \infty} dt \, K_{k_1} (t) K_{k_2} (t) K_{k_3} (t) \nonumber \\ 
 \psi_{\bfk_1 \bfk_2 \bfk_3 \bfk_4}^{(s) \, \prime} &=  i \lambda^2 \int^{t_0}_{- \infty} dt_L \int_{-\infty}^{t_0} d t_R \, K_{k_1} ( t_L ) K_{k_2} (t_L)  \; G_{p_s} ( t_L , t_R  )  K_{k_3} (t_R) K_{k_4} ( t_R )
 \label{eqn:psi3psi4_eg}
\end{align}
at tree level, as shown in the diagram above ($\psi_4 = \psi_4^{(s)} + \psi_{4}^{(t)} + \psi_{4}^{(u)}$, but we will focus on the $s$-channel diagram).  
Note that the two time integrals in $\psi_{4}^s$ are nested: they do not factorise since the bulk-to-bulk propagator $G_{p_s} ( t_L, t_R)$ contains a step function $\theta ( t_L - t_R )$. However, from \eqref{eqn:B2B} we see that the imaginary part,
\begin{align}
 \text{Im} \, G_{p_s} ( t_L , t_R ) = 2 P_{p_s} \;  \text{Im} \, K_{p_s} (t_L)  \; \text{Im} \, K_{p_s} (t_R)
  \label{eqn:ImG}
\end{align}
factorises into separate functions of $t_L$ and $t_R$. Consequently, if we can extract the imaginary part of the internal line in the $\psi_{4}^{(s)}$ exchange diagram, then the two time integrals will factor into a simple product $\psi_3 \times \psi_3$. This is achieved by evaluating $\psi_n$ at a a modified value $\bar{k}$ of the external energies, defined such that $K_{\bar{k}} (t) = K^*_{k} (t)$, and applying a parity transformation on all internal and external spatial momenta, $\{ \bfk, \bfp\} \to \{ - \bfk,-\bfp\}$. For example, for de Sitter mode functions for a massless or conformally coupled field with a Bunch Davies vacuum, one has simply $\bar k = - k $, with the negative real $k$-axis being approached from the lower-half complex plane to guarantee appropriate convergences \cite{Goodhew:2020hob}. Furthermore, to simplify our notation, we'll make often use of the following ``discontinuity" operation\footnote{
We use this terminology by analogy with the amplitude discontinuity $\text{Disc}_s \mathcal{A}_{12 \to 34} = \frac{1}{2i} \left( \mathcal{A}_{12 \to 34} - \mathcal{A}_{34 \to 12}^* \right)$, which appears in the flat space optical theorem. },
\begin{align}
 &  \underset{ k_1 \dots k_j  }{\disc } f(k_1,\dots,k_n;\{p\};\{\bfk \}) \label{defdisc}\\ 
 &\equiv f(k_1,\dots,k_n;\{p\};\{\bfk \}) - f^\ast(k_1,\dots,k_j,-k_{j+1},\dots,-k_n;\{p\};-\{\bfk \}) \,, \nonumber
\end{align} 
where $\{p\}$ denotes internal energies, which are untouched by the $\disc$, and $\{\bfk\}$ are all spatial momenta.
In words, the $\disc$ operation corresponds to subtracting the complex conjugate of $\psi_n$ with all external energies analytically continued to minus themselves \emph{except for those listed in the subscript of} $\text{Disc}$ and all spatial momenta (internal or external) reversed by parity $\bfk \to - \bfk$. For example, no subscript corresponds to replacing all $k_a \to \bar k_a$. This $\text{Disc}$ operation can be used to pick out the imaginary part of the corresponding internal propagator\footnote{
Note that $(-i) \text{Disc} \left[ i \psi_n \right] = i \, \text{Disc}\left[ c_n \right]$ in the notation of \cite{Cespedes:2020xqq}. 
},
\begin{align}
i \,\underset{p_s}{\text{Disc}}  \, \left[ i  \psi^{(s)}_{\bfk_1 \bfk_2 \bfk_3 \bfk_4} \right] &=
  2 \lambda^2 \int^{t_0}_{- \infty} dt_L \int_{-\infty}^{t_0} d t_R \, K_{k_1} ( t_L ) K_{k_2} (t_L)  \; \text{Im} \, G_{p_s} ( t_L , t_R  )  K_{k_3} (t_R) K_{k_4} ( t_R )  \nonumber \\
&=
 \int_{\bfq \bfq' }   \; i \, \underset{q}{\text{Disc}} \, \left[ i \psi_{\bfk_1 \bfk_2 \bfq} \right] \; P_{\bfq \bfq'} \; i \, \underset{q'}{ \text{Disc} } \, \left[ i \psi_{\bfq' \bfk_3 \bfk_4} \right] \; , 
\label{eqn:Disc_psi4s}
\end{align}
where we have used \eqref{eqn:ImG}, and introduced the power spectrum $P_{\bfq \bfq'}$ which includes the momentum conserving $\delta$-function as in \eqref{Pk}.
Note that for translationally invariant interactions each $\psi_n$ also contains an overall momentum conserving $\delta$-function, which in this case can be used to set $\bfq' = - \bfq = \bfp_s$. 

We depict the cutting rule \eqref{eqn:Disc_psi4s} diagrammatically as follows: 
\FloatBarrier
\begin{figure}[htbp!]
\centering
\includegraphics[width=0.7\textwidth]{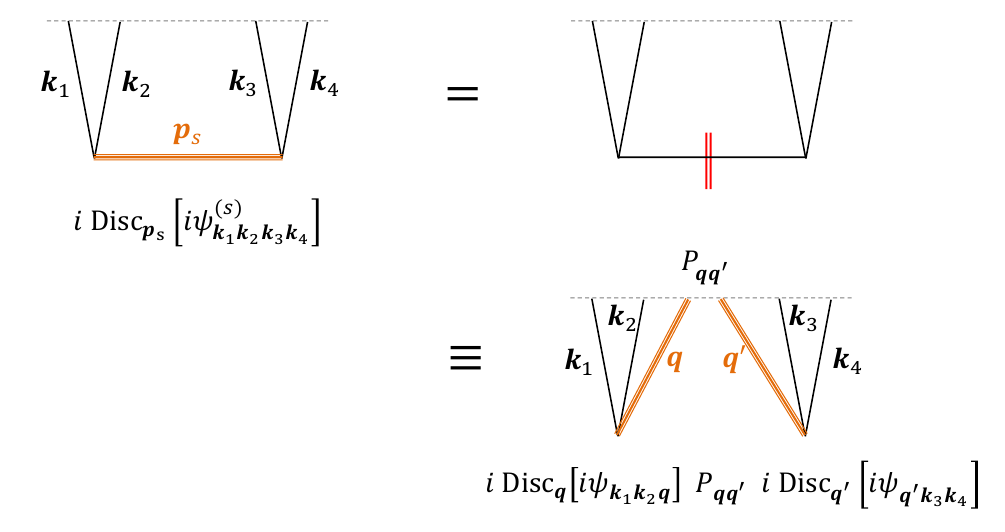}
\end{figure}
\FloatBarrier
\noindent On the left-hand side, we have introduced a highlighted line to represent the propagator of which we are extracting the imaginary part. The energy of the highlighted line appears in the argument of $\text{Disc}$ and so it \textit{not} analytically continued. Our cutting rule then relates this to the first diagram shown on the right, where the red vertical lines indicate which propagators are to be ``cut''. By definition, when a propagator is ``cut'' we replace it with two bulk-to-boundary propagators and insert a factor of the power spectrum, which is shown in the final diagram on the right-hand side.  
This reproduces the Cosmological Optical Theorem of \cite{Goodhew:2020hob} from the point of view of cutting rules. 

An analogous relations holds for the $t$- and $u$-channels, and so the full wavefunction coefficient obeys,
\begin{align}
i \, \underset{p_s p_t p_u}{\text{Disc}} \,\left[ i \psi_{\bfk_1 \bfk_2 \bfk_3 \bfk_4} \right] =  \sum_{\rm perm.}^3 \int_{\bfq \bfq'}  i \, \underset{q}{\text{Disc}} \left[ i \psi_{\bfk_1 \bfk_2 \bfq} \right]\; P_{\bfq \bfq'} \; i \, \underset{q'}{\text{Disc}} \left[ i \psi_{\bfq' \bfk_3 \bfk_4}  \right]
\label{eqn:Disc_psi4_full}
\end{align}
at tree level (note that $\psi_{4}^{(s)}$ depends on $p_t$ and $p_u$ only through analytic combinations like $ \bfp_t \cdot \bfp_t$ and $\bfp_t \cdot \bfp_u$, so $\text{Disc}_{p_s p_t p_u} [ i \psi_{4}^{(s)} ] = \text{Disc}_{p_s} [ i \psi_{4}^{(s)} ]$, and similarly for $\psi_{4}^{(t)}$ and $\psi_{4}^{(u)}$).  

Note that the role of the $\text{Disc}$ combination is to take the imaginary part of the internal lines (bulk-to-bulk propagators) without affecting any external line (bulk-to-boundary propagator). The external lines therefore only appear in our cutting equations as an overall factor. For instance, \eqref{eqn:ImG} is also the relevant cutting rule for \emph{any} diagram in which a single internal line is connected to $n_L$ external legs on the left and $n_R$ external lines on the right, providing we replace $K_{k_1} (t_L) K_{k_2} (t_L)$ with $K_{k_1} (t_L) ... K_{k_{n_L}} (t_L)$ and replace $K_{ k_3} (t_R) K_{ k_4} (t_R)$ with $K_{ k_1'} ( t_R) ... K_{ k_{n_R}'} ( t_R)$. 
The cutting rule for this diagram is the straightforward extension of \eqref{eqn:Disc_psi4s},
\begin{align}
i \, \underset{p}{\text{Disc} } \; \left[ i \psi_{ \{ \bfk \} \{ \bfk' \} }^{(p)} \right]
 =  \int_{\bfq \bfq'} \; i \, \underset{q}{\text{Disc}  } \left[ i \psi_{ \{ \bfk \} \bfq} \right] \;
 P_{\bfq \bfq'} \; i \, \underset{  q'}{ \text{Disc}  } \left[ i \psi_{ \bfq' \{ \bfk' \} } \right] \; , 
 \label{eqn:Disc_psins}
\end{align}
where $\bfp = \sum_{a}^{n_L} \bfk_a = - \sum_{a}^{n_R} \bfk_a'$ is the total momentum flowing from the boundary into (out of) the interaction vertices, i.e. the momentum carried by the internal line, and the $\psi$'s on the right-hand side are contact diagrams with $(n_L+1)$ and $(n_R+1)$ legs, respectively. 
We can therefore focus on only the internal lines (suppressing any external line factors), since this provides more compact expressions which are applicable to a wider range of diagrams (i.e. any diagram in which an arbitrary number of external lines is attached to any of the vertices). 

%%%%
\paragraph{Single-cut rules:} 
%%%%
Finally, note that although we focused above on a simple diagram with only a single internal line, more generally in a diagram with many internal lines we can always use an appropriate $\text{Disc}$ to cut any single propagator.
For instance, for the cubic interaction considered in \eqref{eqn:psi3psi4_eg}, one diagram which contributes to the quintic wavefunction coefficient is given by
\FloatBarrier
\begin{figure}[htbp!]
\centering
\includegraphics[width=0.3\textwidth]{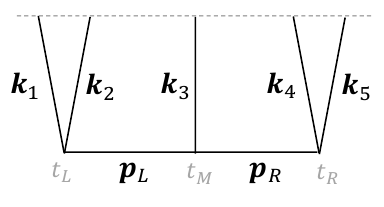}
\end{figure}
\FloatBarrier
\begin{align}
    \psi_{\bfk_1 \bfk_2 \bfk_3 \bfk_4 \bfk_5}^{(p_L, \, p_R) \, \prime} =& i \lambda^2 \int_{-\infty}^{t_0} d t_L  \, dt_M \,  dt_R
K_{k_1} ( t_L) K_{k_2} ( t_L)  G_{p_L} ( t_L , t_M  )\label{eqn:psi5} \\
&\qquad \times K_{k_3} ( t_M) G_{p_R} ( t_M , t_R  ) K_{k_4} ( t_R) K_{k_5} ( t_R)\,, \nonumber
\end{align}
where $p_L = | \bfk_1 + \bfk_2|$, and $p_R = |\bfk_4 + \bfk_5 |$ are the momenta flowing through two internal lines, which connect interaction vertices at times $t_L, \, t_M$ and $t_R$.
To cut the $G_{p_L} (t_L, t_M)$ internal line, we take $\underset{p_L}{\text{Disc}}  \left[ i \psi^{(p_L, \, p_R)}_{\bfk_1 \bfk_2 \bfk_3 \bfk_4 \bfk_5}  \right]$, which extracts the $\text{Im} \, G_{p_L} (t_L, t_M)$ and allows us to use the propagator identity \eqref{eqn:ImG}. 
Diagrammatically, this corresponds to,
\FloatBarrier
\begin{figure}[htbp!]
\centering
\includegraphics[width=0.6\textwidth]{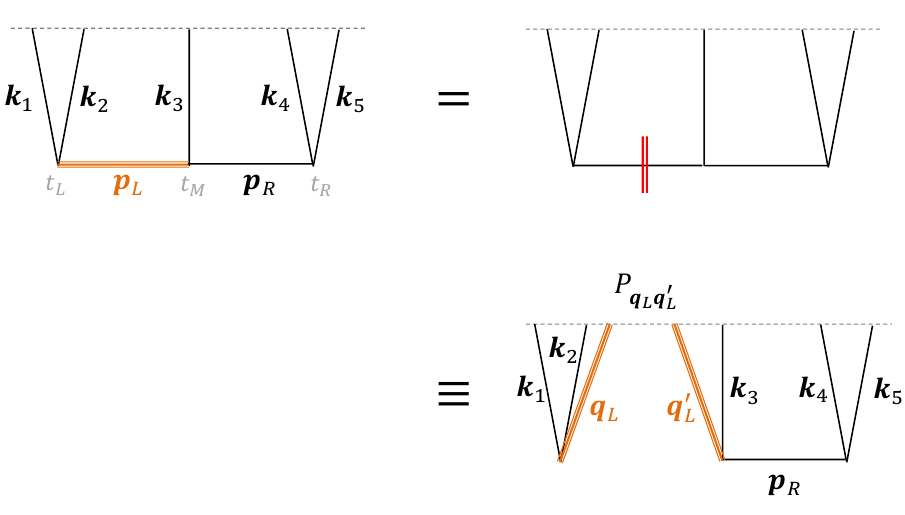}
\end{figure}
\FloatBarrier
\noindent which represents the single-cut rule discussed in \cite{single},
\begin{align}
i \, \underset{p_L}{\text{Disc} } \left[ i \psi_{ \bfk_1 \bfk_2 \bfk_3 \bfk_4 \bfk_5 }^{(p_L, \, p_R)} \right] 
= \int_{ \bfq_L \bfq_L' } \; 
i \, \underset{q_L}{\text{Disc}}  \left[ i\psi_{ \bfk_1 \bfk_2 \bfq_L} \right]
 P_{\bfq_L \bfq_L'} 
i \,  \underset{q_L'}{ \text{Disc} } \left[ i \psi_{ \bfq_L' \bfk_3 \bfk_4 \bfk_5 }^{(p_R)} \right] 
 \label{eqn:psi5_one_cut}
\end{align}
where $\psi_{ \bfq_L' \bfk_3 \bfk_4 \bfk_5 }^{(p_R)}$ is the exchange diagram with $\bfp_R = \bfk_4 + \bfk_5$ flowing through the internal line.
Note that there are three $\delta$-functions on the right-hand-side, which enforce momentum-conservation at each vertex, namely $\bfq_L = - \bfk_1 - \bfk_2$ and $\bfq_L' = -\bfk_3 - \bfk_4 - \bfk_5$, as well as overall momentum conservation $\bfk_1 + \bfk_2 + \bfk_3 + \bfk_4 + \bfk_5 = 0$. We will not discuss these single-cut rules any further here, but refer the reader to \cite{single} for a detailed analysis. Before proceeding, it is worth commenting on the difference between the above single-cut rule and the (multi-cut) cutting rules we will discuss in the rest of this paper:
\begin{itemize}
    \item In single-cut rules we have to analytically continue all internal lines that are not cut, in addition to the external lines. To make this possible, one needs to choose variables such that the energies of all non-cut internal lines appear in the argument of $\psi$, so that they can be analytically continued by $\disc$. Since the energies flowing in the internal lines depend on the specific diagram chosen (i.e. the different channels), \textit{it follows that the choice of variables for single-cut diagrams are diagram dependent}. This is in contrast with the cutting rules we discuss in this paper, in which case we never analytically continue any internal line, and so it does not matter if its energy appears or not as a variable. Indeed, notice that in all our examples, the internal lines are either highlighted, therefore they appear in the argument of $\disc$ and are not analytically continued, or they are cut.
    \item In single-cut rules we can cherry-pick where to cut a given diagram. Conversely, for the cutting rules in this work one has always to sum over all possible cuts, including multiple cuts and no cuts at all. We will see this in the next subsection.
    \item Importantly, in their current formulation, single-cut rules apply only to tree-level diagrams. The reason is that in a loop diagram, the momentum of some internal line is integrated over and so it is not clear how one could analytically continue it by altering the variables of $\psi$. Conversely, \textit{the cutting rules we discuss here apply to diagrams of any loop order}. This is possible because, as we stressed above, we never analytically continue any internal energy. We will see how to deal with loops in Section \ref{sec:cutting_loop}.
\end{itemize}

%%%%%%%%%%%%%%%%
\subsection{Cutting Two Propagators}
\label{sec:cutting_two}
%%%%%%%%%%%%%%%%

Taking a closer look at the quintic wavefunction coefficient \eqref{eqn:psi5}, we see that it contains an integral of the form,
\begin{align}
\int^{t_0}_{- \infty} dt_L \int^{t_0}_{- \infty} dt_M \int^{t_0}_{- \infty} dt_R \,  G_{p_L} ( t_L , t_M  ) G_{p_R} ( t_M , t_R  ) \left[\prod_{a=1}^5 K_{k_a} \right]\; ,
\label{eqn:GG_eg}
\end{align}
which does not factorise due to the pair of $\theta ( t_L - t_M )$ and $\theta (t_M - t_R)$ functions within the bulk-to-bulk propagators. As shown above, by taking a suitable imaginary part of \eqref{eqn:GG_eg} (i.e. the Disc of the corresponding wavefunction coefficient), we can remove at least one of these $\theta$-functions, leading to the single-cut rule in \eqref{eqn:psi5_one_cut}, which now contains only a single exchange integral. 

Remarkably, there is another way to remove at least one $\theta$-function from \eqref{eqn:GG_eg}, and that is to take the imaginary part of \emph{both} propagators:
\begin{align}
& \text{Im} \, \left[ G_{p_L} ( t_L, t_M ) G_{p_R} ( t_M, t_R ) \right] =  \nonumber \\
&+ 2 P_{p_L} \, \text{Im} \left[  K_{p_L} (t_L)    \right] \text{Im} \left[  K_{p_L} (t_M) G_{p_R} (t_M, t_R)   \right]
  \nonumber \\ 
 &+ 2 P_{p_R} \, \text{Im} \left[  K_{p_R} (t_R)    \right] \text{Im} \left[  K_{p_R} (t_M) G_{p_L} (t_L, t_M)   \right]
  \nonumber \\ 
 &- 4 P_{p_L} P_{p_R} \, \text{Im} \, \left[ K_{p_L} (t_L) \right] \text{Im} \left[ K_{p_L} (t_M) K_{p_R} (t_M)  \right]  \text{Im} \left[ K_{p_R} (t_R) \right] \, .
 \label{eqn:ImGG}
\end{align}
This factorises the three nested integrals \eqref{eqn:GG_eg} into the product of two or three lower $n$-point coefficients. 
Using the $\text{Disc}$ to pick out the imaginary part of the two internal propagators, we can use \eqref{eqn:ImGG} to factorise the $5$-point coefficient \eqref{eqn:psi5} into products of lower $n$-point functions, in particular $\psi_4 \times \psi_3$ and $ \psi_3 \times \psi_3 \times \psi_3$, 
\begin{align}
&i\, \underset{ p_L p_R}{ \text{Disc}  } \left[ i \psi_{ \bfk_1 \bfk_2 \bfk_3 \bfk_4 \bfk_5 }^{(p_L, \, p_R)} \right] \nonumber \\
&= \int_{ \bfq_L \bfq_L' } \; 
i \,\underset{ q_L}{ \text{Disc} } \left[ i\psi_{ \bfk_1 \bfk_2 \bfq_L} \right]
 P_{\bfq_L \bfq_L'} 
i \, \underset{ q_L' p_R }{\text{Disc}} \left[ i \psi_{ \bfq_L' \bfk_3 \bfk_4 \bfk_5 }^{(p_R)} \right]
 \nonumber \\
&+\int_{ \bfq_R \bfq_R' } \; 
i \, \underset{ p_L q_R }{\text{Disc}} \left[ i\psi_{ \bfk_1 \bfk_2 \bfk_3  \bfq_R}^{(p_L)} \right]
 P_{\bfq_R \bfq_R'} 
i \, \underset{ q_R'}{ \text{Disc}  } \left[ i \psi_{ \bfq_R' \bfk_4  \bfk_5 } \right]
 \nonumber \\
&- \int_{ \substack{ \bfq_L \bfq_L' \\ \bfq_R \bfq_R'} } \; 
i \, \underset{q_L}{ \text{Disc} } \left[ i\psi_{ \bfk_1 \bfk_2 \bfq_L} \right]
 P_{\bfq_L \bfq_L'} 
i \, \underset{ q_L' q_R }{\text{Disc}} \left[ i \psi_{ \bfq_L' \bfk_3 \bfq_R } \right]
 P_{\bfq_R \bfq_R'} 
i \, \underset{ q_R'}{ \text{Disc} } \left[ i \psi_{ \bfq_R' \bfk_4 \bfk_5 } \right]
 \label{eqn:cut_2_eg}
\end{align}
where $\psi^{(p_L)}_{\bfk_1 \bfk_2 \bfk_3 \bfq_R}$ is the particular exchange contribution to $\psi_4$ in which the internal line carries momentum $\bfp_L = \bfk_1 + \bfk_2$. 
The cutting rule \eqref{eqn:cut_2_eg} corresponds to \emph{summing} over all possible cuts of the internal lines (the left-hand side corresponding to zero cuts), where a cut bulk-to-bulk propagator is replaced by two bulk-to-boundary propagators and a factor of the boundary power spectrum. 

Diagrammatically, we represent the cutting rule \eqref{eqn:cut_2_eg} as:
\FloatBarrier
\begin{figure}[htbp!]
\centering
\includegraphics[width=1.0\textwidth]{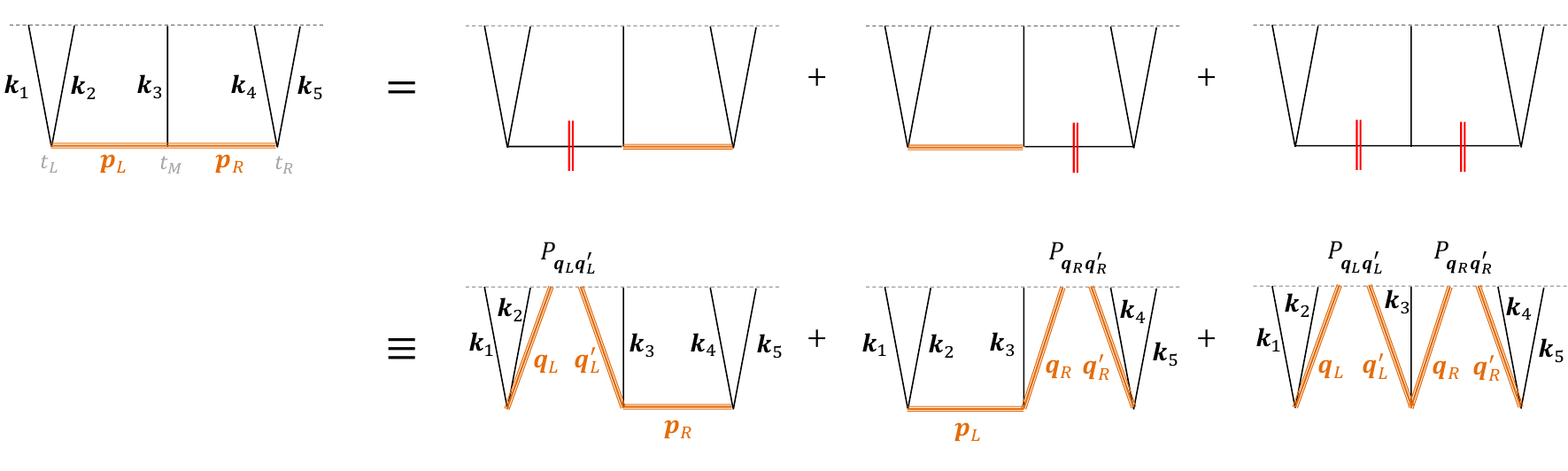}
\end{figure}
\FloatBarrier
\noindent These three cut diagrams on the right-hand-side correspond to taking discontinuities of each disconnected subdiagram (which now contain at most a single bulk-to-bulk propagator), and correspond to the three terms on the right-hand-side of the cutting rule \eqref{eqn:cut_2_eg}.
Note that when we highlight two or more lines in any disconnected subgraph, it corresponds to taking a single $\text{Disc}$ in such a way that that a single imaginary part is taken of the product of the highlighted propagators (and should not be confused with taking multiple discontinuities to extract multiple imaginary parts). For instance, the final diagram on the right-hand-side has three disconnected components (so is the product of three separate $\text{Disc}$'s), and the central subdiagram is given by $\underset{q_L' q_R}{\text{Disc}} \left[ i \psi_{\bfq_L' \bfk_3 \bfq_R} \right]$, which extracts the imaginary part of the product $K_{\bfq_L'} K_{\bfq_R}$.

Just as when cutting a single propagator, here as well it is only the external lines that are analytically continued and not the internal lines. The cutting rule \eqref{eqn:cut_2_eg} can therefore be easily generalised to \emph{any} diagram which contains two internal lines connected in this way. 
For instance, consider the diagram with three interactions vertices, $V_L, V_M$ and $V_R$, shown below. 
A collection of external lines (with momenta $\{\bfk_L\}$) are connected to the left interaction vertex $V_L$, and similarly for vertices $V_M$ and $V_R$. Internal bulk-to-bulk lines  connect $V_L$ to $V_M$ and $V_M$ to $V_R$, and carry momenta $\bfp_L$ and $\bfp_R$ respectively. We denote this particular diagram by $\psi_{ \{ \bfk_L \} \{ \bfk_M \} \{ \bfk_R \}  }^{(p_L, \, p_R)}$, and it contains a triple (nested) time integral of the form \eqref{eqn:GG_eg}. In general this integral can be difficult to perform exactly, however the relation \eqref{eqn:ImGG} allows us to express its discontinuity at fixed $\bfp_L$ and $\bfp_R$ in terms of objects that involve only double time integrals. 
Explicitly, this gives the cutting rule, 
\begin{align}
&i \, \underset{p_L p_R}{\text{Disc}} \left[ i \psi_{ \{ \bfk_L \} \{ \bfk_M \} \{ \bfk_R \} }^{ (p_L \, p_R )} \right] \nonumber \\
&= \int_{ \bfq_L \bfq_L' } \; 
i \, \underset{ q_L }{\text{Disc}} \left[ i\psi_{ \{ \bfk_L \} \bfq_L} \right]
 P_{\bfq_L \bfq_L'} 
i \, \underset{q_L' p_R }{\text{Disc}} \left[ i \psi_{ \bfq_L' \{ \bfk_M \} \{ \bfk_R \} }^{(p_R)} \right]
 \nonumber \\
&+\int_{ \bfq_R \bfq_R' } \; 
i \, \underset{ p_L q_R }{\text{Disc}} \left[ i\psi_{ \{ \bfk_L \}  \{ \bfk_M \} \bfq_R}^{(p_L)} \right]
 P_{\bfq_R \bfq_R'} 
i \, \underset{q_R'}{ \text{Disc} } \left[ i \psi_{ \bfq_R'  \{ \bfk_R \} } \right]
 \nonumber \\
&- \int_{ \substack{ \bfq_L \bfq_L' \\ \bfq_R \bfq_R'} } \; 
i \, \underset{ q_L}{\text{Disc}} \left[ i\psi_{ \{ \bfk_L \} \bfq_L} \right]
 P_{\bfq_L \bfq_L'} 
i \, \underset{ q_L' q_R }{\text{Disc}} \left[ i \psi_{ \bfq_L' \{ \bfk_M \} \bfq_R } \right]
 P_{\bfq_R \bfq_R'} 
 i \, \underset{q_R'}{\text{Disc}} \left[ i \psi_{ \bfq_R' \{ \bfk_R \} } \right] \;  . 
 \label{eqn:cut_2}
\end{align}

%%%%%%%%%%%%%%%%
\subsection{Cutting a Loop}
\label{sec:cutting_loop}
%%%%%%%%%%%%%%%%

The cutting rules \eqref{eqn:Disc_psi4s}, \eqref{eqn:Disc_psi4_full}, \eqref{eqn:Disc_psins}, \eqref{eqn:psi5_one_cut}, \eqref{eqn:cut_2_eg}  and \eqref{eqn:cut_2} shown above are relations among exclusively tree-level wavefunction coefficients. 
We will show how the $\text{Disc}$ operation \eqref{defdisc} can also be used to reduce simple one-loop diagrams to a product of tree-level diagrams.

%%%%
\paragraph{One-Propagator Loop:}
%%%%
The simplest one-loop diagram contains a single internal line, as shown below. 
Unlike the tree-level examples above, this single propagator is evaluated at coincident times, $G_p (t, t)$. In this case, it is not only the imaginary part of the propagator which factorises, but also the real part\footnote{
Note that since we have written the time-ordering in \eqref{eqn:Feyn_rules_1} and \eqref{eqn:Feyn_rules_2} as, $T \hat{\phi}_{\bfk_1} (t_1 ) \hat{\phi}_{\bfk_2} (t_2) = {\theta (t_1 - t_2 ) \hat{\phi}_{\bfk_1} (t_1) \hat{\phi}_{\bfk_2} (t_2)} + \theta ( t_2 - t_1 ) \hat{\phi}_{\bfk_2} (t_2) \hat{\phi}_{\bfk_1} (t_1)$, we are treating $\theta ( 0 ) = 1/2$.
},
\begin{align}
\text{Re} \, \left[ G_{p} ( t , t ) \right] =  P_p \;  \text{Im} \left[   K_{p} (t) K_{p} (t) \right]    \; .
\label{eqn:ReG}
\end{align}
This means that, considering the 1-loop contribution to $\psi_n$ from the interaction $\lambda \phi^{n+2}$,
\begin{align}
 \psi_{\bfk_1 ... \bfk_n}^{\text{1-loop} \;\;  \prime} =  \frac{\lambda}{\color{red} 2} \int_{-\infty}^{t_0} dt \; K_{k_1} (t) ... K_{k_n} (t) \;  \int_{\bfp} \, G_p (t, t)
 \label{eqn:psi_n_loop}
\end{align}
we can use \eqref{eqn:ReG} to write its discontinuity in terms of a tree-level coefficient,
\begin{align}
i\,  \text{Disc} \left[ i \psi_{\bfk_1 ... \bfk_n}^{\text{1-loop}}  \right]  =  \int_{\bfq \bfq'} \;  (-i) \, \underset{ q q'}{\text{Disc}} \left[ i \psi_{ \bfk_1 ... \bfk_n \bfq  \bfq'}^{\rm tree}  \right] \, P_{\bfq \bfq'}
\label{eqn:cut_1loop_1}
\end{align}
where,
\begin{align}
 \psi_{ \bfk_1 ... \bfk_n \bfq  \bfq'}^{\rm tree \;\; \prime} = i \lambda \int_{-\infty}^{t_0} dt \;  K_{k_1} (t) ... K_{k_n} (t)  K_{q} (t)  K_{q'} (t) \; .
 \label{eqn:psi_n_tree}
\end{align}
is the contact contribution to $\psi_{n+2}$. Diagrammatically,
\FloatBarrier
\begin{figure}[htbp!]
\centering
\includegraphics[width=0.55\textwidth]{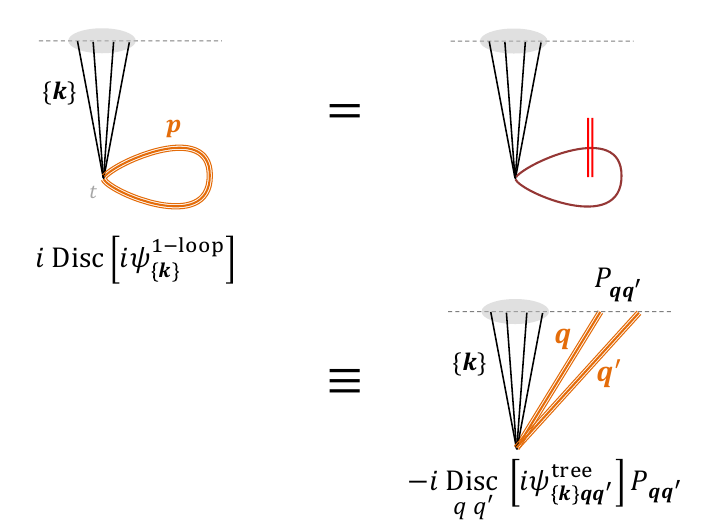}
\end{figure}
\FloatBarrier

Some comments about the cutting rule \eqref{eqn:cut_1loop_1}:
\begin{itemize}

\item[(i)] As in the previous examples, the cutting rule has effectively removed the need to perform an additional time integral---if $\psi_{n+2}^{\rm tree}$ has already been computed using \eqref{eqn:psi_n_tree}, then \eqref{eqn:cut_1loop_1} can be used to infer $\text{Disc} \left[ i \psi_n^{\text{1-loop}} \right]$ without ever carrying out the time integral in \eqref{eqn:psi_n_loop}.

\item[(ii)] Unlike in the tree-level examples, there are no longer enough $\delta$-functions to fix all of the internal momenta in \eqref{eqn:cut_1loop_1}, so one momentum integral is left over. However, unlike in \eqref{eqn:psi_n_loop}, the remaining momentum integral in \eqref{eqn:cut_1loop_1} is now \emph{finite}, and does not require any regularisation.  

\item[(iii)] One could also have used \eqref{eqn:ImG} to write this in terms of the tree-level result, but since \eqref{eqn:ImG} involves two imaginary parts this would result in a cutting rule with multiple (overlapping) discontinuities---these will be discussed separately elsewhere. 

\end{itemize}

%%%%
\paragraph{Two-Propagator Loop:}
%%%%
The next-simplest loop diagram contains a single loop composed of two internal lines, as shown below.
In this case, the diagram contains two time integrals over the product $G_{q_1} (t_1, t_2) G_{q_2} (t_1, t_2)$,
\begin{align}
 \psi^{\text{1-loop}, \, (p) \; \prime  }_{ \{ \bfk \} \{ \bfk' \} } &= \lambda^2 \int_{-\infty}^{t_0} dt_1 \int_{-\infty}^{t_0} dt_2  \prod_{ \bfk_j } K_{k_j} (t_1) \prod_{\bfk_j'} K_{k_j'} (t_2) \\
 &\qquad \times\int_{\bfp_1 \bfp_2} G_{p_1} (t_1, t_2) G_{p_2} (t_1, t_2)  \; \tilde{\delta}^3 ( \bfp_1 + \bfp_2 - \bfp )  \; . \nonumber
\end{align}
where $\bfp = \sum_j \bfk_j$ is the momentum flowing into the loop\footnote{
Note that each of these internal lines may correspond to different fields, but this can be viewed as simply adding additional quantum numbers to the labels $p_1$ and $p_2$, i.e. in \eqref{eqn:ReGG}, the $P_{p_j}$, $K_{p_j}$ and $G_{p_j}$ factors correspond to either exchanged field $1$ or field $2$, and which one can be inferred from their momentum label.   
}.
To factorise this into two separate integrals, one can use the following identity,
\begin{align}
2 \text{Re}  \left[ G_{p_1} ( t_1, t_2 ) G_{p_2} ( t_2, t_1 ) \right]   
&= 2 P_{ p_2} \,  \text{Im} \left[   K_{p_2} (t_1) G_{p_1} ( t_1, t_2 ) K_{p_2} (t_2)   \right] \nonumber  \\
&+  2 P_{ p_1} \,  \text{Im} \left[   K_{p_1} (t_2) G_{p_2} ( t_2, t_1 ) K_{p_1} (t_1)   \right]     \nonumber  \\
&- 4 P_{p_1} P_{p_2}   \text{Im} \left[  K_{p_1} (t_1 ) K_{p_2} (t_1)  \right]  \text{Im} \left[ K_{p_2} (t_2 ) K_{p_1} (t_2)  \right]   \label{eqn:ReGG}  
\end{align}
which relates the real part of the propagators to products of $K_k$ and $G_k$, and which crucially contains imaginary parts acting only on factors evaluated at the same times. 
This allows  one to write  each of the terms on the right-hand side of \eqref{eqn:ReGG} in terms of a single $\text{Disc}$ acting on a tree-level wavefunction coefficient, 
\begin{align}
i\, \underset{p}{\text{Disc}} \left[  i \psi_{ \{ \bfk \} \{ \bfk' \} }^{\text{1-loop}, \, (p)} \right]  
&=
 \int_{ \bfq_2 \bfq_2' } \;  P_{\bfq_2 \bfq_2'} 
(-i) \, \underset{ p_1 q_2 q_2' }{\text{Disc}} \left[ i \psi_{  \{ \bfk \} \bfq_2  \bfq_2' \{ \bfk' \} }^{\text{tree}, \, (p_1) } \right] \nonumber \\ 
& + \int_{ \bfq_1 \bfq_1' } \;  P_{\bfq_1 \bfq_1'} 
(-i) \, \underset{ p_2 q_1 q_1' }{\text{Disc}} \left[ i \psi_{  \{ \bfk \} \bfq_1  \bfq_1' \{ \bfk' \} }^{\text{tree}, \, (p_2)} \right]  \label{eqn:cut_1loop_2}
 \\ \nonumber 
&  + \int_{ \substack{ \bfq_1 \bfq_1' \\ \bfq_2 \bfq_2'} } \; 
i \, \underset{ q_1 q_2 }{\text{Disc}} \left[ i\psi_{ \{ \bfk \} \bfq_1 \bfq_2}^{\rm tree, \, contact} \right]
 P_{\bfq_1 \bfq_1'}  P_{\bfq_2 \bfq_2'} \;
i \,  \underset{ q_1' q_2' }{\text{Disc}} \left[ i \psi_{ \bfq_1' \bfq_2' \{ \bfk' \}  }^{\rm tree , \, contact} \right]
\end{align}
where $\bfp = \sum_{j} \bfk_j$ is the momentum flowing into the loop from the boundary, and $\psi_{  \{ \bfk \} \bfq \bfq' \{ \bfk' \} }^{\text{tree}, \, (p_2) }$ corresponds to the diagram in which a propagator $G_{p_2} (t_1 , t_2)$ connects external legs with momenta $\{ \bfk\}$ and $\bfq$ at time $t_1$ to external legs with momenta $\{ \bfk' \}$ and $\bfq'$ at time $t_2$. 
Diagrammatically,
\FloatBarrier
\begin{figure}[htbp!]
\centering
\includegraphics[width=1.05\textwidth]{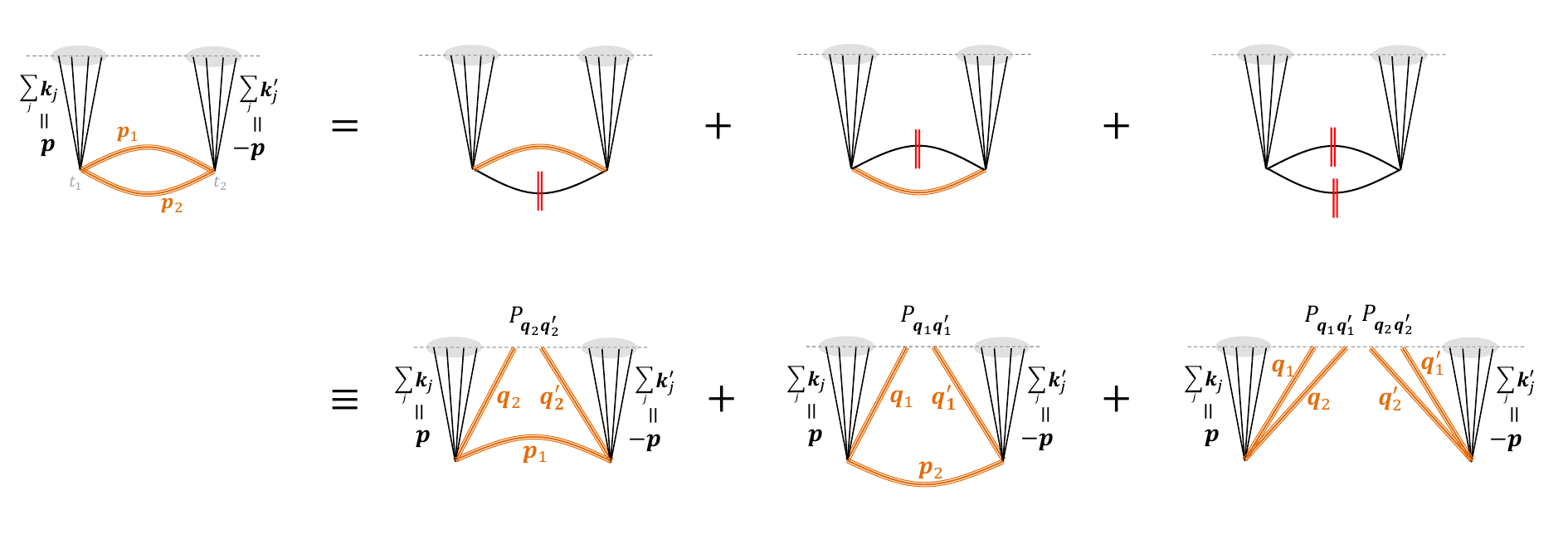}
\end{figure}
\FloatBarrier

Note that at tree-level we had a choice about how many internal lines to highlight with the $\text{Disc}$, leading to the single-cut rules \eqref{eqn:psi5_one_cut} of \cite{single} or our (multiple-cut) cutting rules \eqref{eqn:cut_2_eg}. At loop-level, it is no longer possible to use the $\text{Disc}$ operation to extract arbitrary imaginary parts---for the one-loop example above, the whole loop must be highlighted, since it is not possible to extract $\text{Im} \, G_{p_1} (t_1, t_2)$ alone. This is why going beyond tree-level requires going beyond the cutting of single lines. 
In the following, we will focus only on diagrams in which every internal line is highlighted, i.e. we never analytically continue internal lines

~\\
To sum up, we have used simple algebraic relations between the bulk-to-bulk and bulk-to-boundary propagators to derive powerful cutting rules which relate higher $n$-point wavefunction coefficients to lower $n$-point coefficients, and which crucially can relate 1-loop diagrams to (products of) tree-level diagrams. 
These relations turn out to be surprisingly universal, and we will now show that, faced with \emph{any} $L$-loop diagram, one can take appropriate discontinuities to reduce it to combinations of lower-point $(L-1)$-loop diagrams. 
%Iterating this procedure would be interesting, but we will discuss taking multiple discontinuities elsewhere. 

%%%%%%%%%%%%%%%%
\section{General Cutting Rules for a Single Scalar Field}
\label{sec:theorem}
%%%%%%%%%%%%%%%%

In this section, we begin by stating and proving the cutting rules for a general diagram, with any number of internal/external lines and with any number of loops, but focusing on a single scalar field. We will prove this result with the help of an algebraic identity for the imaginary part of the product of bulk-to-bulk propagators. In the next section we generalize our result to multiple fields with any spin (Section~\ref{sec:multiple}).\\

To help intuition, let's begin with the following simplified statement of the cutting rules:
\begin{align} \label{same}
   i \, \underset{\substack{ \text{internal} \\ \text{lines} } }{\disc }  \left[ i \, \psi^{(D)} \right]     &=  \sum_{\rm cuts}   \left[ \prod_{\substack{\rm cut \\ \rm momenta}} \int P \right]
   \prod_{\rm subdiagrams} (-i) \underset{\substack{ \text{internal } \& \\  \text{ cut lines} } }{\disc } \left[ i \, \psi^{(\rm subdiagram)} \right]\,,
\end{align}
where $D$ is some diagram that is reduced to a number of subdiagrams by cutting one or more internal lines in all possible ways. Notice that in all cases the arguments of $\disc$, i.e. the energies that are not analytically continued, are all the internal lines plus whatever external line resulted from a cut.

We will now make \eqref{same} more mathematically precise. The general cutting rules may be stated in two steps: the first is diagrammatic (how to draw all ``cut'' diagrams), and the second is algebraic (how to evaluate each of the cut diagrams).

\begin{itemize}

\item[Step 1.]
We begin with a connected diagram, $D$, which can be translated using the Feynman rules of Section~\ref{sec:rules} into a contribution $\psi^{(D)}$ to a wavefunction coefficient. 
We denote by $I$ the set of all internal lines in $D$ (of which we are going to extract the imaginary parts), and represent the appropriate $\underset{I}{\text{Disc} } \left[ i \psi^{(D)} \right]$ by highlighting the internal lines (``$I$'' stands for ``Internal''). 
Each of these internal lines can be ``cut'' by replacing them in $D$ with a pair of external lines---i.e. if a line connecting vertices at $t_1$ and $t_2$ is cut, then it is replaced by two external lines that connect $t_1$ to the boundary and $t_2$ to the boundary.
By cutting one or more of the highlighted lines, we produce from the original diagram $D$ a number of ``cut diagrams'', which we denote by $D_{C}$, where $C \subseteq I$ is a list of which internal lines have been cut (``$C$'' stands for ``cut'' and all cut lines are highlighted). 
Notice that, as a result of the cutting, $D_C$ may no longer be connected---we denote by $D_{C}^{(n)}$ the connected subdiagrams contained within $D_C$, and furthermore use $I_n \subseteq I$ to denote which internal lines are contained in $D_C^{(n)}$. 

\item[Step 2.] 
To each cut diagram $D_C$, we associate a function $\tilde{D}_C [ \psi]$ of the wavefunction coefficients in the following way.
First, notice that using the rules of Section~\ref{sec:rules}, we can associate a $\psi^{(D_C^{(n)} )}$ to each connected subdiagram $D_C^{(n)}$. We then take its $\text{Disc}$ with respect to both its internal lines $I_n$ as well as any cut lines. Finally, we replace each cut momenta $\bfp_a$ listed in $C$ with a pair of momenta $\{ \bfq_a, \, \bfq_a'\}$ and a factor of the power spectrum, $P_{\bfq_a \bfq_a'}$, on the boundary. In formulae, this becomes:
\begin{align}
    D_{C}  = \cup_n D_{C}^{(n)}  \;\;\;\; \Rightarrow \;\;\;\;
    \tilde{D}_{C} [ \psi ]  \equiv \left[ \prod_{\substack{ \text{cut lines} \\ a \in C } }^{|C|} \int_{\bfq_a \, \bfq_a'} P_{\bfq_a \bfq_a'} \right] \prod_{\substack{ \text{connected} \\ \text{subdiagrams} \\ n }} \, (-i) \underset{ I_n  \{ q_a\} }{\text{Disc}} \left[ i \, \psi^{ \left( D_C^{(n)} \right) } \right]
    \label{eqn:DC_disconnected}
\end{align}

\end{itemize}
The \textit{general cutting rule} then takes the simple form,
\begin{align}
    \sum_{ \substack{ \text{cuts} \\[2pt] C \subseteq I }}^{2^{|I|}} \, \tilde{D}_C [ \psi ]  = 0 \; , 
    \label{eqn:SumCuts}
\end{align}
where the sum is over all possible ways to cut the internal lines $I$ in the diagram $D$. In particular, since the term $C = \{ \}$ corresponds to not performing any cuts, the corresponding $D_{ \{ \} }$ is simply the original diagram $D$, and so separating this term out we have,
\begin{align}\label{statement}
   i \,\underset{I}{\text{Disc}} \left[ i \, \psi^{(D)} \right] &=  \sum_{ \substack{ C \subseteq I \\ C \neq \{ \} } }^{2^{|I|}-1} \, 
    \left[ \prod_{a \in C}^{|C|} \int_{\bfq_a \, \bfq_a'} P_{\bfq_a \bfq_a'} \right] \prod_{n} \, (-i) \underset{I_n \{ q_a \} }{\text{Disc}} \left[ i \,\psi^{ \left( D_C^{(n)} \right) } \right]
    \; , 
\end{align}
which expresses a particular discontinuity of the diagram $D$ in terms of a sum over diagrams that have at least one line cut. This is a more precise statement of the general cutting relations described in words in \eqref{same}, and is the central result of this work. This result relates the discontinuity of an arbitrary diagram to those of diagrams with fewer loops and/or fewer external legs. We will now prove \eqref{statement}. First, as a lemma we will prove an algebraic identity for the imaginary part of the product of bulk-to-bulk propagators. Second, we will integrate this identity to arrive at \eqref{statement}.

%%%%%%%%%%%%%%%%
\subsection{Lemma: A Propagator Identity} \label{sec:lemma}
%%%%%%%%%%%%%%%%

Our overall strategy is to first consider the \emph{integrands} that appear in each wavefunction coefficient. 
To each diagram $D$ we associate an integrand $\hat{D}$ using the Feynman rules of Section~\ref{sec:rules}, namely a product of bulk-to-boundary and bulk-to-bulk propagators. Since any lines which are not highlighted can be factored out of the sum in \eqref{eqn:SumCuts}, we need only focus on the highlighted lines. The cutting procedure described above corresponds to replacing the $G_p (t_1, t_2)$ from each cut line with,
\begin{align}
    \bfp_a \; \text{line cut} \;\;\;\; \Rightarrow \;\;\;\; G_{p_a} (t, t') \to %\cancel{G}_{p_a} (t, t') = 
    - 2 P_{p_a} K_{p_a} (t ) K_{p_a} (t') \; ,
    \label{eqn:Gcut_def}
\end{align}
where the cut propagator factorises into separate functions of $t$ and $t'$. 
If $D_C$ is disconnected by the cuts, then $\hat{D}_C$ is defined analogously to \eqref{eqn:DC_disconnected}: by taking the product of the imaginary part of each connected subdiagram, after multiplying each by a factor\footnote{
We will see below that this factor arises both because the $\text{Disc}$ in \eqref{defdisc} is related to $\text{Im}$ by a factor of $2i$, and also due to the overall factor of $i^{1-L}$ in the Feynman rules.
} of $(2i)^{L_n}$, where $L_n$ is the number of loops in the subdiagram $\hat{D}_C^{(n)}$,
\begin{align}
    D_{C}  = \cup_n D_{C}^{(n)}  \;\;\;\; \Rightarrow \;\;\;\; \hat{D}_C \equiv  \prod_{ \substack{ \rm connected \\ \text{subdiagrams,} \, n} }\text{Im} \left[ \;  (2i)^{L_n} \hat{D}_{C}^{(n)}  \; \right]  = 0 \; .
\end{align}

We will now prove the following lemma: for any fixed ordering of the vertex times, \eqref{eqn:SumCuts} is obeyed by the integrands, namely,
\begin{align}
\sum_{ \substack{ \text{cuts} \\[2pt] C \subseteq I }}^{2^{|I|}} \hat{D}_C 
% \equiv \sum_{ \substack{ \rm cuts  \\ C }} \; \prod_{ \substack{ \rm disconnected \\ \text{subdiagrams,} \, n} }\text{Im} \left[ \;  (2i)^{L_n} \hat{D}_{C}^{(n)}  \; \right]  
= 0 \; .
\label{eqn:SumCutsG}
\end{align}
Looking ahead, in Section~\ref{sec:proof} we will integrate this lemma over all times and loop momenta to replace each $\hat{D}_C$ with $\tilde{D}_C [ \psi]$, which will hence prove \eqref{eqn:SumCuts}. 

%%%%
\paragraph{Proof:}
%%%%
We begin our proof of \eqref{eqn:SumCutsG} by noting that there is always a \textit{largest time vertex} in the diagram $D$, which we denote by $\bar{t}$ (we assume this is unique, but the same argument works if there are multiple vertices at this largest time). Bulk-to-bulk propagators connected to the largest time vertex simplify because by the definition of $G_p$ we have
\begin{align}\label{largestt}
    G_p(\bar t , t)=2 P_p \, K_p(t) \Im K_p(\bar t) \;\;\;\; \text{when} \;\;\;\; \bar t \geq t \,.
\end{align}
Then, by grouping the terms in \eqref{eqn:SumCutsG} into pairs of cut diagrams which differ by the cutting of only a single line which is connected to $\bar{t}$, we can systematically reduce the number of highlighted lines left to consider. 
For instance, consider two diagrams which differ only in whether the highlighted line between $\bar{t}$ and some other $t_j$ is cut. 
There are only two distinct possibilities: either (i) cutting the $t_j \to \bar{t}$ line separates the diagram into two disconnected pieces, or (ii) the line $t_j \to \bar{t}$ is part of a loop and so cutting this line does not separate the diagram (but does reduce the number of loops by one). These two cases are shown in Figure~\ref{fig:LoopCuttingRuleProof}.  
Considering each in turn:

\begin{figure}
\centering
\includegraphics[width=0.8\textwidth]{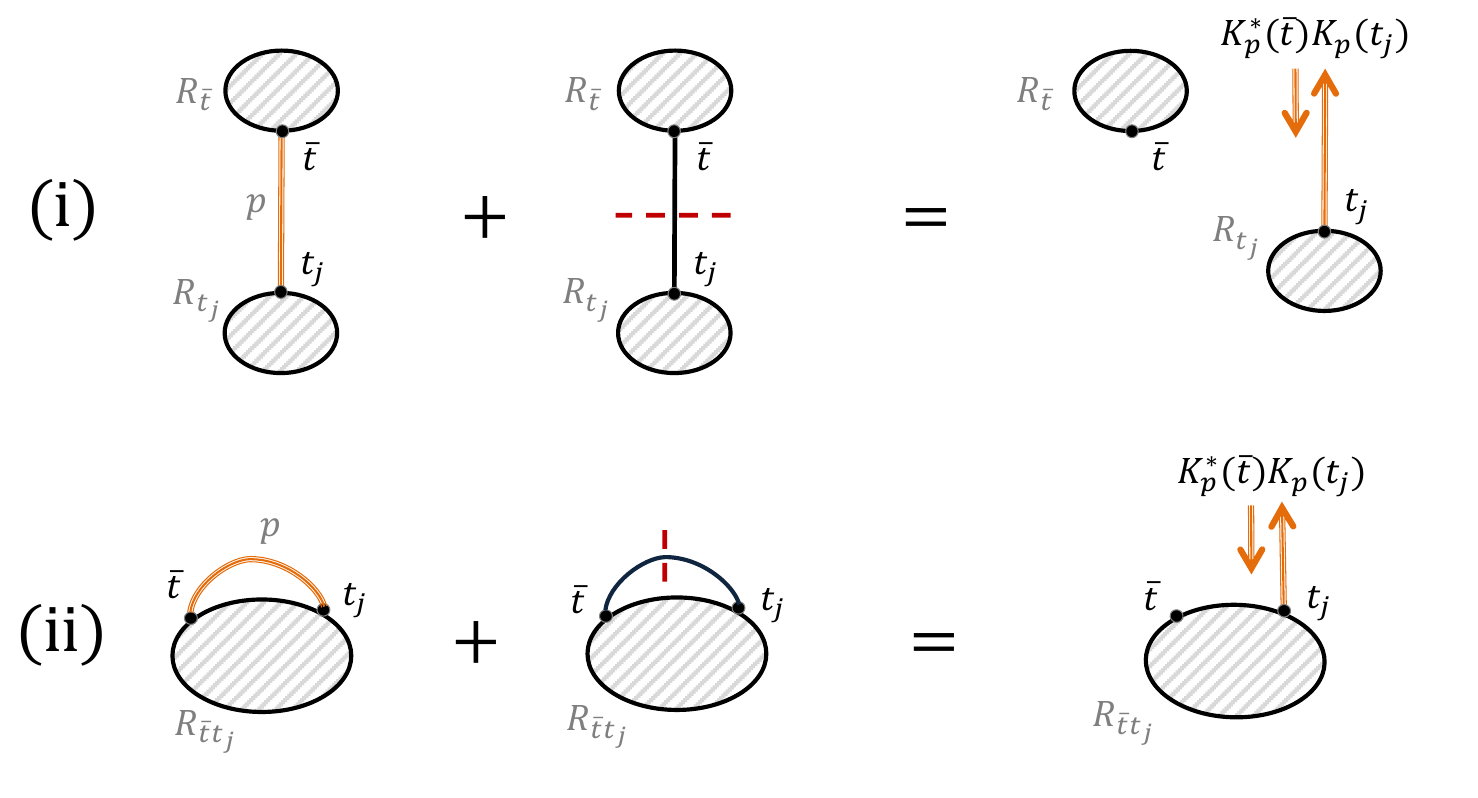}
\caption{An internal line between the largest time vertex $\bar{t}$ and another time $t_j$ is either (i) connecting two otherwise disconnected components, or (ii) forming part of a loop (such that the graph remains connected once it is removed). The pairwise additions shown correspond to \eqref{eqn:PairwiseSum(i)} and \eqref{eqn:PairwiseSum(i)} respectively.}
\label{fig:LoopCuttingRuleProof}
\end{figure}

\begin{itemize}

\item[(i)]  If the two disconnected subdiagrams after the cut are $R_{\bar{t}}$ and $R_{t_j}$, then by using \eqref{largestt} we find
\begin{align}
 \text{Im} \left[ R_{\bar t} \, G_p ( \bar{t} , t_j ) \, R_{t_j} \right] - 2 P_p \text{Im} \left[ R_{\bar t} \, K_p ( \bar t ) \right] \text{Im} \left[ K_p (t_j) R_{t_j}   \right]  \nonumber \\
 \qquad =   -2 P_p \text{Im} \left[ K_p ( \bar t )^* K_p ( t_j ) R_{t_j}   \right]\;  \text{Im} \left[ R_{\bar t} \right]\,.
 \label{eqn:PairwiseSum(i)}
\end{align} 
Hence, we can treat this as an amputation of everything which was connected to the $\bar{t}$ vertex by the $t_j \to \bar{t}$ line. Since the $t_j$ dependence of the right-hand-side has completely factorised, we have extracted the $\text{Im} \left[ R_{\bar{t}} \right]$ of the subdiagram containing $\bar{t}$. This reduces the number of highlighted lines we need to consider by the number of highlighted lines in the amputated $R_{t_j}$.  
%For tree-level diagrams, since every line is in this Case (i), \eqref{eqn:SumCuts} is essentially proven by iterating \eqref{eqn:PairwiseSum(i)} until arriving at $R_{\bar t} = 1$. (see e.g. Figure~\ref{fig:TreeCuttingRuleProofEg}).

\item[(ii)]  If instead the line $t_j \to \bar{t}$ is part of a loop, then if we denote the connected remainder after its removal by $R_{\bar{t} t_j}$, we have that (again using \eqref{largestt})
\begin{align}
 \text{Im} \left[ 2 i G_p ( \bar{t} , t_j ) \, R_{\bar{t} t_j} \right] - \text{Im} \left[ 2 P_p K_p(\bar t) K_p(t_j)  \, R_{\bar{t} t_j}  \right] = - 2 P_p \text{Im} \left[ R_{\bar{t} t_j} K_p (\bar{t})^* K_p (t_j) \right] \; , 
 \label{eqn:PairwiseSum(ii)}
\end{align} 
where there is an additional $2i$ in the first term since before the line is cut there is one additional loop. This pairwise sum has reduced the number of internal lines remaining by $1$, and simply rescales the remaining diagram by a factor of $- 2 P_p \, K_p (\bar{t})^* K_p (t_j)$. 
%Iterating \eqref{eqn:PairwiseSum(ii)} can remove all of the loops from an arbitrary diagram, leaving a sum over tree-level diagrams which cancel as described in Case (i). 

\end{itemize}
  
In either case, the line connecting $\bar{t}$ and $t_j$ has been removed by this pairwise combination and the number of highlighted lines left to consider has decreased. Repeating this for all other highlighted lines which are connected to $\bar{t}$ eventually amputates every highlighted line, leaving a remainder ($R_{\bar{t}}=1$) with vanishing discontinuity ($\text{Im} \left[ R_{\bar t} \right] =  0$ in \eqref{eqn:PairwiseSum(i)}). This proves the claim of lemma \eqref{eqn:SumCutsG}. 
To make this more explicit, we provide two simple examples below.

\begin{figure}[t]
\centering
\includegraphics[width=0.99\textwidth]{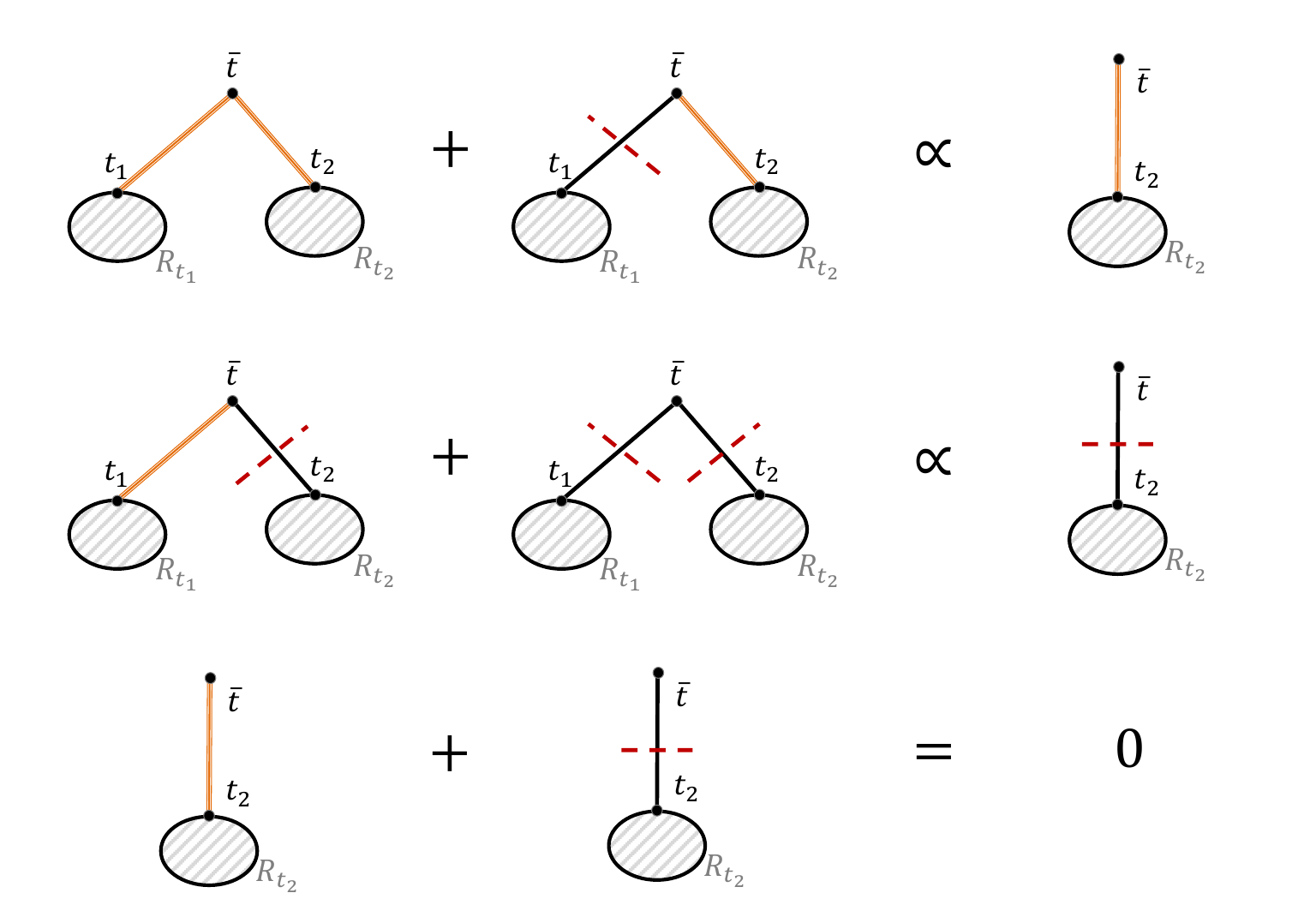}
\caption{
There are four ways to cut two propagators attached to the largest time vertex, $\bar{t}$. They can be paired together as shown in the first and second lines, which amputates all of the $t_1$ dependence. The constants of proportionality are the same, and so adding the two diagrams as on the third line shows that this sum vanishes (see \eqref{eqn:tree_eg_1} and \eqref{eqn:tree_eg_2}).
}
\label{fig:TreeCuttingRuleProofEg}
\end{figure}

%%%%
\paragraph{A Tree-Level Example:}
%%%%
Consider the simple tree-level diagram in which two vertices (at times $t_1$ and $t_2$) are attached by highlighted lines to the largest time vertex $\bar{t}$ ($>$ $t_1$ and $t_2$). 
Focusing on just these two lines, there are four distinct cuts which contribute to the cutting rule \eqref{eqn:SumCuts}.
They can be collected into two pairs, as shown in Figure~\ref{fig:TreeCuttingRuleProofEg}, 
\begin{align}
\hat{D}_{ \{ \} } + \hat{D}_{ \{ p_1 \} } &=\text{Im} \left[ R_{t_1} \, G_{p_1} ( t_1 , \bar{t} ) G_{p_2} ( t_2 , \bar{t} ) \, R_{t_2} \right] - 2 P_{p_1} \text{Im} \left[ R_{t_1} K_{p_1} (t_1) \right]  \text{Im} \left[ K_{p_1} ( \bar{t} ) G_{p_2} (t_2, \bar{t}) \, R_{t_2}   \right]  \nonumber \\
 &\propto  \text{Im} \left[ G_{p_2} (t_2, \bar{t} ) \, R_{t_2} \right]  \label{eqn:tree_eg_1} 
 \end{align}
\begin{align}
\hat{D}_{ \{ p_2 \} } + \hat{D}_{ \{ p_1 ,p_2 \} } &=
-2 P_{p_2} \text{Im} \left[ R_{t_1} \, G_{p_1} ( t_1 , \bar{t} ) K_{p_2} (\bar{t} ) \right] \text{Im} \left[ K_{p_2} ( t_2 ) \, R_{t_2} \right] \nonumber \\
&\qquad\qquad + 4 P_{p_1} P_{p_2} \text{Im} \left[ K_{p_1} (\bar{t} ) K_{p_2} (\bar t) \right] \text{Im} \left[ R_{t_1} K_{p_1} (t_1) \right]  \text{Im} \left[ R_{t_2} K_{p_2} ( t_2 ) \right]  \nonumber \\
 &\propto - 2 P_{p_2} \text{Im} \left[ K_{p_2} (\bar{t} ) \right] \text{Im} \left[ K_{p_2} ( t_2  ) \, R_{t_2} \right]
 \label{eqn:tree_eg_2}
\end{align} 
where the common constant of proportionality is $-2 P_{p_1} \text{Im} \left[ K_{p_1} (t_1) K_{p_1}^* (\bar t) R_{t_1} \right]$, 
which is easily confirmed using \eqref{largestt}. The right-hand-sides of the above equations can be recognised as the discontinuity of diagrams in which $n=1$ vertices are attached to $\bar{t}$, and indeed their sum exactly cancels again by use of \eqref{largestt}. 
Once these integrands are integrated over all times and momenta to make full wavefunction coefficients, this relation effectively reproduces the cutting rule \eqref{eqn:cut_2_eg} given in Section~\ref{sec:cutting_two}.

\begin{figure}[t]
\centering
\includegraphics[width=0.95\textwidth]{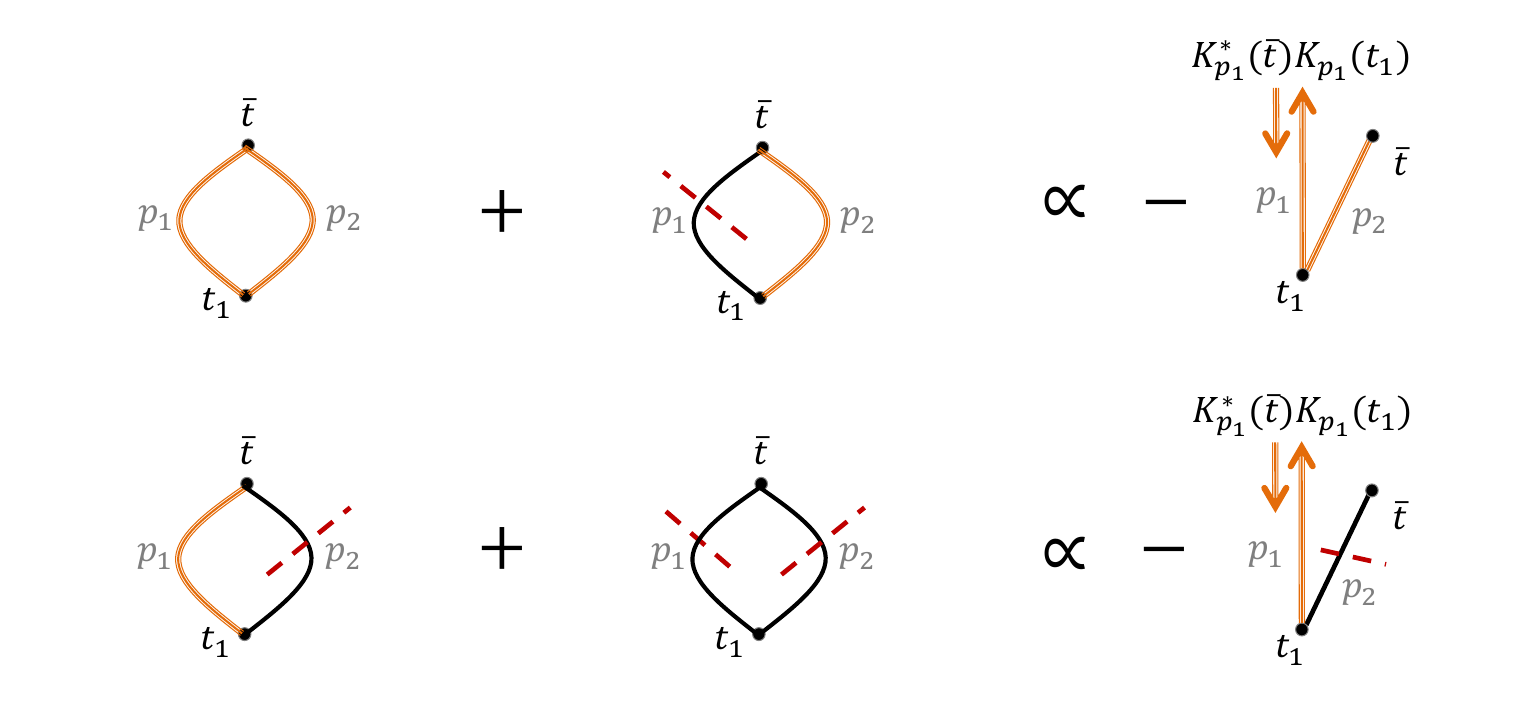}
\caption{
Diagrammatic representation of \eqref{eqn:LoopCuttingRuleProofEg1} and \eqref{eqn:LoopCuttingRuleProofEg}, showing the pairwise sum of two loop diagrams that differ only by the cut of a single line. 
The two terms on the right-hand-sides exactly cancel.
}
\label{fig:LoopCuttingRuleProofEg1}
\end{figure}

%%%%
\paragraph{A One-Loop Example:}
%%%%
Consider the one-loop diagram shown in Figure~\ref{fig:LoopCuttingRuleProofEg1}. Algebraically, the cutting rules associate to each of these diagrams,
\begin{align}
\hat{D}_{\{ \}} + \hat{D}_{\{ p_1 \} } &= \text{Im} \left[ (2i) G_{ p_1} ( \bar{t} , t_1 ) G_{p_2} (\bar{t}, t_1)   \right]
 - 2 P_{p_1} \text{Im} \left[ K_{p_1} (\bar{t} ) K_{p_1} (t_1)  G_{ p_2} (\bar{t}, t_1)   \right]   \nonumber \\ 
&= - 2 P_{p_1} \text{Im} \left[  K^*_{p_1} (\bar t) K_{p_1} (t_1) G_{p_2} (\bar{t},  t_1)   \right]  \label{eqn:LoopCuttingRuleProofEg1}  \\[10pt]
\hat{D}_{\{ p_2 \}} + \hat{D}_{\{ p_1 , p_2 \} } &= -2 P_{p_2} \text{Im} \left[  G_{ p_1 } (\bar{t} , t_1 )  K_{p_2} (\bar{t} ) K_{p_2} (t_1) )    \right]  \nonumber \\
&\qquad\qquad + 4 P_{p_1} P_{p_2} \text{Im} \left[ K_{p_1} (\bar{t} ) K_{p_2} (\bar{t})   \right] \text{Im} \left[ K_{p_1} (t_1) K_{p_2} (t_1) \right]  \nonumber \\
 &= +4 P_{p_1} P_{p_2} \text{Im} \left[ K_{ p_2} (\bar{t})  \right]  \text{Im} \left[  K^*_{p_1} (\bar t) K_{p_1} (t_1) K_{p_2} (t_1)  \right]  \; .
 \label{eqn:LoopCuttingRuleProofEg}
\end{align}
The two terms on the right-hand side sum up to zero by virtue of \eqref{largestt}. 
Once integrated over all times and momenta, this reproduces the cutting rule \eqref{eqn:cut_1loop_2} given in Section~\ref{sec:cutting_loop}.

%%%%%%%%%%%%%%%%
\subsection{Proof of the Cutting Rules}
\label{sec:proof}
%%%%%%%%%%%%%%%%

The lemma \eqref{eqn:SumCutsG} generalises the identities \eqref{eqn:ImG} for $\text{Im} \, G$ (used to cut a single propagator), \eqref{eqn:ImGG} for $\text{Im} \, G_1 G_2$ (used to cut two propagators at tree-level) and \eqref{eqn:ReG} or \eqref{eqn:ReGG} for $\text{Re} \, G$ or $\text{Re} \, G_1 G_2$ (used to cut one or two propagators in a loop) to \emph{any} number of propagators which form \emph{any} number of loops. For convenience we list the first several of these identities in Appendix~\ref{app:propagator}.   
We will now use these general propagator identities to prove the cutting rules \eqref{statement} for an arbitrary $L$-loop diagram. 

First, we can express an arbitrary wavefunction coefficient in terms of an integrand by stripping off all external legs and their associated time integrals, as well as the momentum-conserving $\delta$-functions at each vertex,
\begin{align}
     \psi^{ (D) } =  \left[ \prod_{j=1}^N \int dt_j \prod_{\bfk_a} K_{k_a} (t_j) \right] \hat{\psi}^{(D)} (t_1, ... , t_N) \times \left( \delta \, \text{functions} \right) \; .
     \label{eqn:psihat_def}
\end{align}
For the original connected diagram, the integrand $\hat{\psi}$ contains a product of $|I|$ internal propagators (where $|I|$ is the number of elements in the set $I$), an integral over their momenta (all but $L$ of these integrals may be fixed by the $\delta$-functions), and an overall factor of $i^{1-L}$, as per the Feynman rules of Section~\ref{sec:rules},
\begin{align}
     \hat{\psi}^{ (D) } (t_1, ..., t_N)  =  i^{1-L} \int_{\bfp_1 ... \bfp_{|I|}} G_{p_1} ... G_{p_{|I|} } 
     \label{eqn:psihatD}
\end{align}
where the propagators may depend on any of the times $(t_1, ... , t_N)$.
Using a $\text{Disc}$ to take the imaginary part of this product of propagators, we have, 
\begin{align}
    -i \, \underset{ I }{ \text{Disc} } \left[ i \,  \hat{\psi}^{ (D) } (t_1, ..., t_N) \right] =  \int_{\bfp_1 ... \bfp_{|I|}} (-2)^{1-L} \hat{D}_{\{ \} } \; .
\end{align}
For the cut diagrams $D_C$, the integrand $\hat{\psi}^{(D_C)}$ is given by the analogue of \eqref{eqn:psihatD} with the cut propagators replaced as in \eqref{eqn:Gcut_def}. 
For instance, if after cutting the line $\bfp_1$ the diagram remains connected (case (ii) above), then $D_{\{ \bfp_1 \} }$ has $L-1$ loops and two additional external legs, and so $\hat{D}_{\{\bfp_1\}}$ is related to a wavefunction integrand by,
\begin{align}
    \int_{\bfq_1 \bfq_1'} P_{\bfq_1 \bfq_1'} (-i) \underset{I q_1 q_1'}{\text{Disc}} \left[ i \,  \hat{\psi}^{ (D_{ \{ \bfp_1 \}} ) } (t_1, ..., t_N) \right] =  \int_{\bfp_1 ... \bfp_{|I|} } (-2)^{1-L} \hat{D}_{ \{ \bfp_1 \}  } \; .
\end{align}
On the other hand, if after cutting the line $\bfp_1$ the diagram becomes disconnected (case (i) above), then we have,
\begin{align}
    \int_{\bfq_1 \bfq_1'} P_{\bfq_1 \bfq_1'} (-i) \underset{I_1 q_1}{\text{Disc}}  \left[ i \,  \hat{\psi}^{ (D^{(1)}_{ \{ \bfp_1 \}} ) } (t_1, ..., t_N) \right] &(-i)  \underset{ I_2 q_1' }{\text{Disc}} \left[ i \,  \hat{\psi}^{ (D^{(2)}_{ \{ \bfp_1 \}} ) } (t_1, ..., t_N) \right]  \nonumber \\
    &\qquad=  \int_{\bfp_1 ... \bfp_{|I|} } (-2)^{1-L} \hat{D}_{ \{ \bfp_1 \}  } \; .
\end{align}
Proceeding in this way for diagrams with two, three, ... etc. cuts, we can replace each $\hat{D}_C$ in lemma \eqref{eqn:SumCutsG} with products of $\hat{\psi}^{(D_C)}$ discontinuities, 
\begin{align}
    \int_{\bfp_1 ... \bfp_{|I|}} (-2)^{1-L} \sum_{C \subseteq I  }^{2^{|I|}} \hat{D}_C
    =  \sum_{ \substack{ C \subseteq I } }^{2^{|I|}} \, 
    \left[ \prod_{a \in C}^{|C|} \int_{\bfq_a \, \bfq_a'} P_{\bfq_a \bfq_a'} \right] \prod_{n} \, (-i) \underset{ I_n  \{ q_a \} }{ \text{Disc} } \left[ i \, \hat{\psi}^{ \left( D_C^{(n)} \right) } (t_1, ...,  t_N) \right]
    \; .
    \label{eqn:psihat_cutting_rule}
\end{align}
By our propagator lemma \eqref{eqn:SumCutsG}, the sum on the left-hand-side vanishes. 

The final step is then to multiply by the external propagators and perform the integrals in \eqref{eqn:psihat_def} over the vertices $(t_1, ... , t_N)$ at which they could be attached to the diagram. The crucial property we adopt in this final step is that we can bring the time integrals and the factors of $K_k$ in \eqref{eqn:psihat_def} \emph{inside} the argument of the $\disc$. This is allowed whenever there exists an analytic continuation $\bar{k}$ such that $K_k(\eta)=K^\ast_{\bar k}(\eta)$ for every $\eta$, since then,
\begin{align}
    \disc\left[  R \prod_a^n K_{k_a}\right]&\equiv R \prod_a^n K_{k_a} + \left( R \prod_a^n K_{\bar k_a} \right)^\ast \\
    &=\prod_a^n K_{k_a} \disc \left[ R\right]\,,
\end{align}
for any $R$, as discussed in \cite{Goodhew:2020hob, Cespedes:2020xqq}.
It is not always possible a priori to find such a $\bar{k}$, but a simple solution for $\bar k$ turns out to exists under surprisingly general circumstances \cite{single}. To see this, note that in Minkowski, where $K\sim e^{ikt}$, the above implicit equation for $\bar k$ has solution $\bar k=-k^\ast$, reducing to simply a minus sign for real $k$. In analogy with amplitudes, one can name this property \textit{Hermitian analyticity}, namely $K_k(\eta)=K^\ast_{-k^\ast}(\eta)$. The choice of a Bunch-Davies vacuum enforces $K_k$ on any FLRW spacetime to match the Minkowski result at early times. Then one can prove that, as long as the the coefficients of the linearized equations of motion are not singular in the past, Hermitian analyticity is maintained as time evolves and in particular it remains valid even when the mode function become dramatically different from those in flat spacetime \cite{single}. Indeed, it is easy to see that Hermitian analyticity is satisfied for both massless and conformally coupled scalar fields\footnote{Notice that massless gravitons have the same mode functions as massless scalars, and so they too obey Harmitian analyticity. Also, where the limit is finite we have taken $\eta_0 \to 0$.}
\begin{align}\label{modefct}
    K_k(\eta)&=(1-ik\eta)e^{ik\eta}\,, & \text{(massless scalar)}\\
    K_k(\eta)&=\frac{\eta}{\eta_0} e^{ik\eta}\,, & \text{(conformally coupled scalar)}\,.
\end{align}
The above discussion allows us to promote each $\hat{\psi}^{(D)}$ in \eqref{eqn:psihat_cutting_rule} to $\psi^{(D)}$, and hence proves the general cutting rule \eqref{statement}.

%%%%
\paragraph{A One-Loop Example:}
%%%%
For instance, for the one-loop example given above (see Figure~\ref{fig:LoopCuttingRuleProofEg1}),
%in Section~\ref{sec:cutting_loop}, 
the diagram with zero-, one- or two-cuts corresponds to wavefunction coefficient integrands, 
\begin{align}
\hat{\psi}^{(D)}_{ \{ \bfk \} } (t_1, t_2)  &= \int_{\bfp_1 \bfp_2} G_{p_1} (t_1, t_2) G_{p_2} (t_1, t_2)  \; ,  \nonumber \\
\hat{\psi}^{ \left( D_{ \{ \bfp_1 \} } \right)}_{\{ \bfk \} \bfq_1 \bfq_1' } (t_1, t_2)  &= i \int_{\bfp_2} K_{q_1} (t_1) K_{q_1'} (t_2)  G_{p_2} (t_1, t_2) \; , \nonumber \\
\hat{\psi}^{ \left( D_{ \{ \bfp_1 , \bfp_2\} }^{(1)} \right) }_{\{ \bfk \} \bfq_1  \bfq_2 } (t_1, t_2)  &=  i \, K_{q_1} (t_1) K_{q_2} (t_1)  \;\;\;\; , \;\;\;\; \hat{\psi}^{(D_{ \{ \bfp_1 , \bfp_2\} }^{(2)} )}_{\{ \bfk \} \bfq_1' \bfq_2' } (t_1, t_2)  &=  i \, K_{q_1'} (t_2) K_{q_2'} (t_2) \; .
\end{align}
We can therefore write the $\hat{D}_C$ given in equations \eqref{eqn:LoopCuttingRuleProofEg1} and \eqref{eqn:LoopCuttingRuleProofEg} above in terms of the wavefunction integrands,
\begin{align}
\int_{\bfp_1 \bfp_2} \hat{D}_{ \{ \} } &= \int_{\bfp_1 \bfp_2} \text{Im} \left[ 2 i G_{p_1} ( t_1, t_2 )  G_{ p_2} ( t_1 , t_2 )   \right]  \nonumber \\
&= -i \, \text{Disc}  \left[ i \hat{\psi}^{(D)}_{\{ \bfk \} } (t_1, t_2) \right]  \\[10pt]
\int_{\bfp_1 \bfp_2} \hat{D}_{ \{ \bfp_1 \} } &= \int_{\bfp_1 \bfp_2}  -  2 P_{p_1} \text{Im} \left[ K_{p_1} (\bar{t} ) K_{p_1} (t_1)  G_{ p_2} (\bar{t}, t_1)   \right]  \nonumber \\
&=  \int_{\bfq_1 \bfq_1'}  - P_{\bfq_1 \bfq_1'} i\, \underset{ q_1 q_2}{\text{Disc}} \left[ i \hat{\psi}^{(D_{ \{ \bfp_1 \} })}_{ \{ \bfk\} \bfq_1 \bfq_1' } (t_1, t_2)  \right]   \\[10pt]
\int_{\bfp_1 \bfp_2} \hat{D}_{ \{ \bfp_1, \bfp_2 \} } &= \int_{\bfp_1 \bfp_2} 4 P_{p_1} P_{p_2} \text{Im} \left[ K_{p_1} ( t_1 ) K_{p_2} (t_1) \right] \text{Im} \left[  K_{p_1} (t_2) K_{p_2} (t_2)   \right]  \nonumber \\
&=  \int_{\substack{ \bfq_1 \bfq_1' \\ \bfq_2 \bfq_2' }}  - P_{\bfq_1 \bfq_1'} P_{\bfq_2 \bfq_2'} 
\underset{ q_1 q_2}{\text{Disc}} \left[ i \hat{\psi}^{(D_{ \{ \bfp_1 , \bfp_2 \} }^{(1)})}_{ \{ \bfk\}  \bfq_1 \bfq_2 } (t_1, t_2)  \right] 
\underset{q_1' q_2'}{\text{Disc}} \left[ i \hat{\psi}^{(D_{ \{ \bfp_1 , \bfp_2 \}}^{(2)})}_{ \{ \bfk\}  \bfq_1' \bfq_2' } (t_1, t_2)  \right] \; .
\end{align}
The propagator lemma \eqref{eqn:SumCutsG} for this diagram can therefore be written as,
\begin{align}
    0 &= \int_{\bfp_1 \bfp_2} \left(  \hat{D}_{\{\}} + \hat{D}_{ \{ \bfp_1 \} } + \hat{D}_{ \{ \bfp_2 \} } + \hat{D}_{ \{ \bfp_1 , \bfp_2 \} }  \right) \nonumber \\[10pt]
    &= - i \,  \text{Disc}  \left[ i \hat{\psi}^{(D)}_{\{ \bfk \} } (t_1, t_2) \right]  \nonumber \\
    & + \int_{\bfq_1 \bfq_1'}  P_{\bfq_1 \bfq_1'} (-i) \underset{ q_1 q_2}{\text{Disc}} \left[ i \hat{\psi}^{(D_{ \{ \bfp_1 \} })}_{ \{ \bfk\} \bfq_1 \bfq_1' } (t_1, t_2)  \right] 
    + \int_{\bfq_2 \bfq_2'}  P_{\bfq_2 \bfq_2'}  (-i) \underset{ q_1 q_2}{\text{Disc}} \left[ i \hat{\psi}^{(D_{ \{ \bfp_1 \} })}_{ \{ \bfk\} \bfq_1 \bfq_1' } (t_1, t_2)  \right]  \nonumber \\
    &+ \int_{\substack{ \bfq_1 \bfq_1' \\ \bfq_2 \bfq_2' }}  P_{\bfq_1 \bfq_1'} P_{\bfq_2 \bfq_2'} (-i)
\underset{ q_1 q_2}{\text{Disc}} \left[ i \hat{\psi}^{(D_{ \{ \bfp_1 , \bfp_2 \} }^{(1)})}_{ \{ \bfk\}  \bfq_1 \bfq_2 } (t_1, t_2)  \right] (-i)
\underset{ q_1' q_2'}{\text{Disc}} \left[ i \hat{\psi}^{(D_{ \{ \bfp_1 , \bfp_2 \}}^{(2)})}_{ \{ \bfk\}  \bfq_1' \bfq_2' } (t_1, t_2)  \right] \; .
\label{eqn:pre_cut_eg}
\end{align}
Finally, multiplying by the external propagators and performing the integrals over the times $(t_1, t_2)$ replaces each of these integrands $\hat{\psi}^{(D)}$ with the corresponding coefficient $\psi^{(D)}$, and therefore \eqref{eqn:pre_cut_eg} implies the cutting rule,
\begin{align}
    i \, \text{Disc} \left[ i \psi^{(D)} \right] &=  \sum_{ \substack{ C \subseteq \{ \bfp_1, \bfp_2 \} \\ C \neq \{ \} } }^{3} \, 
    \left[ \prod_{a \in C}^{|C|} \int_{\bfq_a \, \bfq_a'} P_{\bfq_a \bfq_a'} \right] \prod_{n} \, (-i) \underset{ \{ q_a \} }{ \text{Disc} } \left[ i \, \psi^{ \left( D_C^{(n)} \right) } \right]
    \; , 
\end{align}
for this diagram, where in this case the internal momenta are integrated over so the $\text{Disc}$ with no argument on the left-hand-side corresponds to analytically continuing all (and only) the external momenta.

%%%%%%%%%%%%%%%%
\subsection{Extension to Multiple Fields of any Mass and Spin}
\label{sec:multiple}
%%%%%%%%%%%%%%%%

The Cosmological Cutting Rules have been presented so far for a single massless scalar field in de Sitter spacetime with a Bunch-Davies vacuum. However, the same rules apply to the much more general case of any (finite) number of fields of any mass and spin. Here, we only sketch the main argument and refer the reader to \cite{single} for more details on spinning fields and more general FLRW spacetimes.\\

Our proof so far relied on two properties. The first are the propagators identities proven in the lemma in Section \ref{sec:lemma}. These are very general and only rely on the form of the bulk-to-bulk propagator $G_p$ in terms of the bulk-to-boundary propagator $K_k$. The proof assumes nothing about the function $K_k$. This result is therefore valid for any number of fields with any mode functions. It is straightforward to extend this proof to allow for fields of different species/spins. This amounts to decorating the propagators in the lemma \eqref{eqn:SumCutsG} with additional indices that denote any additional quantum numbers. For example, for the cutting rule \eqref{eqn:cut_2}, this amounts to writing,
\begin{align}
& \underset{p_L p_R}{\text{Disc}} \left[ i \psi^{ \{\alpha_L \} \{ \alpha_M \} \{ \alpha_R \} }_{ \{ \bfk_L \} \{ \bfk_M \} \{ \bfk_R \} } \right] \nonumber \\
&= \int_{ \bfq_L \bfq_L' } \; 
\underset{q_L}{\text{Disc}} \left[ i\psi^{\{ \alpha_L \} \beta_L }_{ \{ \bfk_L \} \bfq_L} \right]
 P^{\beta_L}_{\bfq_L \bfq_L'} 
\underset{ q_L' p_R }{\text{Disc}} \left[ i \psi^{ \beta_L \{ \alpha_M \} \{ \alpha_R \} }_{ \bfq_L' \{ \bfk_M \} \{ \bfk_R \} } \right]
 \nonumber \\
&+\int_{ \bfq_R \bfq_R' } \; 
\underset{ p_L q_R }{\text{Disc}} \left[ i\psi^{\{ \alpha_L \} \{ \alpha_M \} \beta_R }_{ \{ \bfk_L \}  \{ \bfk_M \} \bfq_R} \right]
 P^{\beta_R}_{\bfq_R \bfq_R'} 
\underset{  q_R' }{\text{Disc}} \left[ i \psi^{ \beta_R \{ \alpha_R \}  }_{ \bfq_R'  \{ \bfk_R \} } \right]
 \nonumber \\
&+ \int_{ \substack{ \bfq_L \bfq_L' \\ \bfq_R \bfq_R'} } \; 
\underset{ q_L }{\text{Disc}} \left[ i\psi^{\{ \alpha_L \} \beta_L }_{ \{ \bfk_L \} \bfq_L} \right]
 P^{\beta_L}_{\bfq_L \bfq_L'} 
\underset{  q_L' q_R }{\text{Disc}} \left[ i \psi^{ \beta_L \{ \alpha_M \} \beta_R }_{ \bfq_L' \{ \bfk_M \} \bfq_R } \right]
 P^{\beta_R}_{\bfq_R \bfq_R'} 
\underset{ q_R' }{\text{Disc}} \left[ i \psi^{ \beta_R \{ \alpha_R \} }_{ \bfq_R' \{ \bfk_R \} } \right]
\end{align}
where the indices $\alpha$ and $\beta$ collect the other quantum numbers of the fields, such as field type (e.g. flavor), helicity, charges and so on. Notice that these indices are always paired up with the associated momenta. We can therefore omit to write them altogether if we improve our notation to include these indices inside the various $\bfk$'s, $\bfp$'s and $\bfq$'s. The integrals over $\bfq$'s should then be interpreted as having an additional sum over the relevant quantum numbers, for example all the possible helicity of a given spinning field. We refer the reader to \cite{single} for a more explicit discussion and notation.

The second property we needed to translate the propagator identities into equations for the wavefunction coefficient, is that we can find a $\bar k$ such that $K_{\bar k}^\ast (\eta) = K_k(\eta)$ for all times $\eta$. When the fields obey the Bunch-Davies vacuum, this condition is satisfied by $\bar k= -k^\ast$, and we refer to this property of $K_k$ as Hermitian analyticity. In \cite{single} we prove that Hermitian analyticity is valid for fields of any mass and spin on any FLRW spacetime, provided that a weak technical assumption is satisfied by the coefficients of the linearized equations of motion.

%%%%%%%%%%%%%%%%
\section{Inferring Loops from Trees using Perturbative Unitarity}
\label{sec:applications}
%%%%%%%%%%%%%%%%

The general cutting rule derived above allows us to compute the $\text{Disc}$ of a loop-level wavefunction coefficient in terms of simpler tree-level coefficients.
In this section, for a variety of interactions on both Minkowski and de Sitter spacetime backgrounds we will show explicitly how the cutting rules can be used to infer the 1-loop $\text{Disc}$ of the Gaussian width, $\psi_{\bfk_1 \bfk_2}$. 
This provides a new way of estimating when perturbative unitarity breaks down from a purely tree-level calculation. 
Finally, in Section~\ref{sec:physical} we relate these results to the power spectrum. 

%%%%
\paragraph{Perturbative Unitarity:}
%%%%
Unitarity can be used to place a lower bound on the size of loops, given specified tree-level contributions. 
This is familiar from scattering amplitudes on flat space, where the perturbative optical theorem constrains $\text{Im}\, \mathcal{A}^{\text{1-loop}} \geq \int_{\text{Momenta}} | \mathcal{A}^{\rm tree} |^2$.
This can be used to determine at what scale perturbation theory breaks down, in particular when $\int_{\text{Momenta}} | \mathcal{A}^{\rm tree} |^2 \geq | \mathcal{A}^{\rm tree}|$ it signals that $| \mathcal{A}^{\text{1-loop}}|$ must be larger than $| \mathcal{A}^{\text{tree}} |$ if the theory is to remain unitary. 
Our goal in this section is to apply the above cutting rules in a similar spirit, using them to infer the size of 1-loop corrections from purely tree-level calculations.

%%%%
\paragraph{Momenta Integrals:}
%%%%
In the explicit examples below, we will need to evaluate momentum integrals of the form $\int_{\bfq_1 \bfq_2} \delta^3 ( \bfq_1 + \bfq_2 - \bfk  )$ which appear in the cutting rules. Since we have implicitly assumed throughout that the $\text{Disc}$ operation commutes with such integrals, 
\begin{align}
    \text{Disc} \left[ \int_{\bfq_1 \bfq_2} \delta^3 ( \bfq_1 + \bfq_2 - \bfk  ) \, f ( k , q_1, q_2 ) \right] = \int_{\bfq_1 \bfq_2} \delta^3 ( \bfq_1 + \bfq_2 - \bfk  ) \; \underset{q_1 q_2}{\text{Disc}} \left[ f ( k , q_1, q_2 ) \right] \; ,  
\end{align}
we must take care to adopt integration variables which are suitably invariant under $k \to -k$. 
For example, one possible choice of integration variables is,
\begin{align}
\int d^3 \bfq_1 d^3 \bfq_2 \, \delta^3 ( \bfq_1 + \bfq_2 - \bfk  ) \, f ( k , q_1, q_2 ) = 2\pi  \int_0^{\infty} d q_1 \int_{| q_1 - k |}^{|  q_1 + k |} \frac{dq_2}{k} \; q_1 q_2 \; f ( k , q_1, q_2 ) \; ,  
\label{eqn:dp1dp2}
\end{align} 
where the integration limits\footnote{
Note that the limits for $q_2$ follow from $q_2^2 = | \bfk - \bfq_1 |^2$, and it is important to keep the modulus on the $q_1 + k$ upper limit since we allow for $k < 0$ when taking the $\text{Disc}$.  
} are such that the $\text{Disc}\left[ \int_{|q_1-k|}^{|q_1 + k|} \frac{d q_2}{k}  \right] = 0$, and so taking $\text{Disc}$ of this integral amounts to integrating $\underset{q_1 q_2}{\text{Disc}} \,\left[ f (k, q_1 ,q_2) \right]$, as required. 
To simplify the algebra, we will also make use of the following trick: for integrands $f(k,q_1, q_2)$ with the property\footnote{
In general, \eqref{eqn:dp1dp2} corresponds to the integration range $\int_k^{\infty} dq_+ \int_{-k}^{+k} \frac{dq_-}{k}$, which no longer commutes with the $\text{Disc}$ operation. However the property $f(-k,-q_1,-q_2) = -f^* (k,q_1,q_2)$ ensures that $\text{Disc} \left[ \int_0^k dq_+\int_{-k}^{+k} \frac{dq_-}{k} f (k,q_1, q_2) \right] = 0$, which then allows us to write \eqref{eqn:dp+dp-}. 
} that $f(-k,-q_1,-q_2) = -f^*(k,q_1,q_2)$, we can write the $\text{Disc}$ of \eqref{eqn:dp1dp2} as,
\begin{align}
\text{Disc} \left[ \int d^3 \bfq_1 d^3 \bfq_2 \, \delta^3 ( \bfq_1 + \bfq_2 - \bfk  ) \, f ( k , q_1, q_2 ) \right] = \pi \int_0^{\infty} d q_+ \int_{-k}^{k} \frac{d q_-}{k} \; q_1 q_2 \; \underset{q_1 q_2}{\text{Disc}} \left[ f ( k , q_1, q_2 ) \right] \; ,  
\label{eqn:dp+dp-}
\end{align} 
where $q_+ = q_1 + q_2$ and $q_- = q_1 - q_2$. \eqref{eqn:dp+dp-} is often easier to perform since the two integrals are now independent. 

All momentum integrals are to be computed with the prescription $k \to k - i \epsilon$ (i.e. $k$ has a small negative imaginary part) to move poles from the real axis. For instance, using the fact that,
\begin{align}
\int_0^{\infty} \frac{dq}{q} \, \frac{ q^{n} }{ (q + k  )^r } &= \frac{  \Gamma \left( n \right) \Gamma \left( r - n \right) }{ \Gamma (r)  } \; k^{n - r}  \; .
\label{eqn:integral_identity_1}
\end{align}
for all complex $k$, the difference of two such integrals at $k- i \epsilon$ and $e^{-i \pi} (k+ i \epsilon)$ corresponds to,
\begin{align}
\lim_{\epsilon \to 0^+}\int_0^{\infty} \frac{dq}{q} \, \frac{ q^{n} }{ (q^2 + e^{- i \pi} k^2 -i\epsilon )^r } &= \frac{ \Gamma \left( \frac{n}{2} \right) \Gamma \left( r - \frac{n}{2} \right) }{ 2 \Gamma (r)  } \; ( - i k)^{n - 2 r} \;  , 
\label{eqn:integral_identity_2}
\end{align}
where writing $-k^2$ as $e^{-i \pi} k^2$ in the denominator ensures that we have the correct branch of $\sqrt{- k^2} = -i k$ on the right-hand-side. The useful identities \eqref{eqn:integral_identity_1} and \eqref{eqn:integral_identity_2} will be used several times below.

%%%%%%%%
\subsection{On Minkowski}
\label{sec:minkowski}
%%%%%%%%

For a massless scalar field $\phi$ on a Minkowski background, we use mode functions $f_{k} (t) = e^{- i k t}/\sqrt{2k}$, and the corresponding power spectrum is, $P_k = | f_{k} |^2 = 1/(2k)$. 
The bulk-to-boundary and bulk-to-bulk propagators are given by,
\begin{align}
 K_{k} (t) = e^{i k t} \;\;\;\; , \;\;\;\; G_{k} (t_1, t_2 ) = \frac{e^{i k t_2 } }{ k } \sin ( k t_1 ) \theta ( t_1 - t_2 ) + \left( t_1 \leftrightarrow t_2 \right)   \; . 
\end{align}
We will use the cutting rules to compute the one-loop correction to the ($\text{Disc}$ of the) Gaussian width, $\psi_{\bfk_1 \bfk_2}$, from both a quartic $\phi^4$ interaction and a cubic $\phi^3$ interaction.  
Crucially, while the full $\psi^{\text{1-loop}}_{\bfk_1 \bfk_2}$ in both cases is divergent and requires renormalisation, the $\text{Disc} \left[ i\, \psi^{\text{1-loop}}_{\bfk_1 \bfk_2} \right]$ is \emph{finite} and can be inferred directly from the tree-level non-Gaussianities $\psi_{4}$ and $\psi_{3}$.

%%%%%%%%
\subsubsection*{$\phi^4$ on Minkowski} 
%%%%%%%%

For the interaction $\mathcal{L}_{\rm int} = \frac{1}{4!} \lambda \phi^4$, the tree-level quartic wavefunction coefficient is
\begin{align}
 \psi_{\bfk_1 \bfk_2 \bfk_3 \bfk_4}^{\text{tree}} =  \lambda\;  \frac{  \tdelta^3 \left( \bfk_1 + \bfk_2 + \bfk_3 + \bfk_4  \right) }{ k_1 + k_2 + k_3 + k_4 } .
 \label{eqn:psi4tree_eg1}
\end{align}
This is the only simple input required for the cutting rule \eqref{eqn:cut_1loop_1}, which fixes the $\text{Disc}$ of the 1-loop $\psi_{\bfk_1 \bfk_2}$ as,
\begin{align}
    i \, \text{Disc} \left[ i \psi^{\text{1-loop}}_{\bfk_1 \bfk_2} \right]
    =
    \int_{\bfq \bfq'} P_{\bfq \bfq'} \, (-i) \, \underset{q q'}{\text{Disc}} \left[ i \psi_{\bfk_1 \bfk_2 \bfq \bfq'}^{\text{tree}}   \right] \; .
\label{eqn:cuttingrule_eg1}
\end{align}
Explicitly, from \eqref{eqn:psi4tree_eg1} (and the definition \eqref{defdisc} of $\text{Disc}$) we can straightforwardly compute the integrand on the right-hand-side,
\begin{align}
(-i) \underset{q q'}{\text{Disc}} \left[ i \psi_{\bfk_1 \bfk_2 \bfq \bfq'}^{\text{tree}}   \right]
&=   \lambda \frac{ \tdelta^3 \left( \bfk_1 + \bfk_2 + \bfq + \bfq'  \right) }{k_1 + k_2 + q + q'}  + \lambda \frac{\tdelta^3 \left( -\bfk_1 - \bfk_2 - \bfq - \bfq'  \right) }{-k_1 - k_2 + q + q'}   \nonumber \\ 
&=  \lambda \; \frac{2 (q + q')}{ (q + q')^2 - (k_1 + k_2 )^2 } \tdelta^3 \left( \bfk_1 + \bfk_2 + \bfq + \bfq'  \right) \; .
\end{align}
Unlike the loop momentum integral required to evaluate $\psi^{\text{1-loop}}_{\bfk_1 \bfk_2}$ explicitly, the integration on the right-hand-side of \eqref{eqn:cuttingrule_eg1} over external momenta is finite, 
\begin{align}
 \int_{\bfq \bfq'} P_{\bfq \bfq'} (-i) \underset{q q'}{\text{Disc}} \left[ i \psi_{\bfk_1 \bfk_2 \bfq \bfq'}^{\text{tree}}   \right]
 &=  \frac{ \lambda }{ \pi^2 } \int_0^{\infty} dq \; \frac{q^2}{4 q^2 - (k_1 + k_2 {\textcolor{red} - i  \epsilon } )^2} \; \tdelta^3 \left( \bfk_1 + \bfk_2 \right) \nonumber  \\
 &=  \changg{-} \frac{i \lambda}{16 \pi} \;( k_1 + k_2 )  \;\tdelta^3 \left( \bfk_1 + \bfk_2 \right) \;  , 
\end{align}
using the integral identity \eqref{eqn:integral_identity_2}. 
This simple finite integral has computed for us the $\text{Disc}$ of the 1-loop quadratic coefficient,
\begin{align}
 \text{Disc} \left[ i  \, \psi_{\bfk_1 \bfk_2}^{\text{1-loop}} \right] = \changg{-} \frac{\lambda}{16 \pi} \; (k_1 + k_2)  \;\tdelta^3 \left( \bfk_1 + \bfk_2 \right) \; .  
 \label{eqn:Disc_eg1}
\end{align}
We note that \eqref{eqn:Disc_eg1} is consistent with the ``naive dimensional analysis'' (NDA) power counting typically employed for loop amplitudes on flat space\footnote{
Power counting schemes analogous to NDA were developed for inflation in \cite{Adshead:2017srh, Grall:2020tqc} (see also \cite{deRham:2017aoj, Babic:2019ify} for theories with small $c_s$ in particular). 
} \cite{Manohar:1983md}, which keeps track of powers of $4\pi$. 
In this case\footnote{
Note that while NDA was generalised to $d$ dimensions in \cite{Gavela:2016bzc}, this assumed $d$-dimensional Lorentz-invariant kinetic terms for the fields---in our case, although the loop integrals are done over only spatial momenta in $d=3$, the underlying field theory is four dimensional, and so we retain the $4\pi$ counting of $(3+1)$ dimensions. 
}, NDA would give $\psi_{\bfk_1 \bfk_2}^{\text{1-loop}} \sim  \lambda \, k / (4\pi)^2$, and we expect $\text{Disc} \left[ i \psi^{\text{1-loop}}_{\bfk_1 \bfk_2} \right]$ to contain an additional power of $\pi$ since it arises from a logarithmic branch cut. The cutting rules can therefore be viewed as an efficient way of fixing the numerical coefficients, as well as the precise dependence on the momenta, in these power counting formulae. This extends a similar application of unitarity in \cite{Grall:2020tqc}, which focused on scattering amplitudes in the subhorizon limit, to wavefunction coefficients.

%%%%
\paragraph{Comparison with Explicit Computation:}
%%%%
In this simple example, we can check the result \eqref{eqn:Disc_eg1} by performing the loop integral explicitly. 
The quadratic coefficient (with $\delta$-function removed) is given up to $\mathcal{O} (\lambda)$ by,
\begin{align}
 \psi_{\bfk , -\bfk}'  &= - k + \lambda \int_{-\infty}^{0} dt \, K_{k} (t) K_{k} (t) \int_{\bfp} G_{p} (t, t )   \nonumber \\
 &= - k + \frac{ \lambda }{4 k}  \int \frac{d^d \bfp}{(2\pi)^d}  \; \frac{1}{k+p}  \nonumber \\ 
&=  -k  + \frac{ \lambda }{ 4 k } \frac{S_{d-1}}{(2\pi)^d}   \; \, 2 \, \Gamma \left( 1- d \right) k^{d-1}   \nonumber \\
&= - k \left[ 1  + \frac{2 \lambda}{ 16 \pi^2 } \left(  \frac{1}{d-3}  + \log \left(  k \right) + \text{local}  \right)   \right] 
 \; ,
 \label{eqn:psi2_eg1}
\end{align}
where $S_{n}$ is the surface area of the unit $n$-sphere (i.e. $S_2= 4 \pi$)  and we have used \eqref{eqn:integral_identity_1} to evaluate the momentum integral. ``Local'' denotes finite terms which are analytic in $k$, and are therefore sensitive to the renormalisation prescription (i.e. can be fixed by adding local counterterms). In particular, these local terms are purely real. 

The loop integral \eqref{eqn:psi2_eg1} contains a $1/(d-3)$ divergence in dimensional regularisation, but when we take the $\text{Disc}$ we pick out only the (finite) coefficient of the $\log (k)$ running, 
\begin{align}
 \changg{ \text{Disc} \left[ i \, \psi_{\bfk , -\bfk}' \right] =   - i k  \, \frac{2 \lambda}{16 \pi^2}\left(  \log (k ) - [\log ( e^{-i \pi} k )]^* \right)  = - \frac{ \lambda }{16 \pi} \;  2 k }
\end{align}
which indeed agrees with our cutting rule \eqref{eqn:Disc_eg1} on the support of the $\delta$-function.

%%%%
\paragraph{Perturbative Unitarity:}
%%%%
Just as for scattering amplitudes on Minkowski, we can use this $\text{Disc}$ to place a bound on the size of the 1-loop correction. Comparing the tree-level result, $\psi^{\text{tree} \, \prime}_{\bfk ,- \bfk} = -k$, with the loop-level, $\text{Disc} \left[ \psi^{\text{1-loop} \, \prime }_{\bfk , -\bfk} \right]$, we have that $| \lambda | \, \lesssim \, 8 \pi$ is necessary for this interaction to respect unitarity perturbatively. More precisely, while $\psi^{\text{1-loop}}_2$ contains local terms which can be freely fixed by imposing a renormalisation condition (the finite terms in \eqref{eqn:psi2_eg1}), it also contains a non-local $\log (k)$ running, the coefficient of which is an unambiguous prediction of the perturbative theory. Supposing that $|\psi^{\text{1-loop}}_2|$ is initially set to be less than $ |\psi^{\text{tree}}_2|$ at some $k_*$, the condition that $\lambda < 8 \pi$ ensures that $|\psi^{\text{1-loop}}_2|< |\psi_2^{\text{tree}}|$ at scales within an order of magnitude of $k_*$. Interestingly, we note that $\lambda < 8 \pi$ is the same bound that one obtains from the $2\to 2$ scattering amplitude (for which $\text{Im} \, \mathcal{A}^{\text{1-loop}}_{2 \to 2} = \lambda^2/8\pi$ and $\mathcal{A}_{2 \to 2}^{\rm tree} = \lambda$).

%%%%%%%%
\subsubsection*{$\phi^3$ on Minkowski} 
%%%%%%%%

We will now consider a cubic potential interaction for a massless scalar field on Minkowski. Although this potential is not bounded from below, it serves as a useful illustration of how the cutting rules correctly reproduce the $\text{Disc}$ of various 1-loop diagrams.

For the interaction $\mathcal{L}_{\rm int} = \frac{1}{3!} \mu \phi^3$, the tree-level cubic wavefunction coefficient is,
\begin{align}
 \psi_{\bfk_1 \bfk_2 \bfk_3}^{\rm tree} =  \mu \; \frac{ \tdelta^3 \left( \bfk_1 + \bfk_2 + \bfk_3 \right) }{ k_1 + k_2 + k_3 } \; .
 \label{eqn:psi3_eg2}
\end{align}
There are also three tree-level exchange contributions to 
\begin{align}
\psi_{\bfk_1 \bfk_2 \bfk_3 \bfk_4} = \psi_{\bfk_1 \bfk_2 \bfk_3 \bfk_4}^{(p_s)}+ \psi_{\bfk_1 \bfk_2 \bfk_3 \bfk_4}^{(p_t)} + \psi_{\bfk_1 \bfk_2 \bfk_3 \bfk_4}^{(p_u)}\,,    
\end{align}
which are given by
\begin{align}
 \psi^{(p_s) \, \prime}_{\bfk_1 \bfk_2 \bfk_3 \bfk_4} = \frac{\mu^2}{ ( k_1 + k_2 + k_3 + k_4 ) ( k_1 + k_2 + p_s )( k_3 + k_4 + p_s ) } \; ,
 \label{eqn:psi4_eg2}
\end{align}
plus the two permutations of the external legs. These are the only inputs needed to infer the $\text{Disc}\left[ i  \psi^{\text{1-loop}}_{\bfk_1 \bfk_2} \right]$ using the cutting rules.
Explicitly, we require the following $\text{Disc}$ of \eqref{eqn:psi3_eg2} and \eqref{eqn:psi4_eg2}, 
\begin{align}
    (-i) \underset{q_1 q_2}{\text{Disc}} \left[ i \psi^{\rm tree}_{\bfk \bfq_1 \bfq_2} \right] &=     
     \mu  \, \frac{  2 q_+  }{ q_+^2  -  k^2   } \, \tdelta^3 (\bfk_1 + \bfq_1 + \bfq_2 )  \nonumber \\ 
    (-i) \underset{q \, p_0 }{\text{Disc}} \left[ i \psi^{(p_0) \,  \prime}_{\bfk , - \bfk , \bfq, -\bfq } \right] &= 
    \frac{\mu^2}{8} \left( 
    \frac{1}{ k q ( k + q ) } + \frac{1}{- k q ( -k + q ) }
    \right)   \\ 
    (-i) \underset{q \, p}{\text{Disc}} \left[ i \psi^{(p) \, \prime}_{\bfk , \bfq, -\bfq , - \bfk} \right] &= 
    \frac{\mu^2}{2} \left( 
    \frac{1}{ ( k + q ) ( k + q + p )^2 } + \frac{1}{ ( -k+ q) ( - k + q + p)^2} \right)  \nonumber
%
   % (-i) \underset{q \, p_2}{\text{Disc}} \left[ i \psi^{(p_2) \, \prime}_{\bfk , \bfq, -\bfq , - \bfk} \right] &= 
    %\frac{\mu^2}{2} \left( 
    %\frac{1}{ ( k + q ) ( k + q + p_2 )^2 } + \frac{1}{ ( -k+ q) ( - k + q + p_2 )^2}
    %\right)
\end{align}
where here $p_0 = 0$, $p$ is arbitrary and $q_+=q_1+q_2$.

In this theory, there are two diagrams which contribute to $\psi^{\text{1-loop}}_{\bfk_1 \bfk_2}$, 
\FloatBarrier
\begin{figure}[htbp!]
    \centering
    \includegraphics[width=0.5\textwidth]{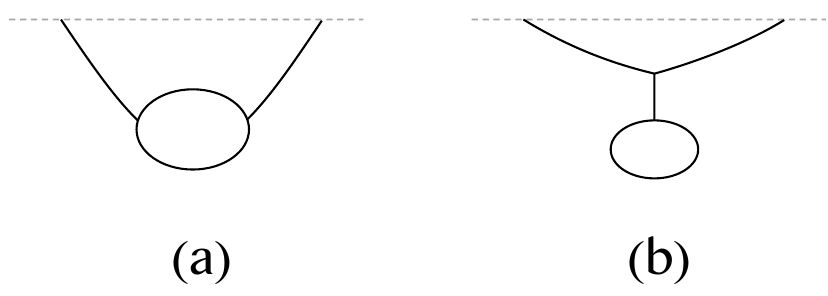}
\end{figure}
\FloatBarrier
\noindent which we label (a) and (b). Applying the cutting rules to diagram (a) we have,
\begin{align}
i\, \text{Disc} \left[  i \psi_{ \bfk_1 \bfk_2 }^{(a)} \right]  
&=
 \int_{ \bfq_2 \bfq'_2 } \;  P_{\bfq_2 \bfq'_2} 
(-i) \, \underset{q_2 q'_2 \, p_1}{\text{Disc}} \left[ i \psi_{  \bfk_1 \bfq_2 \bfq'_2 \bfk_2}^{(p_1)} \right] 
+  \int_{ \bfq_1 \bfq'_1 } \;  P_{\bfq_1 \bfq'_1} 
(-i) \, \underset{q_1 q'_1 \, p_2}{\text{Disc}} \left[ i \psi_{  \bfk_1 \bfq_1  \bfq'_1 \bfk_2}^{(p_2)} \right] 
  \nonumber \\
&  + \int_{ \substack{ \bfq_1 \bfq_1' \\ 
\bfq_2 \bfq_2'} } \; 
(-i) \, \underset{q_1 q_2}{\text{Disc}} \left[ i\psi_{ \bfk_1 \bfq_1 \bfq_2}^{\rm tree} \right]
 P_{\bfq_1 \bfq_1'}  P_{\bfq_2 \bfq_2'} \;
(-i) \,  \underset{ q_1' q_2' }{\text{Disc}} \left[ i \psi_{ \bfq_1' \bfq_2'  \bfk_2}^{\rm tree} \right]
\label{eqn:Discpsia_eg2}\,,
\end{align}
where $p_1= | \bfk_1 + \bfq_2 | $ and $p_2= | \bfk_1 + \bfq_1 |$. Note that the exchange contributions on the first line vanish identically,
\begin{align}
 \int_{\bfq_2 \bfq'_2} P_{\bfq_2 \bfq'_2} \, (-i) \underset{q_2 q'_2 \, p_1}{\text{Disc}} \left[ i \psi^{(p_1) \prime}_{\bfk ,  \bfq_2, \bfq'_2, - \bfk } \right] 
 = 0 \; ,  \qquad
 \int_{\bfq_1 \bfq'_1} P_{\bfq_1 \bfq'_1} \, (-i) \underset{q_1 q'_1 \, p_2}{\text{Disc}} \left[ i \psi^{(p_2) \prime}_{\bfk , \bfq_1 , \bfq'_1 , - \bfk } \right] 
 = 0 \; , 
 \label{eqn:psit0_eg2}
\end{align}
while the $\psi_3 \times \psi_3$ contribution on the second line can be reduced to a single integral using \eqref{eqn:dp+dp-},
\begin{align}
\int_{ \substack{ \bfq_1 \bfq_1' \\ \bfq_2 \bfq_2'} } \; 
(-i) \, \underset{q_1 q_2}{\text{Disc}} \left[ i\psi_{ \bfk \bfq_1 \bfq_2}^{\rm tree} \right]
 P_{\bfq_1 \bfq_1'}  P_{\bfq_2 \bfq_2'} \;
(-i) \,  \underset{ q_1' q_2' }{\text{Disc}} \left[ i \psi_{\bfq_1', \bfq_2' , - \bfk }^{\rm tree} \right]
&=   \frac{\mu^2}{ 16 \pi^2 } \int_{0}^{\infty} d q_+  \; \left(   \frac{ 2 q_+  }{ q_+^2 - k^2  } \right)^2  
\end{align}
which is given by the integral identity \eqref{eqn:integral_identity_1}. So altogether, diagram (a) contributes to the Gaussian width as,
\begin{align} 
  \text{Disc} \left[  i \psi_{ \bfk , - \bfk }^{(a) \prime} \right]     =  \changg{-}  \frac{ \mu^2}{ 16 \pi } \,  \frac{ 1 }{k} \; .
  \label{eqn:psia_eg2}
\end{align}

Now applying the cutting rules to diagram (b) we have,
\begin{align}
i\, \text{Disc} \left[  i \psi_{ \bfk_1 \bfk_2 }^{(b)} \right]  
&=
 \int_{ \bfq \bfq' } \;  P_{\bfq \bfq'} 
(-i) \, \underset{q q' \, p_0}{\text{Disc}} \left[ i \psi_{  \bfk_1  \bfk_2 \bfq_1 \bfq_2}^{(p_0)} \right] \; , 
\label{eqn:Discb_eg2}
\end{align}
where we have used that $\underset{q}{\text{Disc}} \left[ i \psi^{\rm tree}_{\bfk_1 \bfk_2 \bfq} \right] = 0$ (since $\bfk_1 + \bfk_2 = 0$ on imposing the $\delta$-functions) to discard the two diagrams in which the internal line with $\bfp = 0$ is cut. 
This exchange contribution can also be written in the form \eqref{eqn:integral_identity_1},
\begin{align}
 \int_{ \bfq \bfq' } \;  P_{\bfq \bfq'} 
(-i) \, \underset{q q' \, p_0}{\text{Disc}} \left[ i \psi_{  \bfk , - \bfk ,  \bfq , \bfq' }^{(p_0) \, \prime} \right]
=\frac{\mu^2 }{ 16 \pi^2 } \int_0^{\infty} dq \frac{1}{q^2 ( q^2 - k^2 )} 
\end{align}
and so diagram (b) contributes to the Gaussian width as,
\begin{align}
  \text{Disc} \left[  i \psi_{ \bfk , - \bfk }^{(b) \,\prime} \right]     =   \changg{+} \frac{\mu^2 }{ 16 \pi } \, \frac{ 1 }{ 2 k } \; .
  \label{eqn:psib_eg2}
\end{align}
Altogether, unitarity requires that $\text{Disc} \, \left[ i \psi^{\text{1-loop}}_{\bfk_1 \bfk_2} \right]$ is given by the sum of \eqref{eqn:psia_eg2} and \eqref{eqn:psib_eg2},
\begin{align}
 \text{Disc} \, \left[ i  \psi^{\text{1-loop} \, \prime}_{\bfk , -\bfk} \right]  =  \changg{-} \frac{\mu^2}{16 \pi} \frac{1}{2k} \; . 
 \label{eqn:Disc_eg2}
\end{align}
We stress that this required only knowledge of the tree-level coefficients \eqref{eqn:psi3_eg2} and \eqref{eqn:psi4_eg2}, and each momenta integral that we encountered in the cutting rules was manifestly finite (and did not require any regularisation or renormalisation). 
Before we move on to inflationary wavefunction coefficients in the next subsection, let us briefly show how the $\text{Disc}$ \eqref{eqn:psia_eg2} and \eqref{eqn:psib_eg2} could have been computed by instead performing the explicit loop integral.

%%%%
\paragraph{Comparison with Explicit Computation:}
%%%%
Note that diagram (a) can be computed directly using,
\begin{align}
 \psi^{(a) \,  \prime}_{\bfk ,-\bfk} &=  \mu^2 \int_{-\infty}^0 dt_1 \int_{-\infty}^0 d t_2 \, K_{k} (t_1) K_{ k } (t_2) \, \int_{\bfp_1 \bfp_2}  \;  G_{p_1} (t_1, t_2) G_{p_2} (t_1, t_2)  \, \tdelta^3 ( \bfp_1 + \bfp_2 + \bfk  ) \; ,   \nonumber \\
 &= \frac{ \mu^2}{4} \int_{\bfp_1 \bfp_2}  \; \frac{2 k + p_1 + p_2 }{ k ( k+ p_1 ) (k+ p_2) ( k + p_1 + p_2 )^2 }  \; \tdelta^3 ( \bfp_1 + \bfp_2 + \bfk  ) \; , 
\end{align}
This integral can be performed using the $d$-dimensional version of \eqref{eqn:dp+dp-} (which is given in \eqref{eqn:dp1dp2_d}) to replace $d^d \bfp_1 d^d \bfp_2$ with $d p_+ dp_-$. Carrying out the $dp_-$ integral leaves,
\begin{align}
 \psi^{(a) \prime}_{\bfk ,-\bfk}  &=  \frac{ \mu^2}{16 \pi^2} \int_k^{\infty} d p_+  \; \frac{  p_+ + 2 k - 2 (  p_+ + k) \log \left( \frac{ p_+ + 3k}{ p_+ + k }  \right) }{ k ( p_+ + k)^2 } \; \left( p_+^{d-3} +  \mathcal{O} (d-3 )  \right) . 
\end{align}
where we have discarded terms suppressed by $(d-3)$ that do not contribute to the divergence or the logarithmic running. 
In fact, the only term which diverges here is the integral, $\int_k^{\infty} d p_+ \, p_+^{d-2} /( p_+ + k)^2$, which gives,
\begin{align}
 \psi^{(a) \prime}_{\bfk ,-\bfk} &=  \frac{ \mu^2}{16 \pi^2 k}  \left[ \frac{1}{3-d} - \log \left(  k \right) + \text{finite} \right]     \; ,
\end{align} 
and the remaining finite part is purely real. The discontinuity again comes from the logarithmic branch cut,
\begin{align}
\text{Disc} \, \left[ i \psi^{(a)\, \prime}_{\bfk, - \bfk} \right] =  \frac{ \mu^2}{16 \pi^2}  \frac{  \log (k)  -  \log (e^{-i \pi} k ) }{ik} =   \changg{-} \frac{\mu^2}{16 \pi}  \frac{ 1 }{k} \; ,
\label{eqn:c2egExplicit}
\end{align} 
and coincides with the result which we obtained from unitarity \eqref{eqn:psia_eg2}. 

The tadpole diagram (b) is given by,
\begin{align}
 \psi^{(b) \, \prime}_{\bfk, -\bfk} &=  \mu^2 \int_{-\infty}^0 dt_1 \int_{-\infty}^0 dt_2 \, K_{k} (t_1) K_{k} (t_1) G_{ 0 } (t_1, t_2)  \int_{\bfp} G_{p} (t_2 , t_2 )  \nonumber \\
 &=  - \frac{\mu^2}{16 \pi^2} \, \frac{1}{2k } \left[ \frac{1}{3-d}   - \log \left( k \right) + \text{finite}   \right]  \; .
\end{align}
which matches the $\text{Disc}$ inferred using unitarity, 
\begin{align}
 \text{Disc} \left[ i \psi^{(b)\,\prime}_{\bfk , - \bfk} \right] = + \frac{\mu^2}{16 \pi^2} \, \frac{  - \log (k)  +  \log (e^{-i \pi} k ) }{2 i k } =  \changg{+} \frac{\mu^2}{16 \pi} \frac{1}{2k} \; .
 \label{eqn:c2tadpole}
\end{align}

While using the unitarity cuts to compute diagrams (a) and (b) did not provide any information about the divergent part of $\psi^{\text{1-loop}}$, it directly provides the finite $\text{Disc}$ (i.e. the coefficient of the $\log (k)$ running) without the need for laborious loop integrals.

%%%%%%%%
\subsection{On de Sitter}
\label{sec:deSitter}
%%%%%%%%

For a massless scalar field $\phi$ on de Sitter, we use the Bunch-Davies mode function,
\begin{align}
 f_{k} (\eta) =    \frac{ H (1 + i k \eta )  }{ k }   \; \frac{ e^{- i k \eta} }{ \sqrt{2 k} } \;  ,
\label{eqn:ModeMassless}
\end{align}
and the corresponding (late time) power spectrum is, $P_k =| f_k |^2  = H^2/2k^3$. 
In this case, the relevant bulk-to-boundary and bulk-to-bulk propagators are,
\begin{align}
 K_{k} (\eta) &= \left( 1 - i k \eta \right) e^{i k \eta}   \; ,   \label{eqn:deSitterPropagators} \\
 G_{p} ( \eta_1 ,  \eta_2 )  &= \frac{H^2}{p^3} \left[ \left( 1- i p \eta_2 \right) e^{i p \eta_2} 
  (\sin ( p \eta_1 ) - p \eta_1 \cos ( p \eta_1 ) )  \theta ( \eta_1 - \eta_2 )   + \left( \eta_1 \leftrightarrow \eta_2  \right)\right] \; . \nonumber 
\end{align}
We will now use the Cosmological Cutting Rules to compute the one-loop correction to $\text{Disc} \left[ i \psi_{\bfk_1 \bfk_2} \right]$ from the cubic vertices $\dot \phi^3$ and $\dot \phi (\partial_i \phi)^2$.  
The one-loop correction from $\dot \phi^3$ has been computed previously (see e.g. \cite{Senatore:2009cf}), however the unitarity derivation we present here is significantly shorter and less laborious.
To the best of our knowledge, the loop diagrams containing $\dot \phi (\partial \phi )^2$ vertices have not been computed before---likely due to their algebraic complexity---and here we are able to find the $\text{Disc}$ of these diagrams. 
Let us begin with the $\dot \phi^3$ interaction only, and then move on to include the $\dot \phi (\partial \phi )^2$.

%%%%%%%%
\subsubsection*{$\dot{\phi}^3$ on de Sitter} 
%%%%%%%%

For the interaction $\mathcal{L}_{\rm int} = C_{\dot \phi^3} (- H \eta)^{-1}  (\phi')^3$, where $\phi' = \partial_\eta \phi$ in conformal time, the late-time tree-level cubic wavefunction coefficient from the Bunch-Davies initial state is,
\begin{align}
 \psi^{\text{tree} \, \prime}_{\bfk_1 \bfk_2 \bfk_3} = \changg{+} \frac{\changg{12\,} C_{\dot \phi^3}}{H}  \frac{ k_1^2 k_2^2 k_3^2 }{ ( k_1 + k_2 + k_3 )^3 } \; . 
\end{align}
There is also a quartic coefficient $\psi_{4}$ sourced by the $s$, $t$ or $u$-channel exchange of $\phi$, but these turn out not to contribute at this order\footnote{
This can be seen in the following way. At large $q$, the quartic coefficient scales as,
$\psi_{\bfk -\bfk \bfq  -\bfq} \sim  k^4 / q$, in all three channels. The integrand $P_q \psi_{\bfk -\bfk \bfq -\bfq} \sim k^4/q^4$ and therefore $\int_{\bfq \bfq'} P_{\bfq \bfq'} \psi_{\bfk_1 \bfk_2 \bfq  \bfq'}$ does not diverge in $d=3$ dimensions---consequently there is no logarithmic dependence on $k$, and so the $\text{Disc}$ of this integral vanishes}. 

Again there are two diagrams which contribute to $\psi^{\text{1-loop}}_{\bfk_1 \bfk_2}$, which we label (a) and (b) as above. The cutting rule \eqref{eqn:Discpsia_eg2} for $\text{Disc} \left[ i \psi^{(a)}_{\bfk_1 \bfk_2} \right]$ contains exchange terms but once again these integrals vanish identically as in \eqref{eqn:psit0_eg2}. The remaining $\psi_3 \times \psi_3$ integral is given by, 
\begin{align}
& \int_{ \substack{ \bfq_1 \bfq_1' \\ \bfq_2 \bfq_2'} } \; 
(-i) \, \underset{q_1 q_2}{\text{Disc}} \left[ i\psi_{ \bfk \bfq_1 \bfq_2}^{\rm tree} \right]
 P_{\bfq_1 \bfq_1'}  P_{\bfq_2 \bfq_2'} \;
(-i) \,  \underset{ q_1' q_2' }{\text{Disc}} \left[ i \psi_{- \bfk \bfq_1' \bfq_2' }^{\rm tree} \right]
 \nonumber \\
&= \frac{ \changg{36} H^2 C_{\dot \phi^3}^2 }{ 16 \pi^2 } \int_{0}^{\infty} d q_+ \;   \frac{k^4 \left( q_+^3+3 q_+ k^2\right)^2 \left(
15 q_+^4 -10 q_+^2 k^2 +3 k^4\right)}{15
    \left( q_+^2-k^2\right)^6 }   
  \; , 
\end{align}
where we have changed to $(q_+, q_-)$ variables using \eqref{eqn:dp+dp-} and performed the $q_-$ integral. The remaining $q_+$ integral can be carried out using the identity \eqref{eqn:integral_identity_2}, and consequently the cutting rules determine the one-loop discontinuity to be,
\begin{align}
 \text{Disc} \, \left[ i \psi^{(a)\, \prime}_{\bfk - \bfk} \right] =   \changg{-} \frac{H^2 C_{\dot \phi^3}^2}{ \pi}  \, k^3 \, \changg{\frac{3}{10}} \; . 
 \label{eqn:Disca_eg3}
\end{align}

For the one-loop diagram (b), once again we have that $\underset{q}{\text{Disc}} \left[ i \psi_{\bfk_1 \bfk_2 \bfq} \right] = 0$ and so the cutting rule is again simply \eqref{eqn:Discb_eg2}. Unlike for $\phi^3$ Minkowski space, for $\dot \phi^3$  on de Sitter this contribution vanishes because the exchange contribution vanishes,
\begin{align}
    \text{Disc} \left[ i \psi_{\bfk_1 \bfk_2}^{(b)\, \prime} \right] = 0 \; ,
\end{align}
and so the entire discontinuity in $\psi_{\bfk_1 \bfk_2}$ at one-loop is given by \eqref{eqn:Disca_eg3}.

%%%%%%%%
\subsubsection*{$\dot{\phi} (\partial_i \phi)^2$ on de Sitter} 
%%%%%%%%

Now we consider the general cubic interaction, 
\begin{align}
\mathcal{L}_{\rm int} =  -  \frac{ C_{\dot \phi^3} }{ H \eta  } \left( \phi' \right)^3  - \frac{ C_{\dot \phi (\partial \phi)^2}  }{ H \eta }   \phi'  (\partial_i \phi )^2  \; ,
\label{eqn:L3}
\end{align}
which contains both $\dot \phi^3$ and $\dot \phi (\partial \phi )^2$ interactions. 
The corresponding cubic wavefunction coefficient can be found in \cite{Cheung:2007sv}, 
\begin{align}
 \psi_{\bfk_1 \bfk_2 \bfk_3}^{\rm tree} = \frac{ 1 }{ 2 H  k_T^3 } \left[ C_{\dot \phi^3} \, 24  e_3^2 + C_{\dot \phi (\partial \phi)^2}  \left(  24 e_3^2 - 8 k_T e_2 e_3 - 8 k_T^2 e_2^2 + 22 k_T^3 e_3 - 6 k_T^4 e_2 + 2 k_T^6        \right) 
   \right]  
   \label{eqn:c3EFTI}
\end{align}
where 
\begin{align}
    e_3 &= k_1 k_2 k_3\,, & e_2 &= k_1 k_2 + k_1 k_3 + k_2 k_3 & k_T &= k_1 + k_2 + k_3\,.
\end{align}
The exchange contribution to $\psi_4$ is schematically\footnote{
Formally, there is an additional contribution with integrand $\psi_{\bfk, -\bfk, \mathbf{0}} \psi_{\bfq, -\bfq, \mathbf{0}}$, but this vanishes because $\pi$ is derivatively coupled.
},
\begin{align}
 \psi_{\bfk, -\bfk, \bfq, -\bfq}' \sim  \int \frac{d \eta}{\eta} \, \eta^d \;   \psi_{ \bfk ,\bfq , -\bfk-\bfq }' \psi_{ -\bfk ,-\bfq , \bfk+\bfq }'
\end{align}
and so since\footnote{
Although the $C_{\dot\phi (\partial \phi)^2}$ term in $\psi_3$ seems to $\sim q^3$, the numerical coefficients are such that it only $\sim q$ at large $q$, a consequence of the soft theorem for the squeezed bispectrum \cite{Cheung:2007sv}.
} $\psi_{\bfk, \bfq, -\bfk-\bfq}' \sim C_{\dot \phi^3} q^1 + C_{\dot \phi (\partial \phi)^2}  q^1$ at large $q$, we expect $\psi'_{\bfk , -\bfk ,\bfq ,-\bfq} \sim q^{-1} $. This expectation indeed matches the explicit computation. 
The integral $\int_{\bfq \bfq'} P_{\bfq \bfq'} \psi_{\bfk_1 \bfk_2 \bfq  \bfq'}$ therefore does not contain any divergence and so drops out of the cutting rules, and therefore only the twice-cut diagram contributes.
Unitarity therefore fixes the 1-loop $\text{Disc}$ to be, 
\begin{align}
i \text{Disc} \,\left[ i \psi^{(a) \, \prime}_{\bfk_1 \bfk_2} \right] &= 
\int_{ \substack{ \bfq_1 \bfq_1' \\ \bfq_2 \bfq_2'} } \; 
(-i) \, \underset{q_1 q_2}{\text{Disc}} \left[ i\psi_{ \bfk \bfq_1 \bfq_2}^{\rm tree} \right]
 P_{\bfq_1 \bfq_1'}  P_{\bfq_2 \bfq_2'} \;
(-i) \,  \underset{ q_1' q_2' }{\text{Disc}} \left[ i \psi_{- \bfk \bfq_1' \bfq_2' }^{\rm tree} \right] \nonumber \\
 &=   \changg{-}\frac{ H^2 }{ \pi } i k^3 \left[   
 \frac{3}{10} C_{\dot \phi^3}^2  -  \frac{9}{10} C_{\dot \phi^3}   C_{\dot \phi (\partial \phi)^2 }  + \frac{51}{20} C_{\dot \phi (\partial \phi)^2}^2  \right] \; .
 \label{eqn:Disc_eg4}
\end{align}
Although the overall factors of $H$, $k$ and $\pi$ could have been inferred from power counting, the numerical coefficients in \eqref{eqn:Disc_eg4} could not have been. The cutting rules are therefore providing an efficient route to this part (the $\text{Disc}$) of the 1-loop wavefunction, completely removing the need for regularising and performing complicated loop integrals.

%%%%%%%%
\subsection{For the EFT of Inflation}
\label{sec:inflation}
%%%%%%%%

Finally, let's consider inflation. Following the EFT approach of \cite{Cheung:2007st}, we consider the low-energy effective action for perturbations about an expanding FLRW background. 
Although this background introduces an explicit time dependence, temporal diffeomorphisms can be restored (non-linearly realised) by introducing a single scalar degree of freedom, $\pi$, which decouples from the metric perturbations in the so-called decoupling limit ($M_P \to \infty$ with $f_\pi$ fixed). 
In this decoupling limit, the scalar perturbations in the EFT of Inflation are described by, 
\begin{align}
 S [ \pi ]  = \int dt d^3 \bfx \;  a^3  \; \frac{ f_\pi^4 }{c_s^3 } \Big[   \frac{ \dot \pi^2 }{2} - c_s^2 \frac{(\partial_i \pi )^2}{2 a^2}    
 + C_{\dot \pi^3} \dot \pi^3  +  C_{\dot \pi (\partial \pi)^2} \; \dot \pi \frac{(\partial_i \pi )^2}{a^2}            \Big]\,,
 \label{eqn:EFTI}
\end{align}
where $f_{\pi}^4 = 2 c_s M_P^2 | \dot H |  \approx \left( 60 H \right)^4$ is the energy scale associated with the symmetry breaking (fixed by the power spectrum) and $c_s$ is the sound speed. 
The non-linearly realised symmetry fixes $C_{\dot \pi^3}$ and $C_{\dot \pi (\partial_i \pi)^2}$ in terms of $c_s$ and one additional Wilson coefficient, conventionally denoted by $\tilde{c}_3$ \cite{Senatore:2009gt},
\begin{align}
C_{\dot \pi^3} = \frac{1}{2} \left( 1- c_s^2 \right) \left( 1 + \frac{2 \tilde{c}_3}{3 c_s^2} \right)    \quad , \quad
C_{\dot \pi (\partial_i \pi)^2} = - \frac{1}{2} \left( 1 - c_s^2  \right)  \; . 
\end{align} 
$c_s$ and $\tilde{c}_3$ are constrained by the primordial bispectrum \cite{Akrami:2019izv}.

%%%%%%%%
\subsubsection*{Wavefunction at One-Loop} 
%%%%%%%%

We will now use the Cosmological Cutting Rules to compute $\psi^{\text{1-loop}}$ from the EFT of Inflation \eqref{eqn:EFTI}. 
Unlike the $\phi$ of previous subsections, the kinetic term for $\pi$ is not canonically normalised---this is accounted for via the rescaling $\pi = \phi / f_\pi^2$ and $\bfx = c_s \tilde{\bfx}$.
Note that when considering an approximately de Sitter spacetime background, we can write \eqref{eqn:EFTI} in terms of conformal time $H t = - \log ( - H \eta )$ as,
\begin{align}
S [ \phi ] = \int d \eta d^3 \tilde{\bfx} \,  \left[  \frac{1}{2 H^2 \eta^2 } \left[  (\phi')^2 - ( \tilde{\partial}_i \phi )^2 \right]  -  \frac{ C_{\dot \pi^3} }{ f_\pi^2 H \eta  }  (\phi')^3  - \frac{ C_{\dot \pi  (\partial \pi)^2}  }{ f_\pi^2 c_s^2 H \eta }  \phi' ( \tilde{\partial}_i \phi )^2  \right] \; . 
\label{eqn:EFTI2}
\end{align}
where a prime denotes derivatives with respect to conformal time. This expression has the same form  as the interactions \eqref{eqn:L3} considered above, with coefficients,
\begin{align}
 C_{\dot \phi^3} = \frac{ C_{\dot \pi^3} }{ f_\pi^2 } \;\;\;\; \text{and} \;\;\;\;  C_{\dot \phi (\partial \phi)^2} = \frac{ C_{\dot \pi  (\partial \pi)^2} }{ f_\pi^2 c_s^2 } \; . 
\end{align}

Since we have massaged \eqref{eqn:EFTI} into the same form as \eqref{eqn:L3}, we can follow the same steps described in Section~\ref{sec:deSitter} to arrive at the ($\text{Disc}$ of the) one-loop coefficient $\psi_{\bfk_1 \bfk_2}$ of\footnote{In other words, all the wavefunction coefficients we quote refer to the canonically normalized field.} $\phi_{\bfk_1} \phi_{\bfk_2}$,
\begin{align}
    i \text{Disc} \left[ i \psi^{\text{1-loop}}_{\bfk_1 \bfk_2} \right] &=\changg{-}
    \frac{H^2}{f_\pi^4} \frac{i k^3}{480 \pi}  \frac{(1- c_s^2 )^2}{c_s^4} \left[ ( 4 \tilde{c}_3 + 9 + 6 c_s^2)^2 + 15^2
    \right]  
    \label{eqn:Disc_EFTI}   \\
    &=: \changg{-}\frac{H^2}{f_\pi^4} \frac{i k^3}{16 \pi} \; \gamma ( c_s, \tilde{c}_3 ) \; .  \nonumber
\end{align}
Again we note that, while the overall scaling of this quantity could  have been inferred from dimensional analysis alone, the cutting rules have allowed us to go much further by also providing the precise form of the coefficient $\gamma (c_s, \tilde{c}_3 )$. While the $C_{\dot\pi^3}^2$ contribution to $\gamma ( c_s , \tilde{c}_3 )$ could have been extracted from the explicit one-loop computation performed in \cite{Senatore:2009cf}, we are not aware of any previous computation of this general expression (which, without the cutting rules, would require performing the explicit loop integrals with $\dot \pi (\partial_i \pi)^2$ vertices).

%%%%
\paragraph{Quartic Interactions:}
%%%% 
Note that although we only considered the leading cubic interactions in \eqref{eqn:EFTI}, our results are robust against a potentially large quartic corrections.
This is since the one-loop correction from $\dot \pi^4$ does not diverge, as noted previously in \cite{Senatore:2009cf}, and consequently it does not affect the $\text{Disc} \left[ i \psi_{\bfk_1 \bfk_2} \right]$ at one-loop. We can confirm this straightforwardly using the cutting rules.
For the interaction\footnote{
Since this arises from a $d^4 x \; \sqrt{-g} ( g^{\mu\nu} \nabla_\mu \pi \nabla_\nu \pi )^2$, there is no explicit $\eta$ dependence in this interaction. 
} $\frac{1}{4!} C_{\dot \pi^4} (\pi' )^4$, the late-time tree-level quartic wavefunction coefficient from the Bunch-Davies initial state is,
\begin{align}
 \psi^{\text{tree}\; \prime}_{\bfk_1 \bfk_2 \bfk_3 \bfk_4} = - C_{\dot \pi^4}  \frac{ k_1^2 k_2^2 k_3^2 k_4^2 }{ ( k_1 + k_2 + k_3 + k_4)^5 } \; . 
\end{align}
Using the cutting rule \eqref{eqn:cut_1loop_1}, the corresponding contribution to $\text{Disc} \left[ i \psi_{\bfk_1 \bfk_2} \right]$ is given by the integral,
\begin{align}
\int_{\bfq \bfq'} P_{\bfq \bfq'} \; (-i) \underset{q q'}{\text{Disc}} \left[ i \psi_{\bfk , -\bfk, \bfq , \bfq'} \right] = 
\frac{ C_{\dot \pi^4} }{16 \pi^2}  \frac{k^4}{4} \int_0^{\infty}  \frac{dq \;  q^3 }{ (q^2 - k^2)^5 } \left( q^5 + 10 k^2 q^3 + 5 k^4 q \right) \; , \end{align}
which vanishes once evaluated using the integral identity \eqref{eqn:integral_identity_2}.  
This shows that the $\text{Disc}$ of the 1-loop Gaussian width is insensitive to the tree-level trispectrum, at least in the limit where $\dot \pi^4$ dominates over $\dot \pi^2 (\partial_i \pi)^2$ and $(\partial_i \pi )^4$ (which is natural since the latter two interactions are fixed in terms of $c_s$ by the non-linearly realised symmetry).

%%%%%%%%%%%%%%%%%%%%%%%%%%%%%%%%%%%%%%%%%%%%%%%%%%%%%%%%%%%%%%%%%%%%%%%%%%%%%%%
 
\subsection{Physical Interpretation}
\label{sec:physical}

Let us now interpret the physical meaning of the discontinuity \eqref{eqn:Disc_EFTI} in $\psi_2^{\text{1-loop}}$. 

%%%%%%%%
\paragraph{Source of the $\text{Disc}$:} 
%%%%%%%%
In the Minkowski examples in Section~\ref{sec:minkowski}, we found by explicit computation that the $\text{Disc}$ corresponded to the coefficient of the $\log (k)$ running of $\psi^{\text{1-loop}}$. This is exactly analogous to the logarithmic discontinuities encountered in flat space scattering amplitudes at one-loop. It is therefore tempting to conclude from \eqref{eqn:Disc_EFTI} that,  
\begin{align}
    \psi^{\text{1-loop}}_{\bfk_1 \bfk_2} \overset{?}{=}  \frac{H^2}{f_\pi^4} \, \frac{k^3}{16 \pi^2} \;\left[ - \gamma \log \left( k \right) + \text{divergence} + \text{local}    \right]
    \label{eqn:psi_logk}
\end{align}
where the ``local'' remainder is a real analytic function of $k$.
This is indeed the form of the one-loop corrections found in Weinberg's original article \cite{Weinberg:2005vy} (see also \cite{Weinberg:2006ac, Adshead:2008gk}), in which the loop integrals were computed using a certain form of dimensional regularisation (which sends $d^3 p \to d^d p$ but retains $3$-dimensional mode functions). 
However, in \cite{Senatore:2009cf} (see also \cite{Senatore:2012nq, Pimentel:2012tw}), it was pointed out that this $\log (k)$ is absent for other regularisations (including dimensional regularisation with $d$-dimensional mode functions). 
We show explicitly in Appendix~\ref{app:pi3_1loop} that performing dimensional regularisation with $d$-dimensional mode functions introduces an additional $\log ( -ik/H )$ term, and including this contribution gives a 1-loop wavefunction coefficient of the form,
\begin{align}
    \psi^{\text{1-loop}}_{\bfk_1 \bfk_2} =  \frac{H^2}{f_\pi^4} \, \frac{k^3}{16 \pi^2} \;\left[  \changg{+} \gamma \frac{i \pi}{2} + \text{divergence} + \text{local}    \right] \; . 
    \label{eqn:psi_logk_2}
\end{align}
Note that \eqref{eqn:psi_logk_2} and \eqref{eqn:psi_logk} share the same $\text{Disc}$, since \[ \text{Disc} \left[ \changg{-} \log (k) \right] = \text{Disc} \left[ \frac{i \pi}{2} \right] = i \pi\,, \] 
and so both are consistent with our cutting rules. This is to be expected, since the cutting rules use only \emph{tree-level} data, and therefore are not sensitive to how we have chosen to regulate the loop divergences. 
  
Physically, we can trace this additional $\log ( - i H k )$ term in \eqref{eqn:psi_logk_2} back to a logarithmic divergence near the conformal boundary, $\lim_{\eta \to 0} \log  \left( - H\eta \right)$, which arises in $d$ dimensions. 
Such boundary divergences are absent in Minkowski, and so in all of our Minkowski examples the $\text{Disc} \left[ i \psi^{\text{1-loop}} \right]$ implied a $\log \, k$ dependence as in \eqref{eqn:psi_logk}. But in de Sitter, there can be divergences both from the loop momenta $p \to \infty$ (which produce $\log (k)$) and near the boundary $\eta \to 0$ (which produce  $\log ( -i k )$).
The latter do not affect the $\text{Disc}$, and in fact such boundary divergences appear already at tree-level for certain values of the mass \cite{Cespedes:2020xqq}, but since $\text{Disc} \left[ \log (- i k) \right] = 0$ they are consistent with tree-level unitarity \cite{Goodhew:2020hob, Cespedes:2020xqq}.

%%%%
\paragraph{Perturbative Unitarity:}
%%%%
From our cutting rules, which fix the value of $\gamma$ in \eqref{eqn:psi_logk_2}, we can place a lower bound on the size of the one-loop correction to $\psi_{\bfk_1 \bfk_2}$
\begin{align}
    |\psi^{\text{1-loop}}_{\bfk_1 \bfk_2} | \; \geq \; | \text{Im} \, \psi^{\text{1-loop}}_{\bfk_1 \bfk_2} | \; = \; \frac{|\gamma|}{32 \pi} \frac{H^4}{f_\pi^4} | \psi^{\text{tree}}_{\bfk_1 \bfk_2} |
\end{align}
where we have written the imaginary part of \eqref{eqn:psi_logk_2} in terms of $\psi^{\rm tree}_{\bfk_1 \bfk_2} = - k^3/H^2$. Perturbative unitarity therefore requires that the $\gamma (c_s, \tilde{c}_3 )$ defined in \eqref{eqn:Disc_EFTI} is bounded,
\begin{align}
    \frac{ |\gamma ( c_s, \tilde{c}_3 ) | }{f_\pi^4} <  \frac{32 \pi}{H^4} \; .  
    \label{eqn:gamma_unitarity_bound}
\end{align}
Previous bounds on the EFT coefficients using perturbative unitarity have either neglected numerical coefficients (i.e. treating $\gamma$ as simply $\sim \tilde{c}_3^2$) or have worked in a subhorizon limit (i.e. $k \gg H$) where the usual optical theorem  for amplitudes can be applied---see e.g. \cite{Baumann:2011su, Baumann:2014cja, Baumann:2015nta, deRham:2017aoj, Grall:2020tqc} for estimates of the EFT cutoff in that regime. 
By contrast, \eqref{eqn:gamma_unitarity_bound} is the first precise unitarity bound that genuinely incorporates the effects of the expanding spacetime, and therefore applies at values of $k$ which are comparable to $H$. 

Phenomenologically, the bound \eqref{eqn:gamma_unitarity_bound} has already been overtaken by observational constraints on $\tilde{c}_3$ and $c_s$ from the bispectrum. For instance, since $c_s \geq 0.021$ at 95$\%$ confidence \cite{Akrami:2019izv}, $|\gamma| \, \lesssim \, 5 \times 10^7$ for essentially the whole $\tilde{c}_3$ 95$\%$ confidence interval, while $32 \pi f_\pi^4/H^4 \approx 10^9$. 
However, the cutting rules have provided more than the bound \eqref{eqn:gamma_unitarity_bound}: unitarity has completely fixed $\text{Im} \, \left[ \psi_{\bfk_1 \bfk_2} \right]$ at 1-loop, and this has important consequences for the time dependence of the power spectrum.

%%%%%%%%
\subsubsection*{Power Spectrum at One Loop} 
%%%%%%%%

The power spectrum $P_{\bfk_1 \bfk_2} = \langle \hat{\phi}_{\bfk_1} \hat{\phi}_{\bfk_2} \rangle$ can be computed from the wavefunction coefficients in the standard way, by performing an average over all field configurations weighted by the probability density $|\Psi [ \phi ] |^2$, 
\begin{align}
    P_{\bfk_1 \bfk_2}  = \int \mathcal{D} \phi \; | \Psi [ \phi ] |^2 \; \phi_{\bfk_1} \phi_{\bfk_2} \; .
    \label{eqn:Pdef}
\end{align}
At weak coupling, we can expand perturbatively in the non-Gaussian wavefunction coefficients,
\begin{align}
    P_{\bfk_1 \bfk_2} = P_{\bfk_1 \bfk_2}^{\text{tree}} + P^{\text{1-loop}}_{\bfk_1 \bfk_2} + ...  \; ,
\end{align}
where the tree-level result is well-known,
\begin{align}
     P_{\bfk_1 \bfk_2}^{\text{tree}} = - \frac{ \tdelta^3 \left( \bfk_1 + \bfk_2 \right) }{2 \Re \left[\psi^{\text{tree} \, \prime}_{\bfk_1,-\bfk_1} \right]} = | f_{k_1} |^2 \; \tdelta^3 \left( \bfk_1 + \bfk_2 \right) \,,
     \label{eqn:Ptree}
\end{align}
where we have written $\psi_2^{\text{tree}}$ in terms of the free theory mode function $f_k$. 
In every cutting rule derived above, it is this $P^{\text{tree}}_k$ which should be used---for instance, for a massless scalar field on de Sitter this corresponds to $P_k^{\text{tree}} = H^2/2k^3$. 

At 1-loop order, there are corrections from the interactions,
\begin{align}
     P_{\bfk_1 \bfk_2}^{\text{1-loop}} &= 2 | f_{k_1} |^4  \text{Re} \left[ \psi_{\bfk_1 \bfk_2}^{\text{1-loop}} \right]  + 
     \int \mathcal{D} \phi \; e^{- \int_{\bfq} \frac{ \phi_{\bfq} \phi_{-\bfq} }{ 2 |f_q |^2 } } \phi_{\bfk_1} \phi_{\bfk_2} \Bigg[
      \int_{\bfq_1 \bfq_2 \bfq_3 \bfq_4} \frac{ 2 \text{Re} \left[ \psi_{\bfq_1 \bfq_2 \bfq_3 \bfq_4}  \right] }{4!}  \phi_{\bfq_1} \phi_{\bfq_2} \phi_{\bfq_3} \phi_{\bfq_4}
     \nonumber \\ 
     &\qquad\qquad\qquad\qquad+ \int_{ \substack{ \bfq_1 \bfq_2 \bfq_3 \\ \bfq_1' \bfq_2' \bfq_3' }} \frac{ 2 \text{Re} \left[ \psi_{\bfq_1 \bfq_2 \bfq_3}  \right] \text{Re} \left[ \psi_{\bfq_1' \bfq_2' \bfq_3'}  \right] }{ (3!)^2 }  \phi_{\bfq_1} \phi_{\bfq_2} \phi_{\bfq_3} \phi_{\bfq_1'} \phi_{\bfq_2'} \phi_{\bfq_3'} 
     \Bigg] 
     \label{eqn:P1loop}
\end{align}
where the $\int \mathcal{D} \phi$ produces several terms, each of which has a simple diagrammatic interpretation (i.e. all possible pair-wise Wick contractions of the fields). 
Note that the 1-loop corrections (and indeed, all $L$-loop corrections) to the power spectrum are sensitive only to the \emph{real} parts of the wavefunction coefficients, $\text{Re} \left[ \psi_n \right]$. However, above we have shown that unitarity is effectively fixing the \emph{imaginary} parts,  $\text{Im} \left[ \psi_2^{\text{1-loop}} \right]$, in terms of tree-level data.
We will now show that $\text{Im} \left[ \psi_2^{\text{1-loop}} \right]$ plays an important role in determining the \emph{time dependence} of the power spectrum\footnote{
Note that a $\log \, k$ running as in \eqref{eqn:psi_logk} would instead produce a one-loop correction to the spectral tilt of the power spectrum, rather than a one-loop correction to the time dependence. It would be interesting to investigate whether there is some connection between these two effects.
}. This is not surprising, since unitarity is a constraint on the time evolution of the system---in particular see Appendix~\ref{app:schrodinger} where this aspect of the cutting rules is made manifest.

%%%%
\paragraph{Time Derivatives of the Power Spectrum:}
%%%%
Defining the bulk power spectrum $P_k (\eta)$ via \eqref{eqn:Pdef} using the wavefunction $\Psi_{\eta} [ \phi]$ evaluated at a finite conformal time, we can expand near the late-time boundary as,
\begin{align}
 P_k (\eta) = P_k (0) + \eta \partial_\eta P_k (0) + \frac{\eta^2}{2} \partial_\eta^2 P (0) + \frac{\eta^3}{3!} \partial_\eta^3 P (0 ) + \mathcal{O} \left( \eta^4 \right) \; .
\end{align} 
where $P_k (0)$ is given in terms of the boundary wavefunction coefficients\footnote{
Note that for a massless scalar field, the bulk wavefunction coefficient is given by, 
\begin{align}
    \psi^{\text{tree}} =  \frac{i k^2}{H^2 \eta} ( 1 - i k \eta )^{-1} = \frac{i k^2}{ H^2 \eta } - \frac{ k^3 }{ H^2 } + ...
\end{align}
Extracting the late time limit $\lim_{\eta \to 0} \psi^{\text{tree}}_2 (\eta)$ requires a renormalisation of the boundary condition $\eta = 0$ (or equivalently, a Boundary Operator Expansion to replace the bulk $\hat{\phi}$ operator with boundary operators)---see e.g. \cite{Cespedes:2020xqq} for details.
} by \eqref{eqn:Ptree} and \eqref{eqn:P1loop}. 
We will now express the subleading $\partial_\eta^n P_k (0)$ terms in this expansion in terms of the wavefunction coefficients, and see that in fact it is $\partial_\eta^3 P(0)$ that is fixed by $\text{Im} \left[ \psi_2 \right]$. 

Beginning in the Heisenberg picture, $P_{\bfk_1 \bfk_2} (\eta) = \langle \hat{\phi}_{\bfk_1} (\eta) \hat{\phi}_{\bfk_2} (\eta ) \rangle$, the equations of motion can be used to reduce any $\partial_\eta^{n} \hat{\phi}$ to just $\hat{\phi}$ and $\partial_\eta \hat{\phi}$. In particular, near the conformal boundary the only terms which contribute to the correlator of $\partial_\eta^2 \hat{\phi}$ and any other operator $\hat{\mathcal{O}}$ are,
\begin{align}
\lim_{\eta \to 0} \; \langle   \partial_\eta^2 \hat{\phi} \; \hat{\mathcal{O}} \rangle = \lim_{\eta \to 0} \; \left\langle \left[ \frac{2}{\eta} \partial_\eta \hat{\phi} + \partial_i^2 \hat{\phi} - C_{\dot \phi (\partial \phi )^2 } H (\partial_i \hat{\phi} )^2  \right]  \hat{\mathcal{O}} \right\rangle \; ,
\label{eqn:free_eom}
\end{align}
where we have power counted each interaction term using the boundary expansion of the free field profile, 
\begin{align}
\phi \sim \phi^{(0)} ( 1 + \tfrac{1}{2} k^2 \eta^2  + \mathcal{O} ( \eta^3) ) + \phi^{(3)} \eta^3 \left( 1 + \mathcal{O} (\eta^2) \right)    \;, 
\label{eqn:phi_boundary}
\end{align}
for instance the term $C_{\dot \phi (\partial \phi )^2} H \eta \partial_i \left( \phi' \partial^i \phi \right)  \sim \eta^2 \to 0$ near the boundary. The remaining $\partial_\eta \hat{\phi}$ operators may then be written in terms of the conjugate momentum $\hat{\Pi}$ of the Schr\"{o}dinger picture, using,
\begin{align}
\lim_{\eta \to 0} \;  \partial_\eta \phi  =  \lim_{\eta \to 0} \left[ H^2 \eta^2 \; \Pi + C_{\dot \phi (\partial \phi)^2 } H \eta \, (\partial_i \phi )^2  \right] \; . 
\label{eqn:dotphiToPi}
\end{align}
This  then expresses any $\lim_{\eta \to 0} \partial_\eta^n P(\eta)$ in terms of equal-time correlators of $\hat{\phi}$ and $\hat{\Pi}$.

Following \cite{Cespedes:2020xqq}, we can simplify such correlators by performing a unitary transformation of the canonical momentum, $ \tilde{\Pi}_{\bfk} = \hat{\Pi}_{\bfk} + \text{Im} \left[ \psi_{\bfk ,-\bfk}' \right] \hat{\phi}_{\bfk} $, which accounts for the free-field damping and removes any squeezing from the Gaussian state. The $\langle \hat{\phi} \tilde{\Pi} \rangle$ equal-time correlator can be written simply in terms of the wavefunction coefficients as,
\begin{align}
\langle \hat{\phi}_{\bfk_1} \tilde{\Pi}_{\bfk_2} \rangle = \frac{i}{2}  
&+ \int \mathcal{D} \phi \,  e^{- \int_{\bfq} \frac{ \phi_{\bfq} \phi_{-\bfq} }{2 | f_q |^2} } \Bigg[ -  \int_{\bfq_1 \bfq_2 \bfq_3} \frac{ \text{Im} \left[ \psi^{\rm tree}_{\bfq_1 \bfq_2 \bfq_3 \bfk_2} \right] }{3!}  \phi_{\bfq_1} \phi_{\bfq_2} \phi_{\bfq_3} \phi_{\bfk_1} 
\nonumber \\
&\qquad\qquad\qquad+   \int_{\bfq_1' \bfq_2' \bfq_3'} \frac{ \text{Re} \left[ \psi_{\bfq_1' \bfq_2' \bfq_3'}^{\rm tree} \right] }{3} \phi_{\bfq_1'} \phi_{\bfq_2'} \phi_{\bfq_3'}   \int_{\bfq_1 \bfq_2} \frac{ \text{Im} \left[ \psi_{\bfq_1 \bfq_2 \bfk_2}^{\rm tree} \right] }{2!}  \phi_{\bfp_1} \phi_{\bfp_2}  \phi_{\bfk_1}  \Bigg]  \; . 
\label{eqn:phiPi}
\end{align}
Note that both $\text{Im} \left[ \psi_3^{\rm tree} \right]$ and $\text{Im} \left[ \psi_4^{\text{tree}} \right]$ vanish at the boundary, and so $\lim_{\eta \to 0} \langle \hat{\phi}_{\bfk} \tilde{\Pi}_{-\bfk} \rangle'  \to \frac{i}{2}$.

Altogether, this means that we can compute $\lim_{\eta \to 0} \partial_\eta^n P (\eta)$ by first using \eqref{eqn:free_eom} to reduce all $\partial_\eta^n \hat{\phi}$ to $\hat{\phi}$ and $\partial_\eta \hat{\phi}$, and then \eqref{eqn:dotphiToPi} to replace $\partial_\eta \hat{\phi}$ with $\hat{\Pi}$, and finally \eqref{eqn:phiPi} to simplify the mixed $\langle \hat{\phi} \hat{\Pi} \rangle$ correlators. 
%
% In particular, note that if we expand $\text{Im} \left[ \psi_2 \right]$ as a tree-level plus 1-loop correction,
% \eqref{eqn:phiPi} gives, 
% \begin{align}
% \frac{1}{2} \left\langle \left\{ \hat{\phi}_{\bfk} , \hat{\Pi}_{-\bfk}  \right\} \right\rangle &= \left( \frac{k^2}{H^2 \eta} -  \text{Im} \left[ \psi_{\bfk -\bfk}^{\text{1-loop}} \right] + \mathcal{O} ( \eta ) \right) \langle \hat{\phi}_{\bfk} \hat{\phi}_{-\bfk} \rangle
% \end{align}
% near the boundary, where $\text{Im} \left[ \phi_{\bfk -\bfk}^{\text{1-loop}} \right]$ is fixed by unitarity (i.e. the cutting rules).
%
We find that, to quadratic order in the couplings $C_{\dot \phi^3}$ and $C_{\dot \phi (\partial_i \phi )^2}$,
\begin{align}
\lim_{\eta \to 0 } \left[ P_k (\eta)\right]  &=  \frac{H^2}{2 k^3} + P_k^{\text{1-loop}} (0)  \; , \nonumber \\
\lim_{\eta \to 0 } \left[ \partial_\eta P_k (\eta)\right]  &=  0   \; ,\nonumber \\
\lim_{\eta \to 0} \left[ \tfrac{1}{2} \partial_\eta^2 P_k (\eta) \right] &=  \frac{H^2}{2k} + k^2 P_k^{\text{1-loop}} (0)  - C_{\dot \phi  (\partial \phi  )^2}  H  \int_{\bfq_1 \bfq_2} \bfq_1 \cdot \bfq_2 \; B_{q_1 q_2 k}^{\text{tree}} (0)     \nonumber \\
\lim_{\eta \to 0} \left[ \tfrac{1}{3!} \partial_\eta^3 P_k (\eta)  \right] &=  - \frac{ H^4}{3 k^3} \text{Im} \left[ \psi_{\bfk -\bfk}^{\text{1-loop}} \right]  \; ,
\end{align}
where $B_{q_1 q_2 k} (0) = \langle \phi_{\bfq_1} \phi_{\bfq_2} \phi_{\bfk} \rangle |_{\eta \to 0}$ is the boundary value of the bispectrum. So while the coefficients of $\eta^0$ and $\eta^2$ receive divergent loop corrections to their tree-level values (which require renormalisation), the coefficient of $\eta^3$ which has been generated by quantum corrections actually has a fixed \emph{finite} value, determined by $\gamma$.

The fact that it is the third time derivative, $\partial_\eta^3 P$, that is constrained by unitarity seems to be a result of considering a massless scalar field. In particular, using \eqref{eqn:free_eom} and \eqref{eqn:dotphiToPi}, we have that, 
\begin{align}
    \lim_{\eta \to 0} \left\langle \partial_\eta^3 \hat{\phi}_{\bfk} \; \hat{\mathcal{O}} \right\rangle = 2 H^2 \lim_{\eta \to 0} \left\langle  \left(  \tilde{\Pi}_{\bfk} - \text{Im} \left[ \psi^{\text{1-loop} \, \prime}_{\bfk , -\bfk} \right] \hat{\phi}_{\bfk}  \right) \hat{\mathcal{O}} \right\rangle \; .
\end{align}
At tree-level, $\partial_\eta^3 \phi \to  2H^2 \tilde{\Pi}$ as $\eta \to 0$, which comparing with \eqref{eqn:phi_boundary} is the well-known result that $\phi^{(3)}$ plays the role of the momentum conjugate to $\phi^{(0)}$ near the boundary. At 1-loop, $\text{Im} \left[  \psi_2^{\text{1-loop}} \right]$ corrects this relation, effectively mixing some $\phi^{(0)}$ into the conjugate momentum. The cutting rules, which follow from unitarity in the bulk, therefore lead to constraints on the boundary which mix the boundary operator with its shadow (in this case $\phi^{(0)}$ and $\phi^{(3)}$). This is the boundary avatar of bulk unitarity.
It would certainly be interesting to explore this direction further in future.

%%%%%%%%%%%%%%%%
\section{Discussion}
\label{sec:discussion}

In this work, we have derived general Cosmological Cutting Rules for the wavefunction of the universe on FLRW spacetimes, which enforce the constraints of unitarity to each order in perturbation theory. Our results are valid for any number of external legs and to any loop, generalizing previous results obtained in \cite{Goodhew:2020hob,Cespedes:2020xqq} under the banner of the Cosmological Optical Theorem. Our rules take advantage of a set of algebraic relations that rewrite the imaginary part of a product of (bulk-to-bulk) propagators in terms of products of factors with fewer propagators. This reduces the number of nested time integrals that are needed to compute wavefunction coefficient, which are the main computational obstacle in the problem. In words, our rules compute a certain discontinuity of loop diagrams in terms of the discontinuity of diagrams with a lower number of loops. Graphically, our rules consists in noticing that the sum over all possible cuts of a given diagram vanishes. This is very analogous to the flat space cutting rules, but the presence of a boundary term in cosmological (bulk-to-bulk) propagator makes the details quite a bit different.

We also demonstrate how to use the Cosmological Cutting Rules to derive various one-loop corrections to the power spectrum from tree-level results. In particular we consider some simple examples in Minkowski space and then consider the leading cubic and quartic coupling in the Effective Field Theory of Inflation around quasi de Sitter space. In these cases, the discontinuity computed by our rules can be interpreted as the coefficient of a logarithmic corrections to the power spectrum or of its time dependence, depending on the appropriate physical regularization. 

There are a few interesting directions for future research:
\begin{itemize}
    \item We have illustrated the Cosmological Cutting Rules by applying them to the quadratic wavefunction coefficient $\psi_2$, but they can be applied more generally to any non-Gaussian coefficient. Using the general cutting rules presented here to extract information about the one-loop correction to $\psi_4$ would be particularly interesting since in that case unitarity would fix a richer kinematic dependence (in contrast, $\psi_2 \sim k^3$ is fixed by dilations), and in particular could  be combined with a partial wave expansion. This would allow a direct comparison with the existing unitarity bounds on the subhorizon $2 \to 2$ scattering amplitude \cite{Grall:2020tqc}. 
    \item The cutting rules were derived here using unitarity in the bulk, and we have shown that they place constraints on boundary correlators which mix different boundary operators. Further pursuing this connection with the hypothetical conformal field theory at the boundary may shed light on what property of the dual holographic description reproduce unitary dynamics in the bulk. 
    \item In our analysis we restricted to (products of) single discontinuities. But there are additional relations involving multiple discontinuities that can be derived with similar techniques. It would be nice to work these out and study the possible relation to the recently discussed Steinmann relations for the wavefunction \cite{Benincasa:2020aoj}.
    \item Our discussion of the implications of unitarity here and in the previous literature \cite{Goodhew:2020hob,Cespedes:2020xqq} has been perturbative in nature. It would be nice to derive a non-perturbative relation for the full wavefunction of the universe that in perturbation theory reduces to our Cosmological Cutting Rules. This would be useful to derive elastic unitarity bounds and perhaps adapt numerical bootstrap techniques from the amplitude literature (see e.g. \cite{Paulos:2016fap,Guerrieri:2020bto,Guerrieri:2021ivu}.
    \item The analytical structure of cosmological $n$-point function is relatively unexplored. It would be nice to understand what kind of functions can arise in general, for example for a massless scalar or graviton, at tree- and loop level and what their branch points look like. Our results then relates discontinuities found at different perturbative and loop orders.
    \item Finally, the cutting rules for scattering amplitudes play an important role in deriving \emph{positivity bounds}---constraints placed on low-energy EFTs by unitarity (/causality/locality) of the underlying UV completion \cite{Adams:2006sv}. These have recently been developed in a number of ways for scattering amplitudes on Minkowski spacetime \cite{Bellazzini:2016xrt,deRham:2017avq,deRham:2017zjm,Remmen:2020uze,Bellazzini:2020cot,Tolley:2020gtv,Caron-Huot:2020cmc, Li:2021cjv, Arkani-Hamed:2020blm}, and were recently exported to systems without Lorentz boosts in \cite{Grall:2021xxm}. In deriving the Cosmological Cutting Rules, we are now one step closer to realising the same program for cosmological correlators. It would be interesting to now combine the cutting rules presented here with a further study into their analytic structure, developing an analogous set of UV/IR relations which can be used to translate our measurements at the end of inflation into properties of the underlying UV physics.  
\end{itemize}
The cutting rules for amplitudes were derived more than half a century ago while we are only now deriving similar results for cosmological $n$-point functions. There should be many other simple and general results that await discovery in this rather unexplored field. We hope that our progress will bolster the growing interest in this wide open line of research. 

%%%%%%%%%%%%%%%%%%%%%%%%%%%%%%%%%%%%%%%%%%%%%%%

\section*{Acknowledgements}

We would like to thank Harry Goodhew, Mang Hei Gordon Lee, Tanguy Grall, Aaron Hillman, Austin Joyce, Guilherme L. Pimentel, and David Stefanyszyn for useful discussions. 
S.M. is supported by an UKRI Stephen Hawking Fellowship (EP/T017481/1) and partially by STFC consolidated grants ST/P000681/1 and ST/T000694/1.
E.P. has been supported in part by the research program VIDI with Project No. 680-47-535, which is (partly) financed by the Netherlands Organisation for Scientific Research (NWO).

%%%%%%%%%%%%%%%%

\appendix
%%%%%%%%%%%%%%%%
\section{Cutting Rules from the Schr\"{o}dinger Picture}
\label{app:schrodinger}
%%%%%%%%%%%%%%%%

In the main text, we derived cutting rules from properties of the bulk-to-bulk and bulk-to-boundary propagators used in the path integral approach. 
In this Appendix we provide an analogous derivation working directly with the wavefunction in the Schr\"{o}dinger picture. 
This alternative perspective is useful because:
\begin{itemize}

    \item[(i)] It does not refer to any particular diagram, so would be a natural starting point for extending these cutting rules to a non-perturbative statement of unitarity,
    
    \item[(ii)] It extracts from unitarity (namely $\mathcal{H} = \mathcal{H}^{\dagger}$) a number of \emph{conserved quantities}, which can be used to connect the boundary observables at $\eta =0$ with properties of the initial state at $\eta \to - \infty$, 
    
    \item[(iii)] It demonstrates that the propagator identities which we have made use of the main text, rather than being an algebraic accident, are actually inevitable properties of any perturbative solution to the Schr\"{o}dinger equation. 

\end{itemize}

After outlining the general strategy, we will briefly review the tree-level conserved quantities that were derived from unitarity in \cite{Cespedes:2020xqq}. Then we will show how these quantities are corrected at 1-loop.  
In particular, we will show that in a Bunch-Davies initial state these conserved quantities reproduce the cutting rule \eqref{eqn:cut_1loop_2} for the $\text{Disc}$ of $i \psi_2$, in terms of the cubic and quartic wavefunction coefficients, $\psi_3$ and $\psi_4$.

%%%%
\paragraph{Wavefunction Dynamics:}
%%%%
The state of the Universe at time $\eta$ is denoted by $| \Psi_{\eta} \rangle$, with corresponding wavefunction, 
\begin{align}
 \Psi_\eta [\phi ] &= \text{exp} \left( i  \Gamma_\eta [\phi ]   \right) \;\;,  \\
  \Gamma_\eta [\phi]  &=  \int_{\bfk_a} \frac{ i\psi_{\bfk_1 \bfk_2} (\eta) }{2! i}  \phi_{\bfk_1} \phi_{\bfk_2} +  \int_{\bfk_a } \frac{ i \psi_{\bfk_1 \bfk_2 \bfk_3} (\eta) }{3!i } \phi_{\bfk_1} \phi_{\bfk_2} \phi_{\bfk_3}  +   \int_{\bfk_a} \frac{  \psi_{\bfk_1 \bfk_2 \bfk_3 \bfk_4} (\eta) }{4! i}  \phi_{\bfk_1} ... \phi_{\bfk_4}  + ... \nonumber \; . 
\end{align}
In the Schr\"{o}dinger picture, the $\psi_n (\eta)$ coefficients are found by solving their respective equations of motion, using a set of boundary conditions $\psi_n (\eta_0)$ inferred from the initial state $| \Psi_{\eta_0} \rangle$ (e.g. Bunch-Davies at $\eta_0 = \infty$).  
The equation of motion for each non-Gaussian wavefunction coefficient are generated by the functional Schr\"{o}dinger equation,
\begin{align}
- \partial_\eta \Gamma_\eta  = \frac{1}{2 a^{d-1} } \int_{\bfq} \left(   
 \frac{\delta \Gamma_\eta}{\delta \phi_{\bfq}} \frac{\delta \Gamma_\eta}{\delta \phi_{-\bfq}} - i \frac{\delta^2 \Gamma_\eta }{ \delta \phi_{\bfq} \delta \phi_{- \bfq} }
\right) +   \mathcal{H}_{\eta} 
\label{eqn:HJ}
\end{align}
where we have assumed a canonical kinetic term for $\phi$, a conformally flat background spacetime $ds^2 =a^2 (\eta) \left( - d\eta^2 + d\bfx^2\right)$, and $ \mathcal{H}_{\eta} [ \phi ] = \langle \phi | \hat{\mathcal{H}}_{\rm int}   | \Psi_{\eta} \rangle$ is determined by the interaction Hamiltonian, $\hat{\mathcal{H}}_{\rm int} \left[ \hat{\phi} , \hat{\Pi} \right]$, acting on the state.
The second derivative, $\frac{\delta^2 \Gamma_\eta }{ \delta \phi_{\bfq} \delta \phi_{- \bfq} }$, formally diverges (since it requires bringing two operators, $\hat{\Pi}_{\bfx_1}$ and $\hat{\Pi}_{\bfx_2}$, to the same spacetime point), and is responsible for the loop corrections when \eqref{eqn:HJ} is solved perturbatively.

%%%%
\paragraph{Conserved Quantities:}
%%%%
The aim of the game is to manipulate \eqref{eqn:HJ} into the form,
\begin{align}
 \partial_\eta J_\eta =  \mathcal{H}_\eta - \mathcal{H}_\eta^\dagger
\end{align}
where $\mathcal{H}_\eta^\dagger [ \phi ] = \langle \phi |  \hat{\mathcal{H}}_{\rm int}^\dagger  | \Psi_\eta \rangle$. For unitary evolution, $\mathcal{H}_\eta =  \mathcal{H}_\eta^\dagger$, this equation becomes a \emph{consversation law}, $\partial_\eta J_\eta [ \phi ]  = 0$. If we similarly expand the functional $J_\eta [ \phi]$ as,
\begin{align}
J_\eta [ \phi ] = 
\int_{\bfk_a} \frac{ \beta_{\bfk_1 \bfk_2} (\eta) }{2!}  \phi_{\bfk_1} \phi_{\bfk_2} +  \int_{\bfk_a } \frac{ \beta_{\bfk_1 \bfk_2 \bfk_3}  }{3!} \phi_{\bfk_1} \phi_{\bfk_2} \phi_{\bfk_3}  +   \int_{\bfk_a} \frac{ \beta_{\bfk_1 \bfk_2 \bfk_3 \bfk_4}  }{4!}  \phi_{\bfk_1} ... \phi_{\bfk_4} + ... 
\end{align}
then each coefficient $\beta_n$ is constant for any unitary dynamics,
\begin{align}
\text{Unitarity} \;\; \Rightarrow \;\;  \partial_\eta \beta_n = 0 \; . 
\end{align}
This is analogous to how the Hamilton-Jacobi equation is used in classical mechanics to identify constants of motion. 

A procedure for deriving the $\beta_n$ was described in \cite{Cespedes:2020xqq}, where $\beta_3$ and $\beta_4$ were constructed explicitly at tree-level. We will now briefly review that construction, and then extend it to include loops.

%%%%
\subsection*{Tree-Level Constants of Motion}
%%%%

In the context of \eqref{eqn:HJ}, working at ``tree-level'' amounts to discarding the $- i \delta^2 \Gamma_\eta / \delta \phi_{\bfq} \delta \phi_{- \bfq}$ term.  
Taking three or four functional $\delta / \delta \phi$ derivatives of \eqref{eqn:HJ} then produces equations of motion for $\psi_3$ and $\psi_4$,
\begin{align}
\frac{ \partial_\eta \left[ i \psi_{\bfk_1 \bfk_2 \bfk_3} \prod_{a}^3 f_{k_a}^*  \right]}{\prod_{a}^3 f_{k_a}^*}  &= \frac{\delta^3 \mathcal{H}_\eta }{\delta \phi_{\bfk_1} \delta \phi_{\bfk_2} \delta \phi_{\bfk_3}   }  +  \text{loops}    \;\; \label{eqn:psi_eom} \\ 
\frac{ \partial_\eta \left[ i \psi_{\bfk_1 \bfk_2 \bfk_3 \bfk_4} \prod_{a}^4 f_{k_a}^*  \right]}{\prod_{a}^4 f_{k_a}^*}  &= \frac{\delta^4 \mathcal{H}_\eta}{\delta \phi_{\bfk_1} \delta \phi_{\bfk_2} \delta \phi_{\bfk_3} \delta \phi_{\bfk_4} }   -  \sum_{\rm perm.}^3 \frac{1}{ a^{d-1}} \int_{\bfq \bfq'}  \psi_{\bfk_1 \bfk_2  \bfq } \psi_{\bfk_3 \bfk_4 \bfq' }  +   \text{loops}      \;\; \nonumber
\end{align}
where $+ \text{loops}$ denotes the $- i \delta^2 \Gamma_\eta / \delta \phi_{\bfq} \delta \phi_{- \bfq}$ terms we have neglected, and the $f_k^*$ mode functions account for the time-dependence of the field basis, as described in \cite{Cespedes:2020xqq}.  The sum in $\partial_\eta \psi_4$ is over inequivalent permutations of the $\bfk_a$ momenta, effectively producing $s$, $t$ and $u$ exchange contributions.
 
Working at this order, $\mathcal{H}_\eta$ is given by the interaction Hamiltonian with canonical momentum replaced by $- i \psi_2 \phi$, 
\begin{align}
 \mathcal{H}_\eta =  \mathcal{H}_{\rm int} \left[ \phi_{\bfk} , \Pi_{\bfk}  = - i \psi_{\bfk, - \bfk}' \phi_{\bfk}  \right] + \text{loops} \; , 
 \label{eqn:H_on-shell_tree}
\end{align}
where $-i \psi'_{\bfk , -\bfk} = a^{d-1} \partial_\eta f_k^* / f_k^*$ in terms of the free mode function. 
The $\text{Disc}$ operation defined in \eqref{defdisc} has been defined so that $\text{Disc} \left[ f_k^* (\eta) / f_k^* (\eta_0 ) \right] = 0$, and as a consequence,
\begin{align}
 \text{Disc} \left[ \frac{ \delta^n \mathcal{H}_\eta}{\delta \phi_{\bfk_1} \delta \phi_{\bfk_2} ... \delta \phi_{\bfk_n} } \right] =  \frac{\delta^n}{\delta \phi_{\bfk_1} ... \delta \phi_{\bfk_n} } \left[ \mathcal{H}_\eta - \mathcal{H}_\eta^\dagger \right]  \; . 
 \label{eqn:DiscH}
\end{align}

Equations \eqref{eqn:psi_eom} and \eqref{eqn:DiscH} can be immediately combined to give the cubic conserved quantity. For unitary dynamics, in which $\mathcal{H}_\eta - \mathcal{H}^\dagger_\eta = 0$, the $\text{Disc}$ of \eqref{eqn:psi_eom} can be written as,
\begin{align}
\partial_\eta \, \beta_{\bfk_1 \bfk_2 \bfk_3}^{\rm tree}  = 0 \; ,
\end{align}
where, 
\begin{align}
\beta_{\bfk_1 \bfk_2 \bfk_3}^{\rm tree} =  \text{Disc} \left[  i \psi_{\bfk_1 \bfk_2 \bfk_3}  \right] \prod_{a}^3 f_{k_a}^*
\label{eqn:b3}
\end{align}
is the conserved quantity at cubic order at tree-level, where we assume that the overall phase of $f_k^*$ has been chosen so that $\text{Disc} \left[ f_k^* \right] =0$. 

Applying $\text{Disc}$ to the $\partial_\eta \psi_4$ equation of motion does not immediately provide $\beta_4$ because the $\text{Disc} \left[ \psi_3 \times \psi_3 \right]$ is not a total time derivative. 
This is because the basis of $\{ \psi_n \}$ coefficients is not ``diagonal'' in the following sense: if the initial state has $\psi_3 (\eta_0 ) \neq 0$ and $\psi_4 (\eta_0) = 0$, then at later times $\psi_4 (\eta)$ evolves to a non-zero value even in a completely free theory. This mixing can be removed by defining,
\begin{align}
 \tilde{\psi}_{\bfk_1 \bfk_2 \bfk_3 \bfk_4} =  \psi_{\bfk_1 \bfk_2 \bfk_3 \bfk_4} +  \sum_{\rm perm.}^3 \int_{\bfq \bfq'}  \, P_{\bfq \bfq'} \,  \psi_{\bfk_1 \bfk_2 \bfq} \psi_{\bfk_3 \bfk_4 \bfq'} \; , 
\end{align}
which now has the property that setting $\tilde{\psi}_4 (\eta_0 ) = 0$ initially leads to the solution $\tilde{\psi}_4 (\eta) = 0$ for all times in the free theory (irrespective of the initial value of $\psi_3 (\eta_0)$).
The interacting equation of motion \eqref{eqn:psi_eom} for $\tilde{\psi}_4$ is then,
\begin{align}
\frac{ \partial_\eta \left[ i \tilde{\psi}_{\bfk_1 \bfk_2 \bfk_3 \bfk_4} \prod_{a}^4 f_{k_a}^*  \right]}{\prod_{a}^4 f_{k_a}^*}  &= \frac{\delta^4 \mathcal{H}_\eta}{\delta \phi_{\bfk_1} \delta \phi_{\bfk_2} \delta \phi_{\bfk_3} \delta \phi_{\bfk_4} }   - i  \sum_{\rm perm.}^6 \int_{\bfq \bfq'} P_{\bfq \bfq'} \;  \psi_{\bfk_1 \bfk_2 \bfq}  \frac{ \partial_\eta \left[  \psi_{\bfk_3 \bfk_4 \bfq'} f_{k_3}^* f_{k_4}^* f_{q'}^*  \right]   }{ f_{k_3}^* f_{k_4}^* f_{q'}^* }   +   \text{loops}   
\end{align} 
and now the $\text{Disc}$ of the $\psi_3 \partial_\eta \psi_3$ exchange contribution is a total time derivative. For a unitary process, this equation gives,
\begin{align}
\partial_\eta \; \beta_{\bfk_1 \bfk_2 \bfk_3 \bfk_4}^{\text{tree}}  = 0
 \end{align}
where the quartic conserved quantity is,
\begin{align}
\beta_{\bfk_1 \bfk_2 \bfk_3 \bfk_4}^{\text{tree}} = \prod_{a}^4 f_{k_a}^* \Big\{
i \, \text{Disc} \,\left[ i \psi_{\bfk_1 \bfk_2 \bfk_3 \bfk_4} \right] +  \sum_{\rm perm.}^3 \int_{\bfq \bfq'}    \underset{q}{\text{Disc}} \left[ i \psi_{\bfk_1 \bfk_2 \bfq} \right]\; P_{\bfq \bfq'} \;   \underset{q'}{\text{Disc}} \left[ i \psi_{\bfq' \bfk_3 \bfk_4}  \right]
\Bigg\} \; . 
\label{eqn:b4}
\end{align}
Note that for a Bunch-Davies initial state, in which $\beta_4 (\eta_0) = 0$ initially, unitarity requires that this combination of $\text{Disc}$'s vanishes at any later time, $\beta_4 (\eta) = 0$. This reproduces the cutting rule \eqref{eqn:Disc_psi4_full} derived in the main text from the properties of the bulk-to-bulk propagator.  

This is how the tree-level constants of motion \eqref{eqn:b3} and \eqref{eqn:b4} were derived in \cite{Cespedes:2020xqq}. 
We are now going to follow the same procedure but retaining the loop corrections from the $\delta^2 \Gamma / \delta \phi_{\bfq} \delta \phi_{-\bfq}$ term in \eqref{eqn:HJ}. In particular, we will focus on the 1-loop correction to the quadratic coefficient $\psi_2$, and derive the constant of motion $\beta_2^{\text{1-loop}}$.

%%%%
\subsection*{Loop-Level Constants of Motion}
%%%%

We can expand the quadratic wavefunction coefficient as $\psi_{\bfk_1 \bfk_2} = \psi^{\text{tree}}_{\bfk_1 \bfk_2} + \psi_{\bfk_1 \bfk_2}^{\text{1-loop}} + ...$, where $- i \psi^{\text{tree} \, \prime}_{\bfk -\bfk} = a^{d-1} \partial_\eta f_k^* / f_k^*$ is the solution to the Schr\"{o}dinger equation with no $\delta^2 \Gamma / \delta \phi_{\bfq} \phi_{-\bfq}$ term, while $\psi^{\text{1-loop}}_{\bfk_1 \bfk_2}$ is the solution to the equation of motion,    
\begin{align}
\frac{ \partial_\eta \, \left[  i \psi^{\text{1-loop}}_{\bfk_1 \bfk_2} f_{k_1}^* f_{k_2}^*  \right] }{ f_{k_1}^* f_{k_2}^* }  = \frac{\delta^2 \mathcal{H}_\eta }{\delta \phi_{\bfk_1} \phi_{\bfk_2} } - \frac{1}{a^{d-1}} \int_{\bfq} \psi^{\text{tree}}_{\bfk_1, \bfk_2 ,\bfq, -\bfq}  \; . 
 \label{eqn:dc2}
\end{align}
where $\psi^{\rm tree}_4$ satisfies \eqref{eqn:psi_eom} with no loop terms. 
It is not difficult to show that the general solution to this equation can be written in terms of the bulk-to-boundary and bulk-to-bulk propagators of the main text---for example, for a simple $\tfrac{1}{4!} \lambda \phi^4$ interaction in the Lagrangian, the coefficients,
\begin{align}
    \psi^{\text{tree} \, \prime}_{ \bfk_1 \bfk_2 \bfk_3 \bfk_4 } (\eta_0 )  &= + i \lambda \int d \eta \, K_{k_1} (\eta , \eta_0 ) K_{k_2} (\eta, \eta_0) K_{k_3} (\eta , \eta_0) K_{k_4} ( \eta , \eta_0 )  \nonumber \\
    \psi^{\text{1-loop} \; \prime}_{\bfk -\bfk} (\eta_0) &= + \lambda \int d\eta \, K_{k} (\eta , \eta_0) K_{k} (\eta , \eta_0 ) \int_{\bfq} G_q (\eta, \eta, \eta_0 ) \; ,  
\end{align}
satisfy \eqref{eqn:psi_eom} and \eqref{eqn:dc2}, since $\partial_{\eta_0} G_q ( \eta_1 , \eta_2, \eta_0 ) = - a^{1-d} K_k (\eta_1, \eta_0) K_k (\eta_2, \eta_0 )$ (and $\delta^2 \mathcal{H}_{\eta} / \delta \phi^2 = 0$ for this theory since there are no quadratic interactions in $\mathcal{H}_{\rm int}$).
 
The goal is then take the $\text{Disc}$ of \eqref{eqn:dc2}, which effectively removes the interaction term. Note that while there are also loop corrections to the on-shell Hamiltonian \eqref{eqn:H_on-shell_tree}, in particular at one-loop order the canonical momentum should be replaced as $\Pi_{\bfk} = - i \int_{\bfq} \psi_{\bfk \bfq} \phi_{\bfq} - \frac{i}{2} \int_{\bfq_1 \bfq_2} \psi_{\bfk \bfq_1 \bfq_2} \phi_{\bfq_1} \phi_{\bfq_2}$, when $\text{Disc} \left[ \psi^{\text{tree}}_3 \right] = 0$ is fixed by the tree-level $\beta_3$ as in \eqref{eqn:b3} this does not change the fact that $\text{Disc} \left[ \delta^2 \mathcal{H}_\eta / \delta \phi^2 \right]$ vanishes for unitary dynamics.  

However, the $\text{Disc}$ of the $\psi_4^{\text{tree}}$ term in \eqref{eqn:dc2} is not a total derivative, so we have not yet achieved a conservation law. This is because the quartic $\psi_4$ acts as a source for $\psi_2^{\text{1-loop}}$---this means that even in a free theory\footnote{
One may wonder what we mean by $\psi^{\text{1-loop}}_2$ in a free theory. By ``free'', we mean that $\mathcal{H}_{\rm int}=0$ and there are no interactions. By ``one-loop'', in this Appendix we mean next-to-leading order in the small coupling that suppress non-Gaussianities, i.e. we assume that $\psi_n \sim g^{n-2}$ for some small coupling $g$.  
} $\psi_2$ will evolve in time for any initial state with $\psi_4 (\eta_0 ) \neq 0$.  
We can remove this mixing analogously to the $\beta_4^{\text{tree}}$ example above, by defining a new wavefunction coefficient,
\begin{align}
    \tilde{\psi}_{\bfk_1 \bfk_2} = \psi_{\bfk_1 \bfk_2} + \int_{\bfq \bfq'} P_{\bfq \bfq'} \psi_{\bfk_1 \bfk_2 \bfq \bfq'} \; ,
\end{align}
in terms of which \eqref{eqn:dc2} becomes,
\begin{align}
    \frac{\partial_\eta \left[ i \tilde{\psi}^{\text{1-loop}}_{\bfk_1 \bfk_2} f_{k_1}^* f_{k_2}^*  \right] }{ f_{k_1}^* f_{k_2}^* } = \frac{\delta^2 \mathcal{H}_\eta }{ \delta \phi_{\bfk_1} \delta \phi_{\bfk_2} } + i \int_{\bfq \bfq'} P_{\bfq \bfq'} \frac{ \partial_\eta \left[ \psi^{\text{tree}}_{\bfk_1 \bfk_2 \bfq \bfq'} f_{k_1}^* f_{k_2}^* f_{q}^* f_{q'}^* \right] }{ f_{k_1}^* f_{k_2}^* f_{q}^* f_{q'}^* } \; ,
\end{align}
where we have used the following helpful identity \cite{Cespedes:2020xqq},
\begin{align}
    \frac{1}{f^*_q f^*_q } = i a^{d-1} \partial_\eta \left( \frac{f_q}{f_q^*} \right) \; .
\end{align}
Now when we take the $\text{Disc}$ of \eqref{eqn:dc2}, the $\psi_4$ term can be written as a total time derivative using \eqref{eqn:psi_eom}, and therefore we find that the Schr\"{o}dinger equation can be written as,
\begin{align}
\partial_\eta \beta_{\bfk_1 \bfk_2}^{\text{1-loop}} = 0
\label{eqn:b21loop}
\end{align}
for unitary dynamics, where this 1-loop constant of motion is given by, 
\begin{align}
&\frac{ \beta_{\bfk_1 \bfk_2}^{\text{1-loop}} }{ f_{\bfk_1}^* f_{\bfk_2}^*  } = \;\; (-i) \text{Disc} \left[ i \psi_{\bfk_1 \bfk_2}^{\text{1-loop}} \right] + \int_{\bfq \bfq'} P_{\bfq \bfq'} (-i) \underset{q q'}{\text{Disc}} \left[ i \psi^{\text{tree}}_{\bfk_1 \bfk_2 \bfq \bfq'} \right]   \nonumber \\
&\qquad\qquad+   \int_{ \substack{ \bfq_1 \bfq'_1 \\ \bfq_2 \bfq_2'  } } P_{\bfq_1 \bfq'_1} P_{\bfq_2 \bfq_2'} (-i) \underset{q_1 q_2}{\text{Disc}} \left[ i \psi^{\text{tree}}_{\bfk_1 \bfq_1 \bfq_2} \right] (-i) \underset{q_1' q_2'}{\text{Disc}} \left[ i \psi^{\text{tree}}_{\bfk_2 \bfq_1' \bfq_2'} \right] \; . 
\label{eqn:DiscLoopc2}
\end{align}
Since $\beta_2 (\eta_0) = 0$ in the Bunch-Davies initial state, \eqref{eqn:DiscLoopc2} can be used to fix $\text{Disc} \left[ i \psi_2^{\text{1-loop}} \right]$ in terms of $\psi_3^{\text{tree}}$ and $\psi_4^{\rm tree}$. 
This reproduces the combination of both cutting rules \eqref{eqn:cut_1loop_1} and \eqref{eqn:cut_1loop_2}, since it applies to the full $\psi_2$ (which is a sum over both kinds of diagram in general).  

Each of the cutting rules given in Section~\ref{sec:examples} can be derived from the Schr\"{o}dinger picture in this way: by first writing down the Schr\"{o}dinger equation of motion for each $\psi_n$, and then using the $\text{Disc}$ to remove the contribution from $\mathcal{H}_{\rm int}$ (taking care to remove any free-theory mixing which arises between the different wavefunction coefficients). It would be interesting to phrase the general proof of our cutting rules given in Section~\ref{sec:proof} in terms of Schr\"{o}dinger picture dynamics, particularly with regards to formulating a fully non-perturbative unitarity condition.

%%%%%%%%%%%%%%%%
\section{List of Propagator Identities}
\label{app:propagator}
%%%%%%%%%%%%%%%%

In this Appendix we list various identities between the real and imaginary parts of the bulk-to-bulk propagator, $G_{p} (t_1, t_2)$, and the bulk-to-boundary propagator, $K_{p} (t)$. 

%%%%
\subsection{Tree-Level Diagrams}
%%%%
In addition to the relation for a single propagator,
\begin{align}
 \text{Im} \, G_{q} ( t_1, t_2 ) = 2 P_q \text{Im} \, \left[ K_{q} (t_1) \right]  \text{Im} \left[ K_{q} (t_2) \right]
\end{align}
the analogous relation for two propagators is,
\begin{align}
\text{Im} \, \left[ G_{q_1} ( t_1, t_2 ) G_{q_2} ( t_2, t_3 ) \right] &= 
 \sum_{\rm perm.}^2 2 P_{q_1} \text{Im} \left[  K_{q_1} (t_1)    \right] \text{Im} \left[  K_{q_1} (t_2) G_{q_2} (t_2, t_3)   \right]
  \nonumber \\ 
 &- 4 P_{q_1} P_{q_2} \text{Im} \, \left[ K_{q_1} (t_1) \right] \text{Im} \left[ K_{q_1} (t_2) K_{q_2} (t_2)  \right]  \text{Im} \left[ K_{q_2} (t_3) \right] \, .
\end{align}
For three-propagators there are two different diagrams: the four-site chain,
\begin{align}
& \text{Im} \, \left[ G_{q_1} ( t_1, t_2 ) G_{q_2} ( t_2, t_3 ) G_{q_3} (t_3, t_4)    \right] =  \nonumber \\ 
&+ \sum_{\rm perm.}^2 2 P_{q_1}  \text{Im} \left[  K_{q_1} (t_1)   \right]  \text{Im} \left[  K_{q_1} (t_2) G_{q_2} ( t_2, t_3 ) G_{q_3} (t_3, t_4)   \right]  \nonumber \\
&+ 2 P_{q_2} \text{Im} \left[  G_{q_1} (t_1 , t_2) K_{q_2} (t_2) \right] \text{Im} \left[ K_{q_2} (t_2) G_{q_3} (t_3, t_4)   \right] \nonumber \\ 
&- \sum_{\rm perm.}^2 4 P_{q_1} P_{q_2}  \text{Im} \left[  K_{q_1} (t_1)  \right] \text{Im} \left[ K_{q_1} (t_2) K_{q_2} (t_2) \right] \text{Im} \left[  K_{q_2} (t_3) G_{q_3} (t_3, t_4)   \right]  
  \nonumber \\ 
&- 4 P_{q_1} P_{q_3}  \text{Im} \left[  K_{q_1} (t_1)  \right] \text{Im} \left[ K_{q_1} (t_2) G_{q_2} (t_2 , t_3) K_{q_3} (t_3) \right] \text{Im} \left[  K_{q_3} (t_4)  \right]   \nonumber \\
&+8 P_{q_1} P_{q_2} P_{q_3}  \text{Im} \, \left[ K_{q_1} (t_1) \right] \text{Im} \left[ K_{q_1} (t_2) K_{q_2} (t_2)  \right]  \text{Im} \left[ K_{q_2} (t_3) K_{q_3} (t_3) \right] \text{Im} \left[ K_{q_3} (t_4) \right]
\end{align}
and the ``flux-capacitor'',
\begin{align}
& \text{Im} \, \left[ G_{q_1} ( t_1, t_4 ) G_{q_2} ( t_2, t_4 ) G_{q_3} (t_3, t_4)    \right] =  \nonumber \\
&+ \sum_{\rm perm.}^3 2 P_{q_1}  \text{Im} \left[  K_{q_1} (t_1)   \right] \text{Im} \left[  K_{q_1} (t_4)  G_{q_2} ( t_2, t_4 ) G_{q_3} (t_3, t_4)    \right]  \nonumber \\ 
&- \sum_{\rm perm.}^3 4 P_{q_1} P_{q_2}  \text{Im} \left[  K_{q_1} (t_1)  \right] \text{Im} \left[ K_{q_2} (t_2)  \right] \text{Im} \left[ K_{q_1} (t_4) K_{q_2} (t_4) G_{q_3} (t_3, t_4)  \right] 
  \nonumber \\ 
&+ 8 P_{q_1} P_{q_2} P_{q_3}  \text{Im} \, \left[ K_{q_1} (t_1) \right] \text{Im} \left[  K_{q_2} (t_2)  \right]  \text{Im} \left[ K_{q_3} (t_3) \right] \text{Im} \left[ K_{q_1} (t_4) K_{q_2} (t_4) K_{q_3} (t_4)  \right] \; . 
\label{eqn:GGGcutTreeb}
\end{align}

%%%%
\subsection{Single-Loop Diagrams}
%%%%
For the real part of a loop with one, two, three or four propagators, one can use:
\begin{align}
&2 \text{Re} \, \left[ G_{q} ( t , t ) \right] = 2 P_{q}  \text{Im} \left[   K_{q} (t) K_{q} (t) \right]    \label{eqn:Gcut} 
\end{align}

\begin{align}
2 \text{Re} \, \left[ G_{q_1} ( t_1, t_2 ) G_{q_2} ( t_2, t_1 ) \right]     
&= \sum_{\rm perm.}^2 2 P_{q_1} \,  \text{Im} \left[   K_{q_1} (t_2) G_{q_2} ( t_2, t_1 ) K_{q_1} (t_1)   \right]      \nonumber \\ 
&- 4 P_{q_1} P_{q_2}  \text{Im} \left[  K_{q_1} (t_1 ) K_{q_2} (t_1)  \right]  \text{Im} \left[ K_{q_2} (t_2 ) K_{q_1} (t_2)  \right] 
\label{eqn:GGcut}
\end{align}

\begin{align}
& 2 \text{Re} \left[  G_{q_1} (t_1, t_2) G_{q_2} (t_2, t_3) G_{q_3} (t_3, t_1 )  \right]  \label{eqn:GGGcut} \\
&= \sum_{\rm perm.}^3 \,2 P_{q_1} \text{Im} \left[ K_{q_1} (t_2)  G_{q_2} (t_2, t_3) G_{q_3} (t_3, t_1 )   K_{q_1} (t_1)   \right]    \nonumber \\
&- \sum_{\rm perm.}^3 \, 4 P_{q_1} P_{q_2} \, \text{Im} \left[ K_{q_1} (t_2) K_{q_2} (t_2)  \right]         \text{Im} \left[  K_{q_2} (t_3) G_{q_3} (t_3, t_1) K_{q_1} (t_1)   \right]  \nonumber  \\
&+ 8 P_{q_1} P_{q_2} P_{q_3} \, \text{Im} \left[ K_{q_1} ( t_2 ) K_{q_2} (t_2) \right] \text{Im} \left[ K_{q_2} ( t_3 ) K_{q_3} (t_3) \right]    \text{Im} \left[ K_{q_3} ( t_1 ) K_{q_1} (t_1) \right]\nonumber 
\end{align}

\begin{align}
& 2 \text{Re} \left[  G_{q_1} (t_1, t_2) G_{q_2} (t_2, t_3) G_{q_3} (t_3, t_4 ) G_{q_4} (t_4, t_1 )  \right]  \label{eqn:GGGGcut} \\
&= \sum_{\rm perm.}^4 \, 2 P_{q_1}   \text{Im} \left[  K_{q_1} (t_1) K_{q_1} (t_2)  G_{q_2} (t_2, t_3) G_{q_3} (t_3, t_4 ) G_{q_4} (t_4, t_1 )  \right]    \nonumber \\
&- \sum_{\rm perm.}^4 4 P_{q_1} P_{q_2} \, \text{Im} \left[ K_{q_1} (t_2) K_{q_2} (t_2)  \right] \text{Im} \left[  K_{q_2} (t_3)  G_{q_3} (t_3, t_4)  G_{q_4} (t_4, t_1) K_{q_1} (t_1)  \right]          \nonumber  \\
&-\sum_{\rm perm.}^2  4 P_{q_1} P_{q_3} \text{Im} \left[ K_{q_1} (t_2) G_{q_2} (t_2, t_3) K_{q_3} (t_3)  \right]  \text{Im} \left[  K_{q_3} (t_4)  G_{q_4} (t_4, t_1) K_{q_1} (t_1)  \right] \nonumber \\
&+ \sum_{\rm perm.}^4 \, 8 P_{q_1} P_{q_2} P_{q_3} \text{Im} \left[ K_{q_1} ( t_2 ) K_{q_2} (t_2) \right] \text{Im} \left[ K_{q_2} ( t_3 ) K_{q_3} (t_3) \right]  \left[ K_{q_3} ( t_4 ) G_{q_4} (t_4,t_1 ) K_{q_1} (t_1) \right]   \nonumber \\ 
&- 16 P_{q_1} P_{q_2} P_{q_3} P_{q_4} \text{Im} \left[ K_{q_1} ( t_2 ) K_{q_2} (t_2) \right] \text{Im} \left[ K_{q_2} ( t_3 ) K_{q_3} (t_3) \right]  \left[ K_{q_3} ( t_4 ) K_{q_4} (t_4) \right]  \left[ K_{q_4} ( t_1 ) K_{q_1} (t_1) \right]  \nonumber 
\end{align}

%%%%%%%%%%%%%%%%
\section{Explicit One-Loop Computation for $\dot \pi^3$}
\label{app:pi3_1loop}
%%%%%%%%%%%%%%%%

In this Appendix, we describe the computation of the one-loop diagram $\psi^{(a)}_{\bfk_1 \bfk_2}$ from two $\dot \pi^3/3!$ vertices\footnote{As compared to the main text, here we are using a different normalization of the vertex that removes the combinatorial factor $3!$.}. This corresponds to the integral, 
\begin{align}
\psi^{(a) \, \prime }_{\bfk, -\bfk}  
&=  \frac{C_{\dot \phi^3}^2}{H^2} \int_{-\infty}^{0} \frac{d \eta_1}{\eta_1}  \int_{-\infty}^{0} \frac{d\eta_2}{\eta_2} \,  \int_{\bfp_1 \bfp_2} \tdelta^3 \left( \bfp_1 + \bfp_2 - \bfk \right) 
 K'_{k} ( \eta_1 ) K'_{k} (\eta_2 )  G^{(1,1)}_{p_1} ( \eta_1 ,  \eta_2 ) 
 G^{(1,1)}_{p_2} ( \eta_1 ,  \eta_2 )   
\end{align}
where $ K'_k (\eta) = \partial_\eta K_k (\eta)$ and $ G_p^{(1,1)} (\eta_1, \eta_2) = \partial_{\eta_1} \partial_{\eta_2} G_{p} ( \eta_1 ,  \eta_2 )$ are the time derivatives of the propagators \eqref{eqn:deSitterPropagators}. Since this diagram is symmetric in $\eta_1 \leftrightarrow  \eta_2$, we can order $\eta_1 > \eta_2$ and write this as,
\begin{align}
 \psi^{(a) \; \prime }_{\bfk, -\bfk} &=  2 H^2 C_{\dot \phi^3}^2 \int_{-\infty}^{0} \frac{d \eta_1}{\eta_1}  \int_{-\infty}^{\eta_1} \frac{d\eta_2}{\eta_2} \,  \int \frac{d^3 \bfp_1 d^3 \bfp_2 }{(2\pi)^3} \, \tdelta^3 \left( \bfp_1 + \bfp_2 - \bfk \right)   \;  E ( k , p_1, p_2 , \eta_1 , \eta_2)
 \label{eqn:loop_integral}
\end{align}
where $E(k, p_1 , p_2 , \eta_1, \eta_2 ) = k^4 p_1 p_2 \eta_1^2 \eta_2^2  e^{i ( k + p_1 + p_2) \eta_2} e^{i k \eta_1} \sin ( p_1 \eta_1 ) \sin (p_2 \eta_1 )$. 
This integral is divergent, and requires regularisation. 

We will use dimensional regularisation, analytically continuing to $d = 3 + \delta$ dimensions where the integral is formally finite. Unlike for amplitudes on Minkowski spacetime, where this procedure is more or less unique, on de Sitter spacetime care must be taken with specifying precisely how the propagators are to be continued to $d$-dimensions. This subtlety arises because, unlike $e^{\pm i p_\mu x^\mu}$ on Minkowski (which is a good mode function for any spacetime dimensions), the de Sitter mode functions in $d$ dimensions are,
\begin{align}
    f_k  (\eta ) \propto  (-H \eta)^{d/2} H_{\nu}^{(1)} ( - k  \eta )
    \label{eqn:mode_d}
\end{align}
and its complex conjugate, where $H_\nu^{(1)}$ is a Hankel function (of the first kind) and $\nu = \sqrt{ (d/2)^2 - (m/H)^2 }$. Using these mode functions leads to $d$-dependent propagators. 

We will first compute \eqref{eqn:loop_integral} by analytically continuing to $d$ dimensions with the mode functions held fixed (i.e. kept at their $3$-dimensional value \eqref{eqn:ModeMassless}), and then compute it in a scheme which also analytically continues the mode functions. This will result in expressions of the form \eqref{eqn:psi_logk} and \eqref{eqn:psi_logk_2} respectively.

%%%%
\paragraph{$3$-dimensional Mode Functions:}
%%%%
The simplest scheme in which to evaluate \eqref{eqn:loop_integral} is one in which the propagators retain their $3$-dimensional form \eqref{eqn:deSitterPropagators}, and only the momentum integration measures are analytically continued. In this case, the time integrals can be performed straightforwardly as if in $3$ dimensions, giving
\begin{align}
&\int_{-\infty}^{0} \frac{d \eta_1}{\eta_1}  \int_{-\infty}^{\eta_1} \frac{d\eta_2}{\eta_2} \; E ( k , p_1, p_2 , \eta_1 , \eta_2) \nonumber \\ 
&= k^4 p_1 p_2\left( 
     F ( +p_1 +p_2 ) -
     F ( + p_1 - p_2) - F ( -p_1 +p_2 ) +
     F( -p_1 -p_2 )
     \right)
     \label{eqn:Eint}
\end{align}
where, 
\begin{align}
F ( q ) = - \frac{ (k+q)^2 + 5 (k+q) ( k + p_1 + p_2 ) + 10 (k + p_1 + p_2 )^2 }{ (k + p_1 + p_2)^3 ( 2 k + p_1 + p_2 + q )^5 } \; . 
\label{eqn:Fdef}
\end{align}
Now we must integrate this over $\bfp_1$ and $\bfp_2$. The $d$-dimensional integration measure can be written as,
\begin{align}
\int d^d \bfp_1 d^d \bfp_2 \, \delta^d \left( \bfp_1 + \bfp_2 - \bfk \right) \; f ( k ,p_1, p_2) = \frac{S_{d-2}}{2} \int_{k}^{\infty} d p_+ \int_{-k}^{+k} d p_-  \frac{p_1^{d-2} p_2}{k}  \; f ( k ,p_1, p_2)
\label{eqn:dp1dp2_d}
\end{align}
where $p_{\pm} = p_1 \pm p_2$, and $S_{d-2}$ is the surface area of the $(d-2)$-dimensional unit sphere (i.e. $S_1 = 2\pi$). In fact, even before performing these two integrals we can already see the qualitative form of the solution. If we define $\hat{p}_1 = p_1/k$ and $\hat{p}_2 = p_2/k$ (and $\hat{p}_+ = \hat{p}_1 + \hat{p}_2$), then we have,
\begin{align}
     \psi^{(a) \; \prime }_{\bfk, -\bfk} = H^2 C_{\dot \phi^3}^2 \;  \frac{  S_{1+\delta}}{(2\pi)^{3+\delta} } \; k^{3+\delta} \; I ( \delta)
     \label{eqn:psi_I}
\end{align}
where $I ( \delta)$ is the dimensionless integral,
\begin{align}
    I (\delta) = \int_{1}^{\infty} d \hat{p}_+ \int_{-1}^{+1} d \hat{p}_-  \hat{p}_1^{2+\delta} \hat{p}_2^2 \left( 
     \hat{F} (  \hat{p}_1 + \hat{p}_2 ) -
     \hat{F} (  \hat{p}_1 - \hat{p}_2) - \hat{F} ( -\hat{p}_1 +\hat{p}_2 ) +
     \hat{F} ( -\hat{p}_1 -\hat{p}_2 )
     \right)
     \label{eqn:Idef}
\end{align}
and $\hat{F}$ is given by \eqref{eqn:Fdef} with $p_{1,2}$ replaced  by $\hat{p}_{1,2}$ and an overall $k^{-6}$ extracted.
Focussing on only the divergent terms as $\delta \to 0$,
\begin{align}
   I (\delta) &=  \int_1^{\infty} \frac{d \hat{p}_+ \, \hat{p}_+^{d-3}  }{ ( \hat{p}_+  + 1)^6 ( \hat{p}_+ + 3 )^4 } \left( 
\frac{3}{16} \hat{p}_+^{13} + \frac{27}{8} \hat{p}_+^{12}  +
\frac{105}{4} \hat{p}_+^{11} + \frac{921}{8} \hat{p}_+^{10}  + \frac{74 449}{240} \hat{p}_+^9   \right)  + \text{finite} \nonumber \\ 
   &= -\frac{1}{30 \, \delta}  + \mathcal{O} \left( \delta^0 \right) \; . 
\end{align}
This produces,
\begin{align}
     \psi^{(a) \; \prime }_{\bfk, -\bfk} =  -   H^2 \frac{ k^3 }{ 16 \pi^2} \left( 
     \frac{2}{15} C_{\dot \phi^3}^2 \left( \frac{1}{\delta} + \log (k)  \right)  + \text{local}
     \right) \; ,
     \label{eqn:loop_explicit_1}
\end{align}
which, up to the different normalization of the coupling by the combinatorial factor, coincides with \eqref{eqn:psi_logk}, and matches the $C_{\dot \phi^3}^2$ part of $\gamma$ which was inferred from the cutting rules in Section~\ref{sec:deSitter} (see equation \eqref{eqn:Disca_eg3}). 

While this scheme is computationally very simple, one may worry that the $d$ dimensional quantity which we have computed is in fact \emph{not} the wavefunction coefficient of any scalar field theory (it is not a solution of the $d$-dimensional Schr\"{o}dinger equation, since we did not use the $d$-dimensional propagators). Rather, it corresponds to a purely formal manipulation of the integral \eqref{eqn:loop_integral}. 
We are therefore going to consider a second scheme, which also analytically continues the mode functions in such a way that the $d$ dimensional integral that we perform is genuinely computing a wavefunction coefficient of a scalar field in $d$ spacetime dimensions. We will see that in this second scheme, the $\log (k)$ which appears in \eqref{eqn:loop_explicit_1} is absent.

%%%%
\paragraph{$d$-dimensional Mode Functions:}
%%%%
Rather than consider a massless scalar field in $d$ dimensions, for which the mode function \eqref{eqn:mode_d} contains the general Hankel function $H^{(1)}_{d/2} (-k\eta)$, we will instead consider a scalar field of mass $m^2 = H^2 (d^2-9)/4$. This approaches the massless scalar when $d \to 3$, and has the simpler mode function,
\begin{align}
    f_k (\eta) = (- H \eta )^{\delta/2} \; \frac{H ( 1 + i k \eta )}{ k } \frac{ e^{-i k \eta} }{\sqrt{2k}} 
\end{align}
which differs from the $d=3$ mode function for an $m=0$ field only by an overall normalisation of $(-H \eta)^{\delta/2}$. Similarly, de Sitter invariance requires that the interaction vertex be analytically continued to $C_{\dot \phi^3} ( -H \eta)^{-d+2} \left( \phi' \right)^3$. In this scheme, the integral \eqref{eqn:loop_integral} is therefore analytically continued to, 
\begin{align}
 \psi^{(a) \; \prime }_{\bfk, -\bfk} &=  2 H^2 C_{\dot \phi^3}^2 \int_{-\infty}^{0} \frac{d \eta_1}{\eta_1}  \int_{-\infty}^{\eta_1} \frac{d\eta_2}{\eta_2} \,  \int \frac{d^d \bfp_1 d^d \bfp_2 }{(2\pi)^3} \, \tdelta^d \left( \bfp_1 + \bfp_2 - \bfk \right)   \; ( H^2 \eta_1 \eta_2)^{\delta/2} E ( k , p_1, p_2 , \eta_1 , \eta_2)
\end{align}
which differs from the previous scheme by a factor of $(H^2 \eta_1 \eta_2)^{\delta/2}$ in the integrand (note that the propagators also contain additional $\mathcal{O} (\delta)$ suppressed terms which we have neglected since they do not contribute to the divergence). 

Now following the same steps as before, we first perform the $d\eta_2$ and $d\eta_1$ integrals, which gives, 
\begin{align}
&\int_{-\infty}^{0} \frac{d \eta_1}{\eta_1}  \int_{-\infty}^{\eta_1} \frac{d\eta_2}{\eta_2} \; (H^2 \eta_1 \eta_2)^{\delta/2} E ( k , p_1, p_2 , \eta_1 , \eta_2) \nonumber \\ 
&= \frac{ H^\delta k^4 p_1 p_2}{( -i (k + p_1 + p_2 )  )^{\delta}} \left( 
     F ( +p_1 +p_2 ) -
     F ( +p_1 - p_2) - F ( -p_1 +p_2 ) +
     F( -p_1 -p_2 )
     \right) 
\end{align}
where $F(q)$ is given in \eqref{eqn:Fdef}, and again we have discarded $\mathcal{O} ( \delta )$ terms which will not contribute to the divergence. Note that this differs from \eqref{eqn:Eint} by an overall factor of $( -i (k + p_1 + p_2 ) / H )^{-\delta}$.  

Finally, we must perform the $d^d \bfp_1 d^d \bfp_2$ integrals. We can again use \eqref{eqn:dp1dp2_d} to write this in terms of a dimensionless $\int_1^{\infty} d \hat{p}_+ \int_{-1}^{+1} d \hat{p}_-$ integral,
\begin{align}
     \psi^{(a) \; \prime }_{\bfk, -\bfk} = H^2 C_{\dot \phi^3}^2 \;  \frac{  S_{1+\delta}}{(2\pi)^{3+\delta} } \; k^{3} \; \left(  \changg{-}i H \right)^\delta \; I ( \delta)
\end{align}
where we now see that the effect of the additional $(H^2 \eta_1 \eta_2 )^{\delta/2}$ factor in the integrand is to introduce a factor of $\left(  i H / k \right)^\delta$ compared with \eqref{eqn:psi_I}.
Note that, up to $\mathcal{O} (\delta^0)$ finite terms, this $I(\delta)$ coincides with \eqref{eqn:Idef} and in particular shares the divergence $I(\delta) = -1/30 \delta + \mathcal{O}(\delta)$. 
Therefore we arrive at,
\begin{align}
     \psi^{(a) \; \prime }_{\bfk, -\bfk} =     H^2 \frac{ k^3 }{ 16 \pi^2} \left( 
     \frac{2}{15} C_{\dot \phi^3}^2 \left(\changg{-} \frac{1}{\delta} + \log ( i H )  \right)  + \text{local}
     \right) \; ,
\end{align}
in this scheme, which agrees with \eqref{eqn:psi_logk_2}, and again successfully reproduces the $C_{\dot \phi^3}^2$ part of $\gamma$, up to the different vertex normalization, which was inferred from the cutting rules in \eqref{eqn:Disca_eg3}. 
This also agrees with the similar calculation performed in \cite{Senatore:2009cf} of the equal-time in-in correlator. A detailed discussion of this calculation and of the associated renormalization in a variety of schemes can be found in \cite{Jain:2025maa}.

Finally, we note that in this scheme it is crucial to perform the time integrals over $d\eta_1$ and $ d\eta_2$ \emph{before} taking the limit $\delta \to 0$. In particular, had one expanded $(H^2 \eta_1 \eta_2 )^{\delta/2} = 1 + \frac{\delta}{2} \log \left( H^2 \eta_1 \eta_2 \right) + ...$ inside the integral, one would have found that there is a logarithmic boundary divergence, $\lim_{\eta \to 0} \log \left( -H \eta \right)$, as the late-time boundary is approached. In other words, in this scheme, the dimensional regularisation is regulating \emph{both} the $p \to \infty$ UV divergence from the loop integral \emph{and} the $\eta \to 0$ boundary divergence at late times\footnote{
These boundary divergences should not be confused with the IR (secular) divergences that can appear in equal-time correlators \cite{Starobinsky:1986fx, Gorbenko:2019rza, Mirbabayi:2019qtx, Baumgart:2019clc}---the wavefunction coefficients never contain these, they arise only when performing the field average $\int \mathcal{D} \phi | \Psi [ \phi ] |^2$ over field configurations on a fixed $\eta$ hypersurface.
}. These boundary divergences can arise in the non-Gaussian coefficients even at tree-level, and were systematically studied in \cite{Cespedes:2020xqq} for the cubic wavefunction coefficient (see also \cite{Bzowski:2015pba}).

%%%%%%%%%%%%%%%%%%%%%%%%%%%%%%%%
\bibliographystyle{JHEP}
\bibliography{references}
%%%%%%%%%%%%%%%%%%%%%%%%%%%%%%%%

\end{document}